\documentclass[fleqn,usenatbib]{mnras}

\usepackage{graphicx}
\usepackage{microtype}
\usepackage{amsmath,amssymb}
\usepackage{lmodern}
\usepackage{listlbls}
\usepackage{array} 
\usepackage{stfloats}

\title[Shape shifting light dark matter solitons]{Shape Shifting Light Dark Matter Solitons}

\author[D. Ben-Amotz]{Dor Ben-Amotz\thanks{E-mail: dorbenamotz@bennington.edu}
\\
Bennington College, North Bennington, VT 05257, USA
}

\makeatletter
\def\ps@titlepage{%
  \def\@oddhead{}%
  \def\@oddfoot{\hfil\thepage\hfil}%
}
\def\@oddhead{}
\def\@evenhead{}
\makeatother
\begin{document}
\pagestyle{plain}
\maketitle

\begin{abstract}
Dark matter consisting of a Bose-Einstein condensate (BEC) of ultralight particles forms solitons whose cored shape becomes increasingly cusped under the influence of a central point mass, such as a supermassive black hole. Here we present a unified analytic description of the resulting shape changes as a function of soliton mass fraction, spanning the hydrogenic to self-gravitating soliton limits. Solutions of the Schr\"{o}dinger-Poisson equation are expressed as a sum of five Gaussians with numerically optimised coefficients, yielding closed-form expressions for soliton shape-dependent properties. Moreover, new mass-fraction-dependent scaling relations are used to approximate soliton size, density, and total mass directly in terms of observed stellar velocity dispersion and half-light radius. Applications to dwarf spheroidal (dSph) and ultra-faint dwarf (UFD) galaxies -- validated using a spherical-isotropic Jeans analysis -- show that the observed stellar density, velocity and enclosed mass are consistent either with dSph and UFD galaxies having different ultralight dark matter particle masses and no black holes, or with a single universal ultralight dark matter particle mass, requiring the presence of supermassive black holes in many UFD galaxies. These results, combined with a more detailed analysis of the radially resolved stellar velocity dispersions of Draco (dSph) and Segue~I (UFD), are found to be consistent with a universal ultralight dark matter particle mass of $m_0 \approx 1.5\times 10^{-22}$\,eV/c$^2$, to within a factor of~2. The results demonstrate the utility of soliton shape-shifting predictions in constraining dwarf galaxy dark matter profiles and revealing the possible presence of central black holes.
\end{abstract}

%\listoflabels


\section{Introduction}\label{S:Intro}

Interest in dark matter models consisting of a Bose--Einstein condensate (BEC) of scalar 
bosonic particles (often referred to as ``ultralight'', ``wave'', or ``fuzzy'' dark matter 
models) stems in large part from the cored (flat-top) shapes of the predicted dark matter 
``soliton'' density profiles \citep{EbeUlt25,MatSho24,PozDwa24,NieSma20,HuFuz20,BarGal18,
SchCos14,Chavanis25}. Specifically, these predictions better reproduce observed galactic 
velocity dispersions \citep{PozDwa24,PozDet24,RelDar19,MooEvi94} and formation rates 
\citep{EbeUlt25,MatSho24,SchCos14,KauFor93} than the cusped (pointed-top) density profiles 
predicted by canonical cold dark matter (CDM) simulations \citep{HuFuz20,NavStr96}. 
However, definitively distinguishing cored from cusped dark matter profiles remains 
exceptionally challenging \citep{ChaDar21}, particularly given the possible presence of 
supermassive black holes, which can significantly reshape soliton density profiles 
\citep{DavFuz20,Tan25,ChaMas19,KorVor25,Lujan25,Vegetti26,Oguri26_arXiv} and whose 
occurrence in low-mass galaxies is increasingly well supported by large observational 
surveys \citep{PucTri25}. Moreover, evolving tensions and bounds pertaining to the mass 
of dark matter particles $m_0$ and black holes $M_\bullet$ add further uncertainty 
\citep{PozDwa24,GolVia22,RogStr21,SipFuz25,MayUpd25,LiuWar25,Aditya26}. The present work 
contributes to addressing these challenges using the following combined theoretical and 
observational modelling strategy.

The shape-shifting behaviour described here is consistent with previous numerical 
predictions of black-hole-induced soliton shape changes \citep{DavFuz20,Tan25}. Here 
these predictions are extended by deriving accurate closed-form expressions for soliton 
shape as a function of soliton mass fraction \(F\), spanning the self-gravitating 
(\(F=1\)) to hydrogenic (\(F=0\)) limits. Specifically, solutions of the 
Schr\"odinger--Poisson equation are expressed as analytic five-Gaussian (5G) 
shape-shifting functions and used to obtain new $F$-dependent scaling relations, consistent 
with the virial theorem, for the soliton width, velocity dispersion, and binding energy 
(as well as other observable properties). These scaling relations complement earlier 
approximate treatments in which the black hole's influence is approximated by rescaling the 
width of a fixed-shape soliton profile, represented as either a single Gaussian 
\citep{ChaMas19} or a self-gravitating (\(F=1\)) Schive profile 
\citep{SchCos14,Oguri26_arXiv}.

Moreover, previously proposed relations between the velocity dispersion $\sigma$ and 
the mass $M_{1/2}$ enclosed within the half-light radius $R_{1/2}$ \citep{LazAcc20,
WolAcc10} are extended and applied to express soliton size, density, and total mass in 
terms of $m_0$, $M_\bullet$ and $F$. 
A key advantage of the present strategy is that analytical (rather than numerical) $F$-dependent soliton 
shape predictions facilitate efficient comparisons and fits to galactic observations and 
simulations. Such applications are illustrated using 23 dwarf spheroidal (dSph) and 25
ultra-faint dwarf (UFD) galaxy observations. 

The resulting predictions are validated and constrained using 
a spherical-isotropic Jeans analysis to compute aperture-weighted and radially binned 
stellar velocity dispersions for direct comparison with dSph and UFD observations. 
The results are used to test alternative interpretations of the recently discovered 
correlations between stellar core density and radius in dSph and UFD galaxies 
\citep{PozDwa24}. The present predictions are shown to be consistent with the 
interpretation proposed by \citet{PozDwa24}, suggesting that dSph and UFD galaxies 
have different ultralight dark matter particle masses. Alternatively, the present results 
suggest a new interpretation in which all galaxies share the same ultralight 
dark matter particle mass $m_0$, implying that many UFD galaxies contain massive black 
holes, consistent with recently proposed $M_\bullet$ mass bounds \citep{Aditya26} and 
radially resolved velocity dispersion observations of the UFD Segue~I galaxy 
\citep{Simon11,Lujan25}. The latter universal-$m_0$ interpretation implies that 
$m_0 \sim 1.5\times 10^{-22}$\,eV/c$^2$, to within about a factor of two.

The present strategy is confined to predictions pertaining to ground-state BEC solitons, 
and therefore does not treat the excited states that give rise to surrounding NFW-like 
halo tails \citep{SchCos14,PozDet24,LiaDec25,NavStr96}. More specifically, the present 
predictions pertain to galactic models consisting of a dark matter BEC soliton surrounding 
a central point mass, and focus on the internal structure of the soliton core. The 
framework is most directly applicable to dark-matter-dominated dwarf galaxies, including 
those that may host a central black hole \citep{Aditya26,Vegetti26,Oguri26_arXiv,BusDyn21,
ReiHun22,Reines13}, whose stellar distributions are approximately spherical and isotropic, 
largely embedded within the soliton core, and only weakly affected by any outer halo 
component.

The remainder of this paper is organized as follows. Section~\ref{S:SP} introduces the 
coupled dark matter and point-mass Hamiltonian, and defines the associated mass, energy, 
and velocity scaling relations. Section \ref{S:Jeans} describes the Jeans analysis used to predict
stellar velocity dispersion, for direct comparison with aperture weighted and binned velocity measurements.
Section~\ref{S:Soliton1f} presents the properties of 
self-gravitating solitons and compares them with previous analytic approximations and with 
Fornax (dSph) observations. 
  Section~\ref{S:SolitonXf} describes soliton shape changes as 
a function of dark matter mass fraction and derives the resulting relations between dark 
matter particle mass, soliton size, density, total mass, and velocity dispersion. 
Section~\ref{S:dSphUFD} applies the framework to dark-matter-dominated dSph and UFD 
galaxies, leading to alternative interpretations of the observed core density--radius 
scaling relations. The results are summarised and discussed in Section~\ref{S:Discussion}. 
The Appendix provides additional theoretical details, including virial-theorem constraints, 
a description of the numerical optimisation procedures used to obtain analytical (five-Gaussian superposition)
solutions of the Schr\"{o}dinger--Poisson equation, as well as tabulations of the 
resulting Gaussian coefficients, polynomial expansions, and galactic modelling predictions.
\section{Schr\"{o}dinger--Poisson Dark Matter Solitons}\label{S:SP}

The Schr\"{o}dinger equation may be solved to obtain the wavefunctions $\Psi$ and energies $E$ of any system, given its Hamiltonian $\hat{H} = \hat{K}+ \hat{V}$, where $\hat{K}$ and $\hat{V}$ are the kinetic and potential energy operators.
\begin{equation}\label{E:Sch}
\hat{H}\Psi = \left[\hat{K}+ \hat{V}\right]\Psi = E \Psi
\end{equation}
When applied to a system consisting of many low-mass dark matter Bose particles in their ground BEC state, the solution of Eq.~\ref{E:Sch} may be obtained using the Hartree mean-field approximation, which neglects the influence of dark matter particle exchange symmetry. This approximation has been shown to be accurate for Bose particles in the thermodynamic, low-temperature limit that is appropriate for dark matter BEC condensates corresponding to the ground state of a very large number of very low-mass Bose particles \citep{ChaQua11}.

Here we consider model galaxies consisting of a spherically symmetric gravitational potential energy $V(r)$ and a total mass $M$ that may in general contain both a dark matter soliton of mass $M_S=F M$ and a central point mass $M_\bullet=(1-F)M$, where $F$ is the soliton mass fraction. The resulting spherically symmetric soliton wavefunction $\Psi(r)$ and probability density $\rho(r) = |\Psi(r)|^2$ are obtained as ground-state solutions of the following Schr\"{o}dinger--Poisson equation, expressed in terms of the dimensionless distance and energy variables $x = r/a_0$ and $\epsilon = E/\epsilon_0$ (as further described below and in Appendix~\ref{app:A}).

\begin{equation}\label{E:SP}
\left[-\frac{1}{x}\frac{\partial^2}{\partial x^2} x +\frac{V(x)}{\epsilon_0}\right]\Psi(x) = \epsilon\Psi(x)
%= \frac{E}{\epsilon_0}
\end{equation}
The length and energy scaling constants $a_0$ and $\epsilon_0$ are functions of $M$ and $m_0$.
\begin{equation}\nonumber
a_0 \equiv \frac{\hbar^2}{G M m_0^2} = \frac{\hbar}{\sqrt{2 m_0 \epsilon_0}}
\end{equation}
\begin{equation}\label{E:a0e0}
\epsilon_0 \equiv  \frac{\hbar^2}{2 m_0 a_0^2} = \frac{G M m_0}{2 a_0} = \frac{m_0^3}{2}\left(\frac{G M}{\hbar}\right)^2
%= \frac{1}{2} m_0 v_0^2
\end{equation}
These expressions may be rearranged in various ways to, for example, obtain the following three equivalent expressions for the dark matter particle mass $m_0$.
\begin{equation}\label{E:m0}
m_0 = \frac{\hbar}{\sqrt{GM a_0 }}= \frac{\hbar^2}{2 \epsilon_0 a_0^2} = \frac{\hbar}{a_0 v_0}
\end{equation}
The third equality introduces the velocity scaling factor, $v_0$, in terms of which the energy scaling factor may be expressed as $\epsilon_0 = \frac{1}{2} m_0 v_0^2$.
\begin{equation}\label{E:v0}
v_0 = \sqrt{\frac{2 \epsilon_0}{m_0}}= \frac{G M m_0}{\hbar}=\frac{\hbar}{m_0 a_0}
\end{equation}
%
%$v_0= \hbar/m_0 a_0 = G M m_0/\hbar$, %= \sqrt{GM/a_0}$
%
Note that the above identities imply the following intriguing relation, reflecting the fundamentally quantum mechanical nature of solitonic dark matter.
\begin{equation}\label{E:m0a0v0}
\hbar = m_0 v_0 a_0
\end{equation}

Since the scaling constants $a_0$, $\epsilon_0$ and $v_0$ depend only on $M$ and $m_0$, they are independent of $F$ and the shape of the dark matter density distribution, $\rho(x)$, which would not be the case for other length and energy scaling parameters, such as, for example, the soliton ground-state energy or radius. In the hydrogenic limit ($F\rightarrow 0$) the scaling constants $a_0$, $\epsilon_0$ and $v_0$ become equal to the Bohr radius, ground-state energy and root-mean-squared velocity of the dark matter particles, respectively. In the self-gravitating limit ($F\rightarrow 1$) the soliton size increases as the associated particle binding energy and velocity decrease (as quantitatively described in Section~\ref{S:SolitonXf}).

The Poisson equation yields the following relation between the dark matter probability density, $\rho(x)= |\Psi(x)|^2$,
% = \rho(r) a_0^3$,
and the potential energy, $V_S(x)$, of a dark matter particle of mass $m_0$ in a self-gravitating soliton of mass $M$ (as further described in the Appendix~\ref{app:A}).

\begin{equation}\label{E:SGPE}
V_S(x) = -2 \epsilon_0 \left[\frac{1}{x}\int_0^x \rho(x) 4\pi x^2 dx + \int_x^\infty \rho(x) 4\pi x dx\right]
\end{equation}

The potential energy of a dark matter particle interacting with a central point mass $M$ is $V_\bullet(x) = -2\epsilon_0/x$, and thus a galactic system with a dark matter soliton mass of $M_S = FM$, and a central point mass $M_\bullet = (1-F)M$ has a total mass $M$ and the following potential energy.
\begin{equation}\label{E:Vx}
V(x) = F V_S(x) + (1-F) V_\bullet(x)
\end{equation}
%_\bullet
%
Note that, since the soliton density profile $\rho(x)$ is determined by the full potential energy, $V(x)$, the soliton shape is in general expected to be $F$-dependent, as is $V_S(x)$.

The expectation values of the potential and kinetic energies of the soliton (ground state) may be obtained as follows.
\begin{eqnarray}\nonumber
\langle{V}\rangle/ \epsilon_0 &=&\int_0^\infty \Psi(x) \hat{V}\Psi(x) 4 \pi x^2 dx\\\label{E:Vave} &=&\int_0^\infty \rho(x) V(x) 4 \pi x^2 dx\\\nonumber
\langle{K}\rangle/\epsilon_0 &=& \int_0^\infty \Psi(x) \hat{K}\Psi(x) 4 \pi x^2 dx~~~~~~~~~~~~~~~~~~~~~~~~~~~~~~
\\\label{E:Kave}
&=&-\int_0^\infty \Psi(x) \left(\frac{1}{x}\right)\frac{\partial^2}{\partial x^2} \left[x \Psi(x)\right] 4 \pi x^2 dx
\end{eqnarray}
Thus, $E = \langle K\rangle+\langle V\rangle$ is the total energy of a soliton dark matter particle, and so $\epsilon = E/\epsilon_0 =\langle K\rangle/\epsilon_0+\langle V\rangle/\epsilon_0$.

In the hydrogenic limit all of the above expressions may be evaluated analytically (rather than numerically). This limit pertains to a vanishingly small amount of dark matter surrounding a central point mass. The resulting soliton is isomorphic with a hydrogen $1s$ state, for which $\rho(x) = \rho_0e^{-2x}$, $\langle V\rangle + \langle K\rangle = -\epsilon_0$, and $\langle K\rangle = \epsilon_0$, consistent with the virial theorem requirement that $\langle V\rangle/\langle K\rangle = -2$ \citep{BenUnd14}.

The enclosed mass of the soliton $M_S(x)$ is obtained as follows, where $f_S(x)$ is the fraction of the soliton mass within a sphere of radius $r = x a_0$.

\begin{equation}\label{E:MDMx}
f_S(x) = \frac{M_S(x)}{M_S} = \int_0^x \rho(x)4\pi x^2 dx
\end{equation}
The total enclosed mass of a system with central point mass $M_\bullet$ and a soliton mass fraction $F$ is 
\begin{eqnarray}\nonumber
M(x) &=& M_S f_S(x) + M_\bullet \\\label{E:Mx}
&=& M\left[F f_S(x) + (1-F)\right]\\\nonumber
&=& M f_T(x).
\end{eqnarray}
Thus, $f_T(x)=1+ F [f_S(x)-1]$ is the fraction of the total mass within a sphere of radius $r$, including the soliton and the central point mass.

Although the soliton ground state has zero angular momentum, a tracer star or gas molecule within the soliton may occupy a stable circular orbit with a rotational velocity of $v_{\rm{rot}}(x)$, obtained by equating the gravitational and centrifugal forces on the tracer star \citep{MemHar89}.
\begin{equation}\label{E:vmax}
\nu_r \equiv \frac{v_{\rm{rot}}(r)}{v_0}= \frac{\sqrt{G M(r)/r}}{\sqrt{G M/a_0}} = \sqrt{\frac{f_T(x)}{x}}
\end{equation}
The rotational velocity $v_{\rm{rot}}(r)$ of a tracer in a self-gravitating soliton has a maximum value of $v_{\rm{max}}/v_0 \approx 0.375$, which is similar in magnitude to the velocity dispersion of the corresponding self-gravitating soliton dark matter particles $\nu_S \equiv \sqrt{\langle v_S^2\rangle}/v_0 \approx 0.3294$, which is related as follows to the soliton kinetic energy.

\begin{equation}\label{E:RMSv}
\nu_S \equiv \frac{\sigma_S}{v_0} =
\frac{\sqrt{\langle v_S^2\rangle}}{v_0}=
\sqrt{\frac{\langle K\rangle}{\epsilon_0}}
\end{equation}
Note that $\langle K\rangle = \frac{1}{2} m_0 \langle v_S^2\rangle$, and $\sigma_S^2 = \langle v_S^2\rangle$ is the mean-squared velocity of the dark matter particles.

The following additional approximations, which play a central role in applications of present results to dwarf galaxy observations, relate the observed stellar (or gas) line-of-sight velocity dispersion, $\sigma \equiv \sigma_{\rm{los}}$, in a dark matter dominated galaxy to the corresponding 3D stellar velocity dispersion $\sigma_\ast$, tracer rotational velocity $v_{\rm{rot}}(r_{{-}3})$ and dark matter particle velocity dispersion $\sigma_S$.

\begin{equation}\label{E:vsSigma}
\sqrt{3}\,\sigma \approx \sigma_\ast \approx v_{\rm{rot}}(r_{{-}3}) \approx \sigma_S% = \sigma_S v_0
\end{equation}
The above approximations are obtained assuming that the stellar distribution is spherically symmetric and has an isotropic velocity dispersion \citep{LazAcc20,WolAcc10}, such that the mean-squared stellar velocity is $\langle v_\ast^2\rangle \equiv \sigma^2_\ast = 3 \sigma^2$. The second approximation in Eq.~\ref{E:vsSigma} pertains to the predicted circular rotational velocity of a hypothetical tracer star (or gas molecule) measured at a radius of $r_{{-}3}$, at which the logarithmic slope of the stellar density profile is equal to $-3$ \citep{LazAcc20,WolAcc10}. This approximation is consistent with the observed ratio $v_{\rm{max}}/\sigma \approx 1.3\pm0.4$ of the maximum gas rotational velocity and stellar velocity dispersion in early-type (elliptical and lenticular) galaxies with stellar masses between $10^{10}$\,$\rm{M}_\odot$ and $10^{11}$\,$\rm{M}_\odot$ \citep{SerLin16}. 

In applying Eq.~\ref{E:vsSigma} to dark matter dominated galaxies it is further assumed here that $v_{\rm{rot}}(r_{{-}3})$ pertains to the $r_{{-}3}$ radius of the dark matter soliton density profile (rather than the stellar density profile). Although  $\sigma_\ast \approx v_{\rm{rot}}(r_{{-}3}) \approx \sigma_S$ is only expected to be appropriate for galaxies with similar stellar and dark matter profiles, while $v_{\rm{rot}}(r_{{-}3}) \approx v_{\rm max}$ is more generally appropriate for any dark matter dominated galaxy.  The third approximation in Eq.~\ref{E:vsSigma} is predicted to be appropriate for any dark-matter dominated galaxies, regardless of the relative sizes of the stellar and dark matter distributions, as further described in Section~\ref{S:SolitonXf} (and demonstrated by the similarity of the solid and dashed blue curves in Fig.~\ref{F:Layout_vs_f}a). 

However, in galaxies with significantly different stellar and dark matter density profiles, $\sigma_S$ and $\sigma_\ast$ are expected to differ as they pertain to different length scales.
Nevertheless, $\sigma_\ast \approx \sigma_S$ is a convenient first approximation that makes it possible to estimate
soliton properties directly from the observed velocity dispersions $\sigma$.
%, and is expected to be appropriate for galaxies with dark matter and stellar distributions of similar size. 
Predictions obtained assuming $\sigma_\ast \approx \sigma_S$ are shown to be consistent with observations of Fornax, as described at the end of Section~\ref{S:Soliton1f}.  More generally, observationally consistent predictions for a wider range of dSph and UFD galaxies, obtained using both dual-$m_0$ and universal $m_0$ models, imply that the ratio $\sigma_S/\sigma_\ast$ remains near one, to within about 20\% for most of the dSph galaxies and within about a factor of 2 for most of the UFD galaxies (as further described in Sections~\ref{S:dSphUFD} and \ref{S:Discussion}, and shown in Appendix  Tables~\ref{T:Dual_m0} and \ref{T:1x5m0_1SE}).

The comparisons with dSph and UFD observations are facilitated by incorporating the following observational estimate of the total mass $M_{1/2}$ enclosed within the observed (2D projected) half-light radius $R_{1/2}$ \citep{WolAcc10}.
\begin{equation}\label{E:MhalfWolf}
M_{1/2} \approx \frac{4 \sigma^2 R_{1/2}}{G}
\end{equation}
The above expression is obtained assuming a spherical, equilibrium, dispersion-supported (non-rotating) stellar population \citep{WolAcc10}, which is considered to be appropriate for most dSph and UFD galaxies \citep{SimFai19}.  Thus, Eq.~\ref{E:MhalfWolf} can provide a powerful self-consistency criterion, imposed by requiring that the observationally derived $M_{1/2}$, obtained using Eq.~\ref{E:MhalfWolf}, is equal to the predicted enclosed mass $M(r_{1/2})$, obtained using Eq.~\ref{E:Mx}, where $r_{1/2} \approx \frac{4}{3} R_{1/2}$ is the 3D half-light radius corresponding to the observed (2D projected) half-light radius $R_{1/2}$ \citep{WolAcc10}.

The wavefunctions, $\Psi(x)$, and energies, $\epsilon$, that solve Eq.~\ref{E:SP} must in general be determined numerically (except in the hydrogenic limit). The resulting self-gravitating soliton predictions (for which $F=1$ and $M = M_S$) are described in Section~\ref{S:Soliton1f}, and those pertaining to the $F$-dependent soliton shape changes are described in Section~\ref{S:SolitonXf}.

\section{Jeans Analysis}\label{S:Jeans}

The spherical Jeans equation relates the three-dimensional stellar
tracer density $\nu_\star(r)$, radial velocity dispersion $\sigma_r(r)$,
and anisotropy parameter $\beta(r) = 1 - \sigma_t^2/\sigma_r^2$
(where $\sigma_t^2 \equiv (\sigma_\theta^2+\sigma_\phi^2)/2$)
to the total enclosed gravitating mass $M(<r)$.
\begin{equation}
\frac{d}{dr}\!\left[\nu_\star(r)\,\sigma_r^2(r)\right]
+ \frac{2\beta(r)}{r}\,\nu_\star(r)\,\sigma_r^2(r)
= -\nu_\star(r)\,\frac{G M(<r)}{r^2}.
\label{E:jeans}
\end{equation}
Given $\nu_\star(r)$, $\beta(r)$, and $M(<r)$,
Equation~\ref{E:jeans} may be solved to obtain $\sigma_r^2(r)$.
\begin{equation}
\sigma_r^2(r)=\frac{1}{\nu_\star(r)I(r)}
\int_r^{\infty} \nu_\star(s)\,I(s)\,\frac{G M(<s)}{s^2}\,ds,
\label{eq:sigr}
\end{equation}
\begin{equation}
I(r)=\exp\!\left[\int^r \frac{2\beta(t)}{t}\,dt\right].
\end{equation}

Unless stated otherwise, all the results in this work are obtained assuming isotropic
stellar orbits, $\beta(r)=0$, so that $I(r)=1$.  The enclosed mass
$M(<r)$ is taken to consist of the dark matter soliton and central black hole,
$M(<r) \equiv M(r/a_0)$ as obtained using Eq.~\ref{E:Mx}, since the stellar
mass contributes negligibly to $M(<r)$ in the dwarf galaxy systems of interest.
However, the stellar tracer density $\nu_\star(r)$ in Eq.~\ref{E:jeans}
remains essential, and is represented as a central soliton (Schive) profile plus an NFW tail, 
fit to the observed stellar distribution
as described by \citet{PozDwa24} (using the parameters provided in Appendix Table~\ref{T:dSph_UFD_data}).

The projected line-of-sight velocity dispersion is obtained by
integrating $\sigma_r^2(r)$ along the line of sight.
\begin{equation}
\sigma_{\rm los}^2(R)=\frac{2}{\Sigma_\star(R)}
\int_R^{\infty}\!\left[1-\beta(r)\frac{R^2}{r^2}\right]
\frac{\nu_\star(r)\,\sigma_r^2(r)\,r\,dr}{\sqrt{r^2-R^2}},
\label{eq:siglos}
\end{equation}
where $\Sigma_\star(R)$ is the projected stellar surface density.
\begin{equation}
\Sigma_\star(R)
= 2\int_R^{\infty} \nu_\star(r)\,\frac{r\,dr}{\sqrt{r^2-R^2}}.
\end{equation}
The observed aperture-averaged velocity dispersions compiled by \citet{PozDwa24}
are compared with the following tracer-weighted mean dispersion.
\begin{equation}\label{E:Jeans_sigma_ave}
\langle\sigma_{\mathrm{los}}\rangle(R_{\max})
= \left[\frac{\displaystyle\int_0^{R_{\max}}
\Sigma_\star(R)\,\sigma_{\mathrm{los}}^2(R)\,2\pi R\,dR}
{\displaystyle\int_0^{R_{\max}}
\Sigma_\star(R)\,2\pi R\,dR}\right]^{1/2}.
\end{equation}
Unless stated otherwise, $R_{\max}$ is set equal to the approximate
extent of the stellar distribution for each galaxy, as determined by the 
radial range Figs. 10-25 in \citet{PozDwa24} (and provided in
 Appendix Table~\ref{T:dSph_UFD_data} as $R_{\rm max}$).

\section{Self-Gravitating Solitons}\label{S:Soliton1f}

\begin{figure*}
\begin{center}
\includegraphics[width=4.5 in]{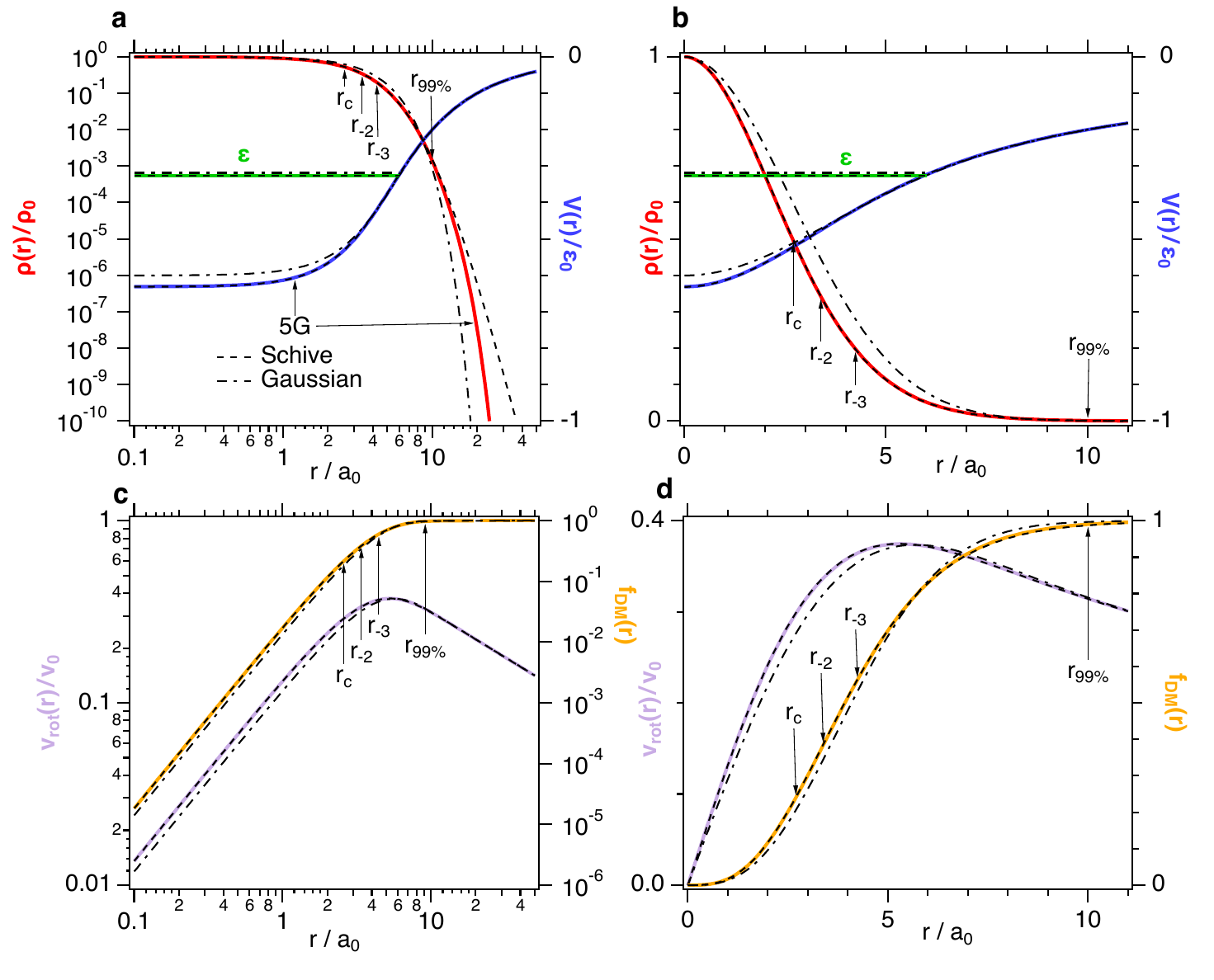}\
\end{center}
\caption{The top two panels (a) and (b) show the potential energy (blue curves, left axis) and mass probability density (red curves, right axis) for a self-gravitating soliton composed of ultralight particles of mass $m_0$ in a system with a total mass $M$, plotted using either (a) log-log scales (for the bottom and left axes) or (b) linear scales (for all axes). The solid, dashed, and dot-dashed curves pertain to the 5G, Schive, and Gaussian approximations, respectively. The green horizontal lines indicate the total binding energy $\epsilon$ of an ultralight soliton particle. The lower two panels (c) and (d) show the tracer rotational velocity (purple curves, left axis) and integrated soliton mass fraction (orange curves, left axis), again plotted using either (c) log-log or (d) linear scales. The arrows mark the locations of the half-density radius $r_c$ and the radius $r_{99\%}$ containing $99\%$ of the total soliton mass, as well as the radii $r_{{-}2}$ and $r_{{-}3}$ at which the logarithmic slopes of the soliton density are $-2$ and $-3$, respectively. The radius $r_{{-}3}$ is also close to the radius $r_{50\%} = 3.925\,a_0$ containing $50\%$ of the total soliton mass, as well as to the location of the maximum rotational velocity.}
\label{F:VrRhor}
\end{figure*}

The shape and energy of a self-gravitating BEC soliton are equivalent to the quantum mechanical ground state of a system consisting entirely of dark matter Bose particles of mass $m_0$ with a total mass $M = M_S$. The resulting numerical solution was first obtained over 50 years ago \citep{Rufsys69,MemHar89}. More recently, Schive, Chiueh, and Broadhurst \citep{SchCos14} have suggested the following appealingly simple analytical approximation to the soliton probability density $\rho(r) =\rho_{\rm{Schive}}(r)$.
\begin{equation}\label{E:SPiSchive}
\rho_{\rm{Schive}}(x)
= \frac{\rho_0}{\left[1+b(x/x_c)^2\right]^8}
\end{equation}
The maximum probability density is $\rho_0 = \rho_{\rm{Schive}}(0)$ and the constant $b = 2^{1/8}-1 \approx 0.090508$ (often rounded to $0.091$) is that required to assure that $x_c = r_c/a_0$ is equal to the radius at which $\rho$ attains half its maximum value, $\rho_{\rm{Schive}}(r_c) =\frac{1}{2} \rho_0$. The above approximation accurately represents the true $\rho(r)$ out to about $r \approx 4 r_c \approx 10 a_0$, beyond which Eq.~\ref{E:SPiSchive} overshoots the more accurate numerical solution, as shown in Fig.~\ref{F:VrRhor} and further described below.

The following single Gaussian functional form is another simple, reasonably accurate, approximation for the self-gravitating soliton probability density, $\rho(r) = \rho_{\rm{Gaussian}}$, that has been widely used in variational treatments of bosonic systems and self-gravitating Bose--Einstein condensates \citep{ChaMas11,ChaMas19,Chavanis19,ChaPre19,Alcubierre18,Rinpar23}.
\begin{equation}\label{E:SPiGaussian}
\rho_{\rm{Gaussian}}(x)
= \rho_0\,e^{-(x/x_0)^2}
\end{equation}
Note that in this case $x_0 = r_0/a_0 = x_c/\sqrt{\ln{2}} \approx 1.201 \,x_c$.

A more accurate approximation may be obtained by representing the soliton ground-state wavefunction as a sum of five Gaussian functions, $\Psi_{\rm{5G}}(x)$, whose amplitude and width coefficients, $c_i$, are numerically optimised so as to minimise the ground-state energy while maintaining self-consistency with the virial theorem (as further described in the Appendix~\ref{app:B}). Although similar Gaussian expansion methods have been applied in other Bose--Einstein condensate studies \citep{Hiyama03,Perez96}, the following five-Gaussian (5G) function is the first such representation of dark matter solitons, not only in the self-gravitating ($F=1$) limit, but also over the full range of soliton mass fractions $0 \le F \le 1$, as further described in Section~\ref{S:SolitonXf}.
\begin{equation}\nonumber
\Psi_{\rm{5G}}(x)
=\sqrt{\rho_0}\,\left(\frac{\sum_{j=0}^{4} c_{2j} e^{-(x/c_{2j+1})^2}}{\sum_{j=0}^{4}c_{2j}}\right)
\end{equation}
and 
\begin{equation}\label{E:5G}
\rho_{\rm{5G}}(x) = \rho_0\left(\frac{\sum_{j=0}^{4} c_{2j} e^{-(x/c_{2j+1})^2}}{\sum_{j=0}^{4}c_{2j}}\right)^2
\end{equation}

Figure~\ref{F:VrRhor} shows $\rho(r)$, $V(r)$, $M(r)$, and $v_{\rm{rot}}(r)$ predictions obtained using $\rho_{\rm{5G}}$ (coloured curves), $\rho_{\rm{Schive}}$ (dashed black curves), and $\rho_{\rm{Gaussian}}$ (dot-dashed black curves). The results on the left- and right-hand panels are the same, plotted either on logarithmic or linear axis scales, respectively. The green horizontal lines in the upper two panels mark the soliton energy, $\epsilon$, in relation to its potential energy, $V(r)/\epsilon_0$. Note that beyond $r/a_0 > 6$ the soliton probability density tail results from quantum mechanical tunnelling into the classically forbidden region in which $V(r)/\epsilon_0 > \epsilon$.  The radii $r_c \approx r_{-1.32} \approx 2.68 a_0$, $r_{{\mbox{-}}2} \approx 3.39 a_0$, $r_{{\mbox{-}}3} \approx 4.33 a_0$, and $r_{99\%} \approx r_{-9.61} \approx 9.95 a_0$ shown in Fig.~\ref{F:VrRhor} are the values of $r$ at which $\rho(r)$ is half its maximum value, $d\ln{M(r)}/d\ln{r} = -2$, $d\ln{M(r)}/d\ln{r} = -3$, and $f_S(r)= 0.99$, respectively.

The probability densities in Figs.~\ref{F:VrRhor}a and \ref{F:VrRhor}b are divided by $\rho_0 = \rho(0) = [\Psi(0)]^2$. Thus, $M_S\, \rho_0$ is the maximum dark matter mass density per $a_0^3$, $(M_S/m_0)\,\rho_0$ is the maximum number of dark matter particles per $a_0^3$, and $(M_S/a_0^3)\,\rho_0$ is the maximum density of the soliton per unit volume (in whatever mass and volume units are used to express $M$ and $a_0^3$). The value of $\rho_0$ is determined by normalising $\rho(r)$ such that $\int_0^\infty \rho(x) 4\pi x^2 dx =1$. The resulting values of $\rho_0$ for the Schive, Gaussian, and 5G approximations are 0.004400, 0.003379, and 0.004397, respectively (all pertaining to a self-gravitating soliton with $F=1$). More generally, at any value of $F$, the normalisation constant of $\rho_{\rm{5G}}$ is expressible explicitly in terms of the coefficients $c_i$ (as further described in the Appendix~\ref{app:A}).

The $\Psi_{\rm{5G}}$ ground-state energy, indicated by the green horizontal lines in Figs.~\ref{F:VrRhor}a and \ref{F:VrRhor}b, may be expressed as follows in terms of $\epsilon_0$, $m_0$, $a_0$, and $v_0$.
\begin{eqnarray}\nonumber
E_{\rm{5G}} = \epsilon_{\rm{5G}} \epsilon_0 &\simeq& - 0.32554\,\epsilon_0\\\nonumber
&=& - 0.32554 \left(\frac{\hbar^2}{2 m_0 a_0^2}\right)\\\label{E:E}
&=& -0.16277 \,m_0 v_0^2
\end{eqnarray}
This energy is slightly lower than that obtained using the previous two approximations, thus ensuring that it is closer to the true ground-state energy, in keeping with the variational theorem requirement that the true ground-state wavefunction is that which minimises the ground-state energy. More specifically, $\epsilon_{\rm{Schive}} \approx -0.3253\epsilon_0$ (with $x_c = r_c/a_0 \approx 2.69$) and $\epsilon_{\rm{Gaussian}} \approx -0.3183\epsilon_0$ (with $x_0 = r_0/a_0 \approx 5.32$).

The accuracy of the optimised $\Psi_{\rm{5G}}$ is further verified by noting that an essentially identical energy and wavefunction shape are obtained when approximating $\Psi$ as the sum of either 4 or 6 Gaussian components (with re-optimised coefficients), thus confirming that a five-Gaussian basis set is sufficient to accurately represent the exact wavefunction $\Psi(r)$ at any value of $0\le F\le 1$ (as demonstrated in Section~\ref{S:SolitonXf}). Moreover, all three of the above approximate solutions of Eq.~\ref{E:SP} are essentially perfectly self-consistent with the virial theorem (as further described in the Appendix~\ref{app:A}).

The shape of $\rho(r)$ makes it possible to determine the total mass of a self-gravitating soliton, $M$, from the mass contained within $r_c$, $r_{{\mbox{-}}2}$, or $r_{{\mbox{-}}3}$.
\begin{equation}\label{E:MfromRx}
M \approx \frac{M(r_c)}{0.2357} \approx \frac{M(r_{{\mbox{-}}2})}{0.3864} \approx \frac{M(r_{{\mbox{-}}3})}{0.5817}
\end{equation}
The following expressions for $m_0$ in terms of $r_c$, $M_c = M(r_c)$, and $\sigma$ may be obtained from Eqs.~\ref{E:m0} and \ref{E:vsSigma}.
\begin{equation}\label{E:m0_Sigma}
m_0 = 0.795 \frac{\hbar}{\sqrt{G M_c r_c }}
\approx 0.566 \frac{\hbar}{r_c \,\sigma}
\end{equation}
The first identity is equivalent to the following previously obtained scaling relation \citep{SchCos14}, expressed in observational units of $M_c$\,(M$_\odot$), $r_c$\,(kpc), and $m_0$\,(eV/c$^2$), here obtained using the 5G predicted values of $r_c \approx 2.68\,a_0$ and $M_c = M(r_c) \approx 0.2357 M$ for the self-gravitating soliton.
\begin{equation}\label{E:Mc}
M_c\,({\rm{M}}_\odot) = \frac{5.47\times 10^9}{r_c ({\rm{kpc}}) [m_0/10^{23} ({\rm{eV}}/{\rm{c}}^2)]^2 }
\end{equation}
The second approximate equality in Eq.~\ref{E:m0_Sigma} is obtained using Eq.~\ref{E:vsSigma}, combined with the present self-gravitating soliton prediction that $v_{\rm{rot}}(r_{{-}3})/v_0 \approx 0.366$. Thus, Eq.~\ref{E:m0_Sigma} may also be re-expressed as follows, in observational units.
\begin{equation}\label{E:m0Mcrc}
m_0\,({\rm{eV}}/{\rm{c}}^2) \approx \frac{7.35\times 10^{-19}}{\sqrt{M_c({\rm{M}}_\odot)\,r_c ({\rm{kpc}})}}
\approx \frac{1.11\times 10^{-21}}{r_c ({\rm{kpc}}) \sigma ({\rm{km/s}})}
\end{equation}

The Fornax (dSph) galaxy is a good test case for self-gravitating dark matter predictions, as it is nearly entirely composed of dark matter \citep{PasAct18}, and so its soliton shape should be consistent with the above predictions. Its observed velocity dispersion, $\sigma \approx \sigma_{\rm{los}} \approx 12$ (km/s) \citep{PasAct18}, and the Schive et al.\ predictions \citep{SchCos14}, $r_c \approx 0.93$\,kpc, $M_c \approx 9.2 \times 10^7$ M$_\odot$, $M \approx 3.7 \times 10^8$ M$_\odot$, combined with the two expressions in Eq.~\ref{E:m0Mcrc}, yield remarkably consistent values of $m_0 \approx 10^{-22}$ eV/c$^2$. More specifically, the first expression in Eq.~\ref{E:m0Mcrc} predicts $m_0 \approx 7.9\times 10^{-23}$ eV/c$^2$, consistent with the Schive value of $\sim 8\times 10^{-23}$ eV/c$^2$ \citep{SchCos14}, and the second expression predicts $m_0 \approx 9.9\times 10^{-23}$. Moreover, the predicted central dark matter mass density of $\rho_\odot \approx 3\times 10^7$ (M$_\odot$/kpc$^3$), obtained assuming $F=1$ and $m_0 = 1\times 10^{-22}$ eV/c$^2$, is in good agreement with the value of $\rho_\odot \approx 4 \times 10^7$ (M$_\odot$/kpc$^3$) obtained both by Schive et al.\ \citep{SchCos14} (using Eq.~4 in the Supplementary Information) and from the ``best model'' fit to Fornax (dSph) structural and kinematic properties reported by Pascale et al.\ \citep{PasAct18}. These agreements confirm the approximate validity of the $\sigma_S \approx \sigma_\ast$ approximation in Eq.~\ref{E:vsSigma}, as well as the self-consistency of the present soliton predictions with those obtained by Schive et al.\ \citep{SchCos14}. Additional comparisons with the measured properties of other dSph and UFD galaxies, including predictions obtained when relaxing the assumption that $\sigma_S \approx \sigma_\ast$, are described in Sections~\ref{S:Observational_comparisons} and \ref{S:Discussion} (and the predicted values of $\sigma_S/\sigma_\ast$ obtained using various modeling assumptions are provided in Appendix Tables~\ref{T:optimizem0}-\ref{T:F0}).

\section{Shape Shifting Solitons}\label{S:SolitonXf}

To illustrate the influence of baryonic coupling on the shape of a dark matter soliton, one may introduce a central point mass into the centre of a dark matter soliton, with a mass fraction $F$, where $F=1$ corresponds to a self-gravitating soliton and $F=0$ corresponds to a central point mass (or black hole) surrounded by a vanishingly small amount of dark matter. The following results are obtained by solving Eq.~\ref{E:SP} for systems with various values of $0 \le F \le 1$, to obtain the corresponding ground-state eigenfunction $\Psi_{\rm{5G}}(r)$ and energy $\epsilon = E/\epsilon_0$, where $\Psi_{\rm{5G}}(r)$ is again represented using Eq.~\ref{E:5G} with $F$-dependent coefficients (obtained as further described in the Appendix~\ref{app:A}).

Figure~\ref{F:Rhovsf} shows how $F$ influences the shapes of (a) $\rho(r)$, (b) $V(r)$, (c) $v_{\rm{rot}}(r)$, and (d) the integrands used to obtain $\langle V\rangle$ and $\langle K\rangle$. The inset panels in (a), (b), and (c) show the same results plotted on logarithmic scales. The $F = 1$ results in Fig.~\ref{F:Rhovsf} are equivalent to the self-gravitating soliton results shown in Fig.~\ref{F:VrRhor}. The $F \rightarrow 0$ predictions approach the hydrogenic limit at which the gravitational potential is dictated entirely by the central point mass, $V(x)/\epsilon_0 \rightarrow -2/x$. The dashed and dotted curves in Fig.~\ref{F:Rhovsf} compare the exact $\rho_{1s}(r) = \rho_0 e^{-2x}$ and numerical $\rho_{\rm{5G}}(r)$ hydrogenic predictions, respectively. The good agreement between the two $F=0$ predictions confirms that representing $\rho(r)$ as the sum of five Gaussians is sufficient to produce accurate predictions over the entire $0\le F \le 1$ range. Note that the kinetic energy integrand in Fig.~\ref{F:Rhovsf}d contains substructure resulting from its representation using five Gaussian components. This substructure, which is not present in the exact $F=0$ kinetic energy integrand (dashed curve), does not significantly influence the resulting kinetic (or total) energy expectation values, as evidenced by the less than $0.1\%$ difference between the exact and 5G predicted values of $\langle K\rangle$ and $\epsilon$ in the hydrogenic ($F=0$) limit.

\begin{figure*}
\begin{center}
\includegraphics[width=4.5 in]{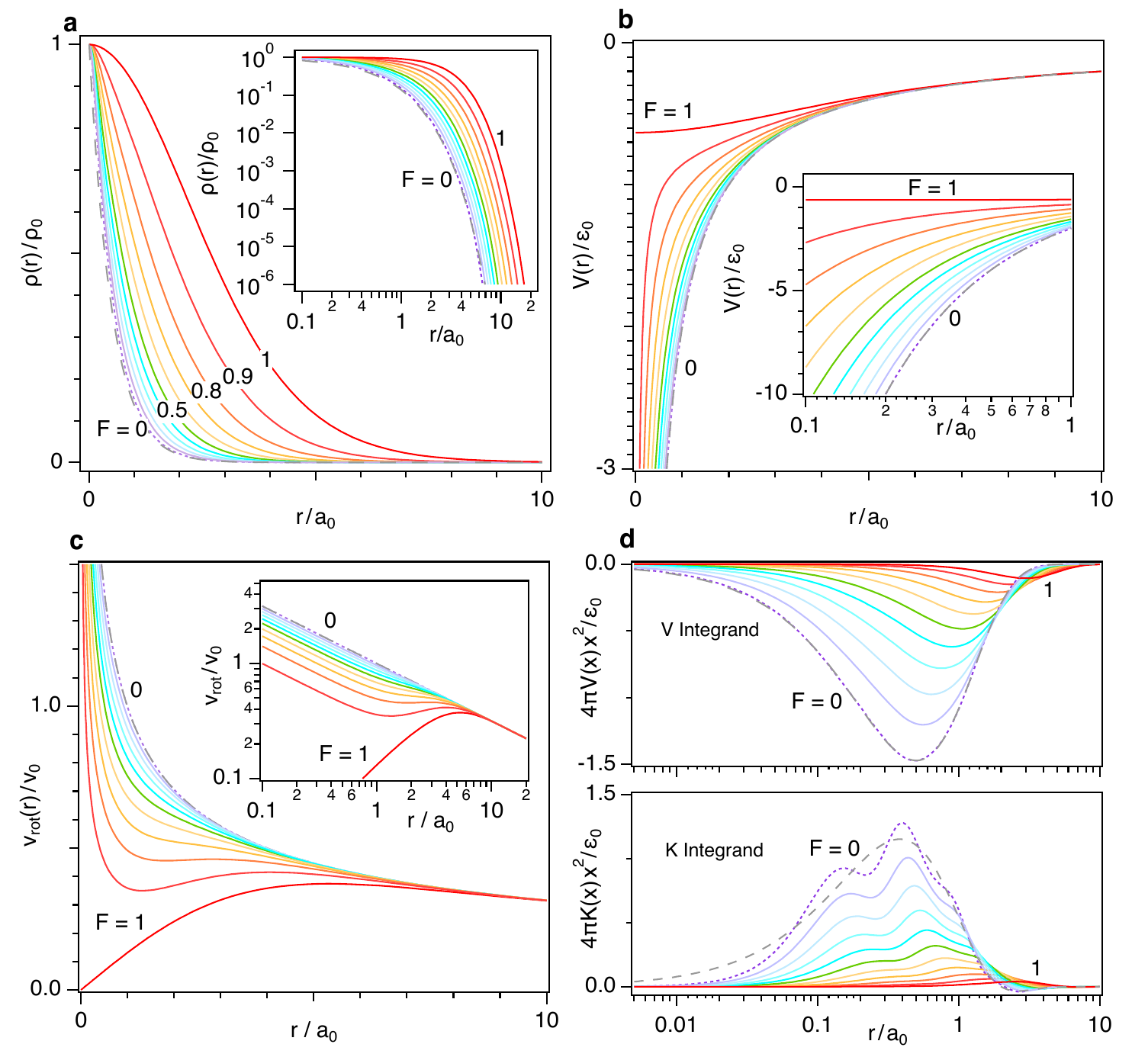}
\end{center}
\caption{Shape shifting soliton predictions as a function of soliton mass fraction $F$. (a) Soliton probability density, (b) potential energy, (c) tracer rotational velocity, and (d) the integrands of Eqs.~\ref{E:Vave} and \ref{E:Kave}. The inset panels in (a)--(c) show the same results plotted on a logarithmic scale. The dashed and dotted $F=0$ curves pertain to the exact hydrogenic and approximate 5G predictions, respectively.}
\label{F:Rhovsf}
\end{figure*}

The $\rho(r)$ predictions in Fig.~\ref{F:Rhovsf}a show how the cored (flat-top) shape of self-gravitating solitons becomes increasingly cusped (peaked-top) when a central point mass is introduced into the soliton. More specifically, the initial slope, $S_0$, of the soliton density profile may be defined as follows in terms of the logarithmic density difference between $r= a_0/10$ and $r=a_0$.
\begin{equation}\label{E:S0}
S_0 \equiv \frac{\log{\rho (a_0)}-\log{\rho (a_0/10)}}{\log{(a_0)}-\log{(a_0/10)}} =
\frac{\log{[\rho (a_0/10)/\rho (a_0)]}}{ \log{[1/10]}}
\end{equation}
When defined in this way, the self-gravitating soliton ($F=1$) has a nearly zero slope of $S_0 \approx -0.018$, consistent with its cored shape, while at the other extreme limit ($F \rightarrow 0$) the soliton has a cusped shape with a slope of $S_0 \approx -1.3$. The latter cusped slope is comparable to that of an NFW cold dark matter density profile for which $ -3 < S_0 < -1$, where $\rho_{\rm NFW} = \rho_0/[y(1+y)^2]$, $y=r/r_S$, and $r_S$ is the NFW scale radius at which $S_0 = -2$ (between the limits of $S_0=-1$ when $y\ll1$ and $S_0 = -3$ when $y\gg1$). The shape of the soliton is dictated by $F$ and its width is inversely proportional to $m_0$. For a given value of $m_0$, the soliton narrows and becomes more cusped with decreasing $F$, due to the increasing influence of the central point mass. 

For a self-gravitating ($F=1$) soliton, Eq.~\ref{E:m0Mcrc} may be used to estimate
%obtain 
$m_0$ eV/c$^2$ from observationally derived values of $\sigma$ (km/s) and $r_c$\,kpc. More generally, the following expression may in principle be used to estimate $m_0$ from $\sigma$ and $r_c$, combined with and an estimated value of $F$. The $F$-dependent functions in square brackets are shown in Fig.~\ref{F:Layout_vs_f}a and further described below (and in the Appendix~\ref{app:A}).
\begin{equation}\label{E:m0SigmaRc}
%m_0\, ({\rm{eV/c}^2}) \approx 1.107\times 10^{-21} \,\frac{[x_c(F)\,\nu_{{-}3}(F)]}{r_c\,({\rm{kpc}})\,\sigma\,({\rm{km/s}})}
m_0\, ({\rm{eV/c}^2}) \approx 1.107\times 10^{-21} \,\frac{[x_c(F)\,\nu_S(F)]}{r_c\,({\rm{kpc}})\,\sigma\,({\rm{km/s}})}
\end{equation}
For a given value of $m_0$ and $F$, the approximate identities in Eq.~\ref{E:vsSigma} may be used to express all the observable properties of the soliton directly in terms of the corresponding velocity dispersion, $\sigma$. For example, the following equations (derived from Eqs.~\ref{E:a0e0}--\ref{E:v0} and \ref{E:vmax}--\ref{E:vsSigma}) predict the soliton radius $r_c$\,kpc, soliton core density $\rho_\odot$ (M$_\odot$/kpc$^3$), and total galactic mass $M$ M$_\odot$, as functions of velocity dispersion $\sigma$ (km/s), for given values of $m_0$ eV/c$^2$ and $F$.
\begin{eqnarray}\label{E:RcSigma}
r_c\,({\rm{kpc}}) & \approx& 1.107\times 10^{-21}\,\frac{\left[x_c(F)\,\nu_S(F)\right]}{m_0\,\sigma} \\\label{E:RhoSigma}
\rho_\odot\,({\rm{M}}_\odot/{\rm{kpc}}^3) & \approx& 5.692\times 10^{47}\left[\frac{F\rho_0(F)}{[\nu_S(F)]^4}\right]\,m_0^2\,\sigma^4 \\\label{E:MSigma}
M \,({\rm{M}}_\odot) & \approx&7.718\times 10^{-16}\left[\frac{1}{\nu_S(F)}\right]\frac{\sigma}{m_0}
\end{eqnarray}
Note that Eqs.~\ref{E:m0SigmaRc}--\ref{E:MSigma} may be rearranged in various ways to, for example, express $m_0$ as a function of $\rho_\odot$ and $\sigma$ (or $M$ and $\sigma$), or to express $\rho_\odot$ as a function of $M$ and $\sigma$, and so on. The $F$-dependent functions in Eqs.~\ref{E:m0SigmaRc}--\ref{E:MSigma} are shown in Fig.~\ref{F:Layout_vs_f}a and provided as polynomial expansions in Appendix~\ref{app:A}, along with several other soliton properties, including $r_{50\%}$, $r_{99\%}$, $\langle K\rangle$, and $\langle V\rangle$.

\begin{figure*}
\begin{center}
\includegraphics[width=4.5 in]{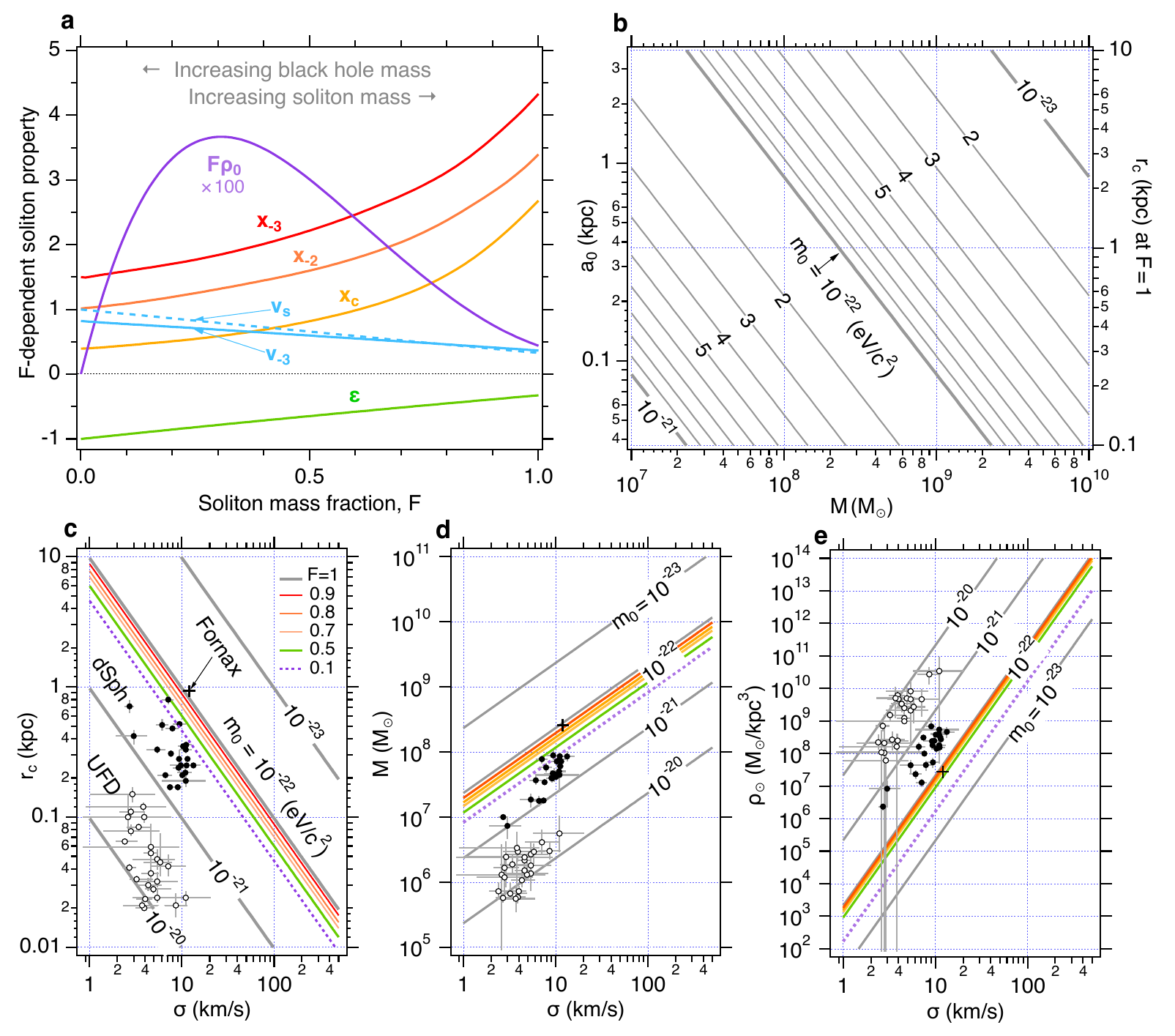}
\end{center}

\caption{Soliton property predictions as a function of (a) soliton mass fraction $F$, (b) total galactic mass $M$, and (c)-(e) stellar velocity dispersion $\sigma$. 
The curves in (a), which may be reproduced using $F$-dependent polynomial expansions provided in Eq.s~\ref{E:PolyF} of Appendix~\ref{app:A}, and used in Eqs.~\ref{E:m0SigmaRc}--\ref{E:MSigma}. The diagonal lines in (b) relate $a_0$ and $r_c$ to $M$ and $m_0$, as further described in the text.  The diagonal lines in (c)-(e) relate $\sigma$ to $r_c$, $M$, $\rho_\odot$ and $F$, and the points pertain to dSph and UFD galactic observations, as further described in the text.
 .}
%  ... as further described in the text
\label{F:Layout_vs_f}
\end{figure*}

Figure~\ref{F:Layout_vs_f} shows predicted soliton properties in both dimensionless and observational units. The dimensionless shape-dependent properties in Fig.~\ref{F:Layout_vs_f}a are plotted as a function of the soliton mass fraction, $F$. The purple $F \rho_0(F)$ curve (which appears in Eq.~\ref{E:RhoSigma}) is proportional to the mass density at the core of the soliton, $\rho_\odot = F \rho_0 M /a_0^3$, which is plotted in observational units in panel (e) and further described below. The two blue curves in Fig.~\ref{F:Layout_vs_f}a show the similarity between the predicted tracer rotational velocity $\nu_{-3} = v_{\rm{rot}}(r_{{-}3})/v_0$ (solid blue curve) and the dark matter velocity dispersion $\nu_S = \sigma_S/v_0$ (dashed blue curve), consistent with the third approximate equality in Eq.~\ref{E:vsSigma}. The red and orange curves in Fig.~\ref{F:Layout_vs_f}a show the soliton dimensionless radii $x_c = r_c/a_0$, $x_{-2} = r_{-2}/a_0$, and $x_{-3} = r_{-3}/a_0$ pertaining to its density profile (excluding the central point mass). These radii can be converted to kpc units by using Eq.~\ref{E:a0e0} to obtain $a_0$ as a function of $M$ and $m_0$, whose predictions are shown in Fig.~\ref{F:Layout_vs_f}b.  Note that the right-hand axis of Fig.~\ref{F:Layout_vs_f}b shows $r_c$\,kpc pertaining to a self-gravitating solition ($F=1$).  The values of $r_c$ at other values of $F$ can be obtained by multiplying $a_0$ on the left-axis of Fig.~\ref{F:Layout_vs_f}b by the corresponding values of $x_c$ in Fig.~\ref{F:Layout_vs_f}a, which indicate, for example, that $r_c \approx a_0$ when $F\approx 0.67$.  

The diagonal lines in Figure~\ref{F:Layout_vs_f}c show how the soliton radius $r_c$ depends on $\sigma$ for different values of $m_0$ and $F$.  The coloured and dotted lines pertain to predictions obtained assuming $F<1$.  The + point pertains to the Fornax (dSph) galaxy, for which $r_c = 0.93$\,kpc \citep{SchCos14} and $\sigma = 12$ (km/s) \citep{PasAct18} yield the dark matter particle mass of $m_0 \approx 10^{-22} $\,eV/c$^2$ that is consistent with a self-gravitating soliton shape.  Thus, if the value of $m_0$ differed significantly from $\sim$$10^{-22}$ eV/c$^2$ then either $\sigma$ or $r_c$ would have to change proportionately, as Eq.~\ref{E:m0SigmaRc} indicates that $m_0$ is inversely proportional to both $\sigma$ and $r_c$.  Note that the red ($F=0.9$) line in Fig.~\ref{F:Layout_vs_f}c indicates that if $10\%$ of the total mass of Fornax (dSph) consisted of a central black hole, its $r_c$ would only decrease by about $\sim$$22\%$ relative to its value in a purely self-gravitating soliton (with $F=1$), thus illustrating the robustness of observationally inferred $m_0$ values. In other words, a value of $m_0$ obtained assuming that dark matter has a self-gravitating soliton shape would remain accurate to within about $\pm20 \%$ as long as a central black hole and stars contributed less than about $10\%$ to the total mass of a galaxy. 

The closed and open circle points in Figure~\ref{F:Layout_vs_f}c represent the observed stellar $r_c$ and $\sigma$ values of various dSph and UFD galaxies, respectively (as obtained from Tables II and III of \citealt{PozDwa24}, and also provided here in Appendix~\ref{app:B} Tables 1 and 2). The analysis described by \citet{PozDwa24} assumed that the stellar and dark matter distributions have the same shape. If so, then the locations of the points in Figure~\ref{F:Layout_vs_f}c imply that the dSph and UFD have different dark matter particle masses with $10^{-22}\ge m_0 \ge  10^{-21}$ eV/c$^2$ for dSph galaxies and  $10^{-21}\ge m_0 \ge  10^{-20}$ for UFD galaxies.  

The diagonal lines in Figures~\ref{F:Layout_vs_f}d and e show how the total galactic mass $M$ and the maximum soliton density $\rho_\odot = F \rho_0 M /a_0^3$ depend on $\sigma$, for dark matter particles of mass $m_0$. The colour diagonal lines in Fig,~\ref{F:Layout_vs_f}d reveal the decrease M as $F$ decreases, for given values of $m_0$ and $\sigma$.  One might expect the dark matter core density, $\rho_\odot$, to decrease more strongly with decreasing $F$, since the soliton mass $M_S = F M$ is proportional to $F$.  However,  the narrowing of the soliton with decreasing $F$ (associated with its increasingly cusped shape), conspires to counter the soliton's decreasing total mass, thus making the soliton's core density relatively insensitive to $F$ down to $F=0.5$, as indicated by the nearly overlapping coloured curves in Fig.~\ref{F:Layout_vs_f}e.  At very low $F < 0.1$, the soliton maximum density $\rho_\odot$ drops precipitously, since $M_S\rightarrow 0$ as $F\rightarrow 0$. The points in Figs.~\ref{F:Layout_vs_f}d and e show dSph and UFD predictions obtained using Eqs.~\ref{E:m0SigmaRc}-\ref{E:MSigma}, assuming $F=1$, as further discussed in the next section, which also includes alternative interpretations of the same observations.

%\newpage
%
\begin{figure*}
 \begin{center}
  \includegraphics[width=4 in]{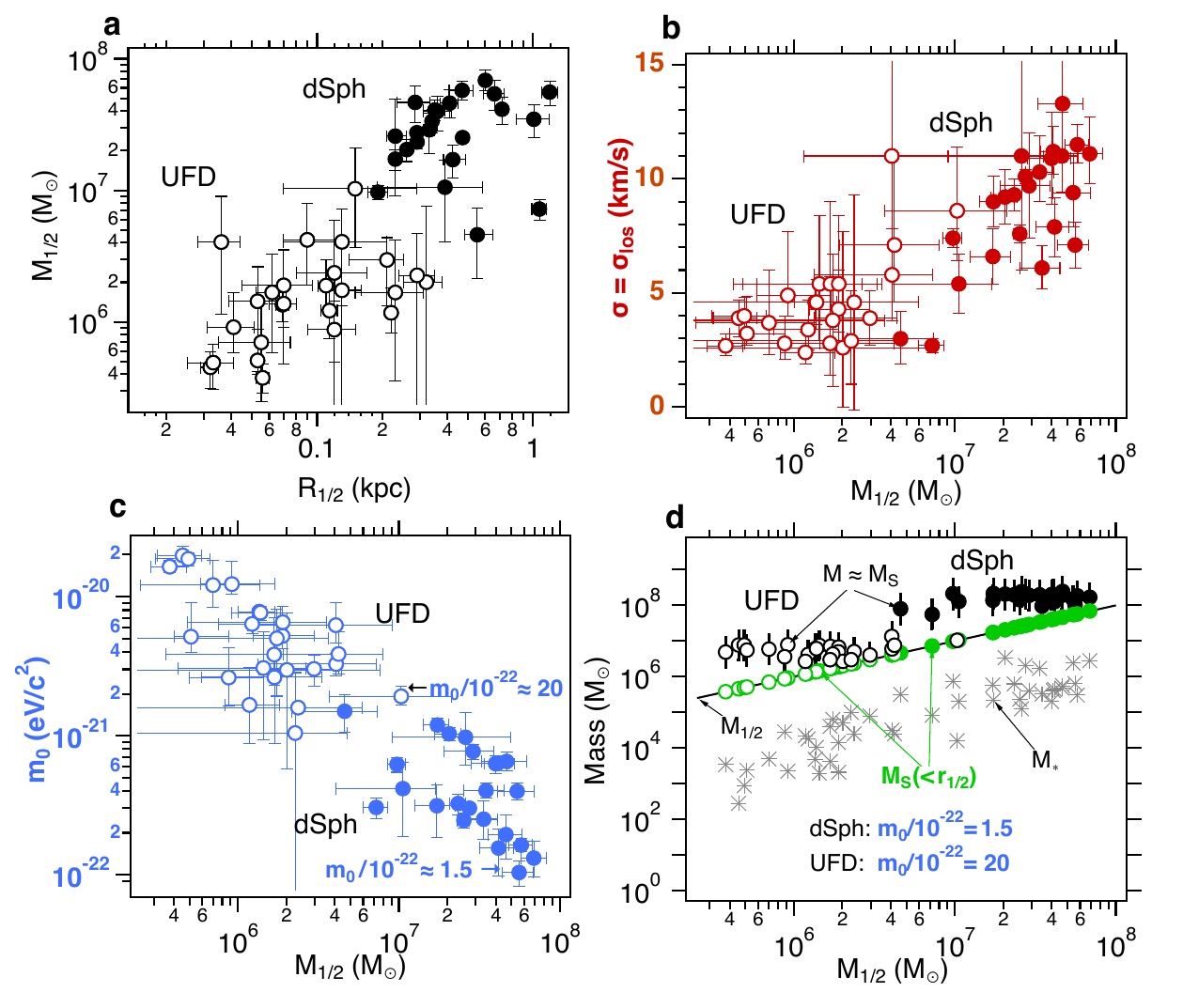}
 \end{center}
\caption{Comparison of the observed and predicted properties of dSph (closed points) and UFD (open points) galaxies.
(a) The enclosed mass $M_{1/2}$ within the half-light radius $R_{1/2}$, obtained from the
 observed $\sigma$ and $R_{1/2}$ \protect\citep{PozDwa24}, 
using the Wolf mass estimate Eq.~\ref{E:MhalfWolf}
\protect\citep{WolAcc10}. (b) Dependence of the observed $\sigma$ on $M_{1/2}$. 
(c) The apparent soliton particle mass $m_0$ obtained assuming $F=0$ and
$\sigma_S=\sigma_\ast$, where $m_0$ is optimised to obtain observationally consistent $M_{1/2}$ predictions.
The two value of $m_0$ pointed to in the figure are those that most accurately reproduce the
observed $\sigma$ on $M_{1/2}$ of the dSph and UFD galaxies, as further described in the text.
(d) The predicted total mass $M \approx M_S$ (black points) and soliton mass with the half-light radius 
$M_{1/2,S}$ (green points), obtained using the lighter and heavier optimal $m_0$ value for the dSph and UFD 
galaxies, respectively. The $\ast$ points correspond to the stellar mass of each galaxy, obtained from its 
observed luminosity $L_{\rm obs}$ ($L_\odot$) assuming a mass-to-light ratio of 1. The corresponding
Jeans analysis demonstrates the observational consistency of these predictions with the observed velocity
dispersions $\sigma$ of each galaxy, as shown in Table~\ref{T:Dual_m0} of Appendix~\ref{app:B}, as
predictions with $|\Delta_\sigma|\le1$ are observationally consistent, as further described in the text (and Appendix~\ref{app:B}.}
 \label{F:Expt_Layout}
\end{figure*}

\section{Observational Comparisons and Interpretations}\label{S:Observational_comparisons}\label{S:dSphUFD}

\begin{figure*}
 \begin{center}
  \includegraphics[width=6 in]{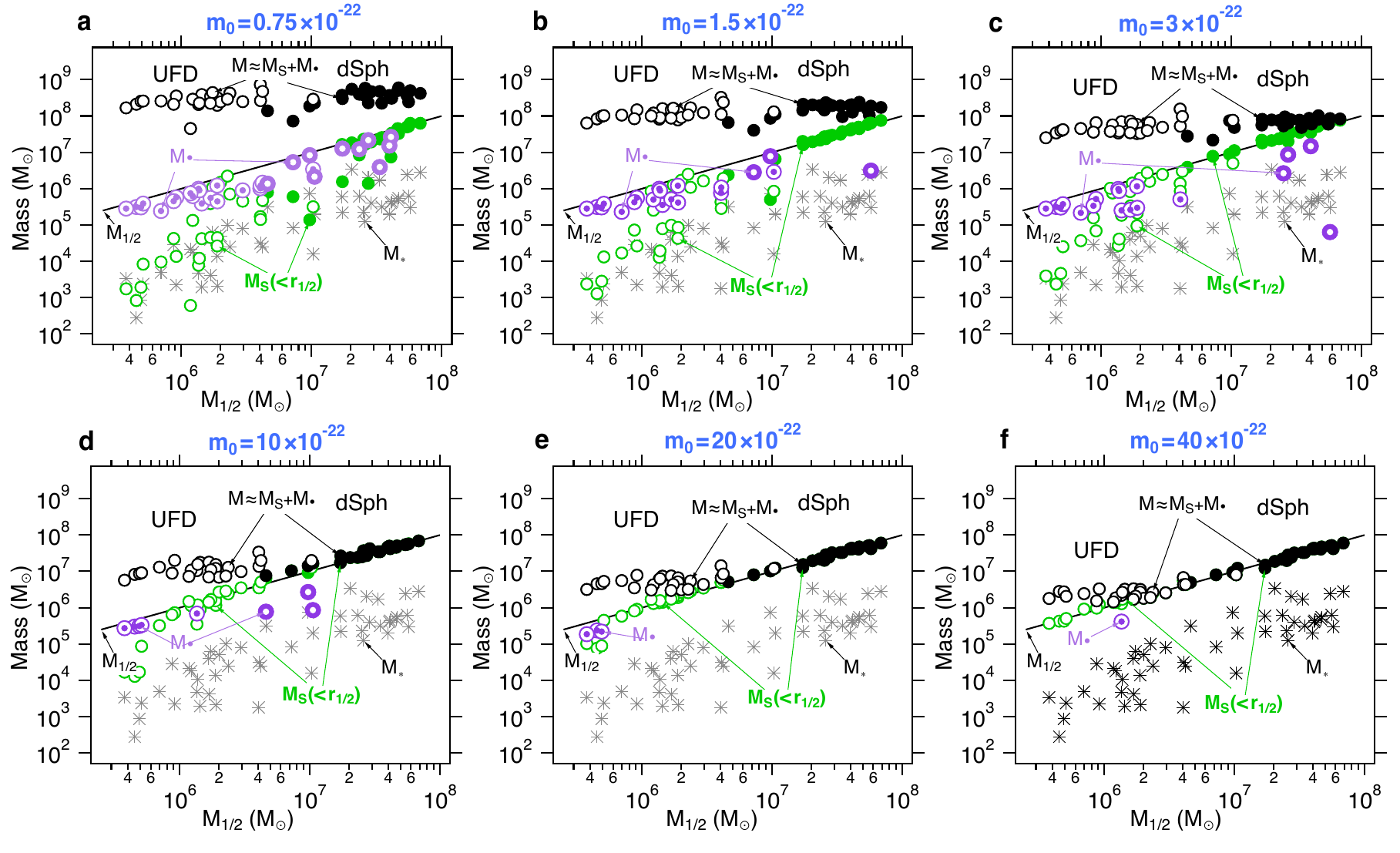}
 \end{center}
\caption{Predictions obtained assuming that all the dSph and UFD galaxies contain the same type of dark matter particle of mass $m_0$, where each panel contains predictions obtained assuming a different value of $m_0$, either within a factor of two of $m_0 = 1.5\times 10^{-22}$\,eV/c$^2$ (upper three panels) or within a factor of two of $m_0 = 20\times 10^{-22}$\,eV/c$^2$ (lower three panels).  The symbols in each panel are the same as those in panel d of Fig.~\ref{F:Expt_Layout}, but now with additional purple dotted-circle points corresponding to the black hole mass $M_\bullet$ predicted to be present in some of the galaxies (the dark and light purple points pertain to dSph and UFD galaxies, respectively).  The predictions pertaining to the results in panels (a)-(c) and (e) are provided in Tables~\ref{T:1x5m0_1SE}--\ref{T:20m0_1SE} in Appendix~\ref{app:B}. The most observationally consistent predictions are those in panel (b), although the predicted number of galaxies containing massive black holes decreases as the observational uncertainty constraints are relaxed, as further described in the text (as well as in Section~\ref{S:Discussion} and Tables~\ref{T:20m0_1SE}--\ref{T:1x5m0_3SE} of  Appendix~\ref{app:B}).}
 \label{F:Expt_Layout_B}
\end{figure*}

The following comparisons illustrate applications of the above shape-shifting predictions, 
highlighting degeneracies between alternative interpretations of the dSph and UFD 
observations shown in Fig.~\ref{F:Layout_vs_f}. The results are found to both validate 
the dual $m_0$ interpretation described by \citet{PozDwa24} (as illustrated in 
Fig.~\ref{F:Expt_Layout}), and demonstrate that the same observations are consistent with 
a single universal $m_0$, implying that many of the UFD galaxies contain massive black 
holes (as illustrated in Fig.~\ref{F:Expt_Layout_B}).

The 5G and spherical-isotropic Jeans analysis results shown in Fig.~\ref{F:Expt_Layout} 
demonstrate the consistency of these predictions with the \citet{PozDwa24} interpretation 
of the dSph and UFD observations, extended to ensure that the predictions are self-consistent with 
the Wolf estimate of the mass $M_{1/2}$ enclosed within the observed (2D projected) 
half-light radius $R_{1/2}$.  

The upper two panels in Figure~\ref{F:Expt_Layout} show the observed (a) $M_{1/2}$ plotted 
as a function of $R_{1/2}$, and (b) $\sigma$ plotted as a function of $M_{1/2}$ for the 
23 dSph and 25 UFD galaxies described by \citet{PozDwa24}, represented by closed and open 
circular points, respectively. The Wolf $M_{1/2}$ values are obtained 
using Eq.~\ref{E:MhalfWolf}, given the observed $\sigma$ and $R_{1/2}$ 
values reported by \citet{PozDwa24}, (which are also provided in Appendix Table~\ref{T:dSph_UFD_data}).

Figure~\ref{F:Expt_Layout}c shows the apparent values of $m_0$ obtained assuming $F=1$ and 
optimising $m_0$ to yield predictions that are self-consistent with the observed $M_{1/2}$. More 
specifically, for each $m_0$, the total galactic mass $M$ is determined using 
Eq.~\ref{E:MSigma}, from the observed value of $\sigma$, with $F=1$. The points 
in Fig.~\ref{F:Expt_Layout}c show the values of $m_0$ for which the 5G-predicted 
$M(<r_{1/2})$, obtained using Eq.~\ref{E:Mx} with $r_{1/2}=\frac{4}{3}R_{1/2}$, equals 
the observationally derived $M_{1/2}$ from Eq.~\ref{E:MhalfWolf}.

\citet{PozDwa24} have suggested that all the dSph and UFD observations can be well 
approximated by assuming that there are only two distinct dark matter particle masses in 
dSph and UFD galaxies, with the UFD $m_0$ about ten times larger than the dSph $m_0$. The 
present 5G and self-consistent $M_{1/2}$ predictions are consistent with that conclusion, 
and further yield optimal values of $m_0 \approx 1.5\times 10^{-22}$\,eV/c$^2$ and 
$m_0 \approx 20\times 10^{-22}$\,eV/c$^2$ for the dSph and UFD galaxies, respectively 
(as indicated in Fig.~\ref{F:Expt_Layout}c). These two values of $m_0$ are those for which 
the spherical-isotropic Jeans predictions minimise the average deviation between the 
predicted and observed velocity dispersions $\sigma$ of the dSph and UFD galaxies (with 
$F=1$). More specifically, the resulting $\sigma$ errors are $0.1\pm 1.1$\,(km/s) for the 
dSph galaxies (with $m_0 \approx 1.5\times 10^{-22}$) and $-0.1\pm 1.3$\,(km/s) for the 
UFD galaxies (with $m_0 \approx 20\times 10^{-22}$). When the assumed values of $m_0$ are 
either increased or decreased by a factor of two, the resulting total errors (mean plus standard deviation)
 approximately double.

Figure~\ref{F:Expt_Layout}d shows 5G predictions obtained 
using the above two $m_0$ values. The black solid and open circle points are the predicted 
total masses $M \approx M_S$ of the dSph and UFD galaxies, respectively. The green solid and 
open circle points are the predicted $M_S(<r_{1/2})$ soliton masses enclosed within the 
stellar half-light radius, which are consistent with the observed values of $M_{1/2}$ 
represented by the solid black line. The $\ast$ points in Fig.~\ref{F:Expt_Layout}d are the 
total stellar masses $M_\ast$, estimated using the observed luminosities \citep{PozDwa24}, 
assuming a stellar mass-to-luminosity ratio of 1. Note that these predictions imply that 
$M_S \gg M_\ast$, and thus are consistent with the often noted conclusion that all the 
dSph and UFD galaxies are dark matter dominated.

Figure~\ref{F:Expt_Layout_B} shows results pertaining to alternative interpretations 
of the same dSph and UFD observations, obtained assuming that all the dSph and UFD 
galaxies have a single universal dark matter particle mass $m_0$ and a variable soliton 
mass fraction $F$, thus allowing for the possible presence of black holes of mass 
$M_\bullet = (1-F) M$ in some of the galaxies. All the predictions in 
Fig.~\ref{F:Expt_Layout_B} were obtained 
by varying both $F$ and $M$ so as to optimise the agreement between the predicted and 
observed values of both $\sigma$ and $M_{1/2}$. More specifically, the predictions in 
Fig.~\ref{F:Expt_Layout_B} were obtained by minimising the maximum value of $|\Delta_\sigma|$ and $|\Delta_M|$, 
defined as $\Delta_\sigma \equiv (\sigma_\text{model} - \sigma_\text{los,obs})/
|\delta\sigma_\text{obs}|$ and $\Delta_M \equiv (M(<r_{1/2}) - M_{1/2,\text{obs}})/
|\delta M_{1/2,\text{obs}}|$, where $\delta\sigma_\text{obs}$ and 
$\delta M_{1/2,\text{obs}}$ are the asymmetric observational uncertainties of $\sigma$ and $R_{1/2}$
 reported by \citet{PozDwa24}, and propagated to 
$M_{1/2}$ using Eq.~\ref{E:MhalfWolf} (as further described in Section~\ref{S:Jeans} and 
Appendix~\ref{app:jeans}).

Each panel in Fig.~\ref{F:Expt_Layout_B} contains results obtained assuming a different 
universal value of $m_0$. The upper three panels pertain to lighter particle masses within 
a factor of two of $m_0 \sim 1.5\times 10^{-22}$\,eV/c$^2$, and the lower three panels 
pertain to heavier masses within a factor of two of $m_0\sim 20\times 10^{-22}$\,eV/c$^2$. 
The symbols in Fig.~\ref{F:Expt_Layout_B} have the same meanings as those in 
Fig.~\ref{F:Expt_Layout}d, with additional purple dotted-circle points corresponding to 
the predicted black hole masses (with dark and light purple points for dSph 
and UFD galaxies, respectively).

\begin{figure*}
 \begin{center}
\includegraphics[width=5 in]{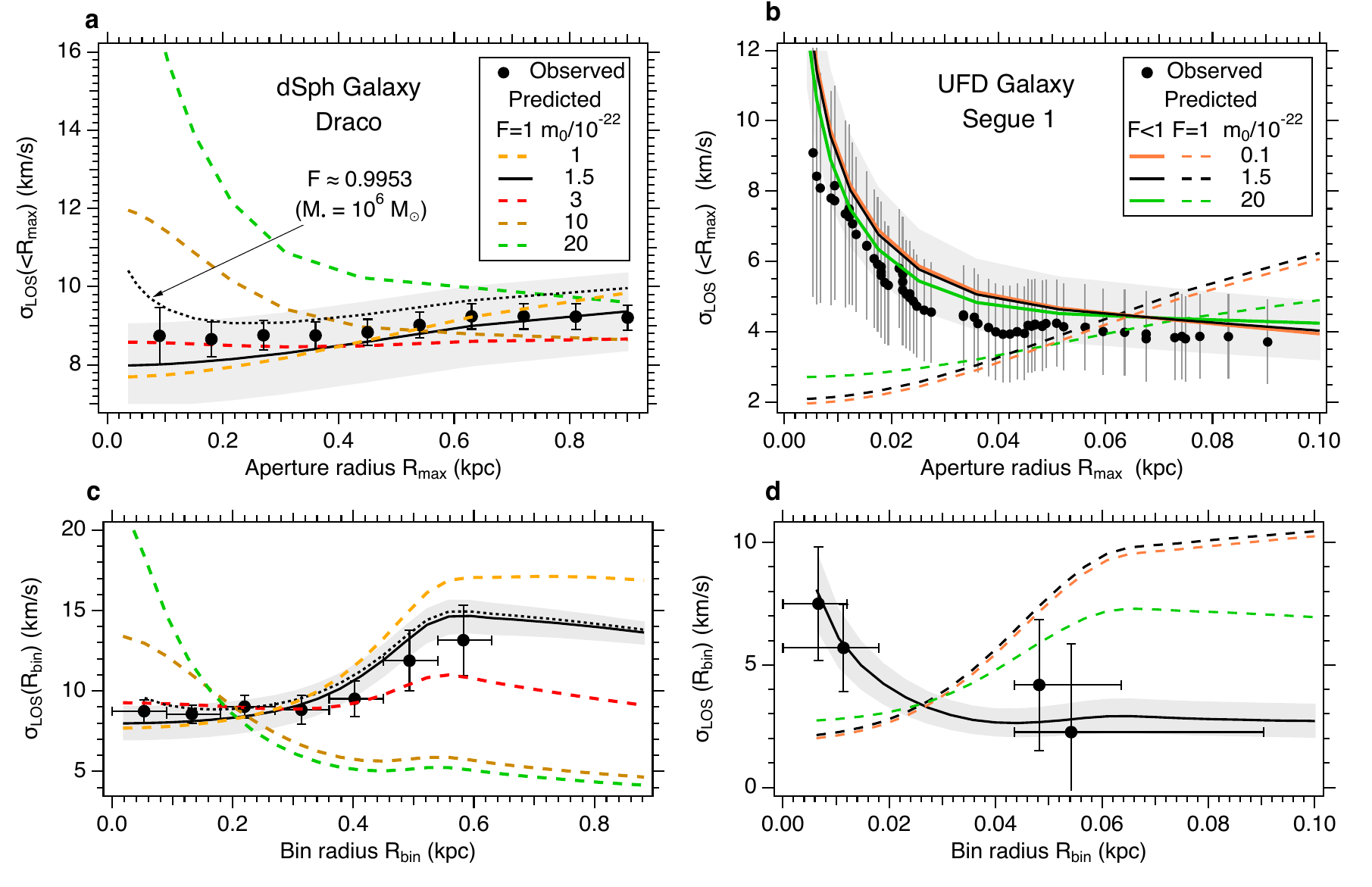}
  \end{center}
 \caption{Comparisons of the observed and predicted Jeans analysis of the radially 
resolved velocity dispersion of Draco (dSph) and Segue~I (UFD) galaxies. The points in 
the upper two panels are accumulated aperture-weighted velocity dispersions plotted as a 
function of aperture radius $R_{\rm max}$, and those in the lower two panels are binned 
velocity dispersions as a function of bin radius $R_{\rm bin}$, obtained from observations 
reported by \citet{Walker15} (for Draco) and \citet{Martinez11} (for Segue~I), as further 
described in the text (as well as Section~\ref{S:Jeans} and Appendix~\ref{app:jeans}). 
Note that the Draco observations are consistent with predictions obtained using 
$m_0\sim 1.5\times 10^{-22}$\,eV/c$^2$, and clearly inconsistent with 
$m_0 \gtrsim 3\times 10^{-22}$\,eV/c$^2$. Most of the predictions for Draco in panels 
(a) and (c) pertain to $F=1$, indicating that these observations do not require the 
presence of a central black hole, although the dotted black curve in panel (a) indicates 
that the observations are also consistent with $m_0\sim 1.5\times 10^{-22}$\,eV/c$^2$ 
and a black hole of mass up to $M_\bullet \le 10^6$\,M$_\odot$. The Segue~I results in 
panels (b) and (d) are clearly inconsistent with $F=1$ predictions (dashed curves), and 
predict the presence of a black hole of mass $M_\bullet \sim 4\times 10^5$\,M$_\odot$ 
(solid curves), in good agreement with a recent independent analysis by \citet{Lujan25}.}
\label{F:Draco_Segue_Fig}
\end{figure*}

All the dSph and UFD galaxies are again predicted to be dark matter dominated with 
$M_S > M_\ast+M_\bullet$. However, these universal $m_0$ 
predictions imply that the observed 
$M_{1/2}$ (solid black diagonal line) is in some cases 
due primarily to dark matter (green points), while in other cases $M_{1/2}$ is predicted 
to be dominated by the central black hole mass $M_\bullet$ (purple points).   The results shown 
in Fig.~\ref{F:Expt_Layout_B} further 
indicate that number of predicted black holes in the dSph and UFD galaxies decreases with increasing $m_0$.  
Predictions obtained assuming the heavier dark particle masses (lower three panels)
 
imply that the soliton radii of some of the galaxies are smaller than the corresponding stellar half-light radii
$R_{1/2}$. More specifically, this is the case for the dSph galaxies with the largest $M_{1/2}$ masses, for 
which $M$ coincides with $M_{1/2}$, implying that essential all the soliton mass is contained within 
the stellar half-light radius.    Although such a  tightly localized soliton is not necessarily impossible the more detailed radially 
resolved velocity dispersion results shown in Fig.~\ref{F:Draco_Segue_Fig} (and further described below) suggest that it is inconsistent with the observed properties of the Draco dSph galaxy.   If such  tightly localized soliton distributions are more generally rejected as physically unrealistic, then that implies that the universal value of $m_0$ must be smaller than about 
$3\times 10^{-22}$\,eV/c$^2$ in order to be compatible with all of the dSph and UFD 
galactic observations.

Figure~\ref{F:Draco_Segue_Fig} contains radially resolved velocity dispersion observations 
and 5G predictions for the Draco dSph galaxy (in the left-hand two panels) and the 
Segue~I UFD galaxy (in the right-hand two panels). The upper two panels contain 
aperture-weighted velocity dispersions accumulated as a function of aperture radius 
$R_{\rm max}$, and the lower two panels contain binned velocity dispersions at various bin 
radii $R_{\rm bin}$. The points are obtained from radially resolved observations of Draco 
\citep{Walker15} and Segue~I \citep{Martinez11}, selected, accumulated and binned as 
described in Appendix~\ref{app:sigma_profiles}. The curves in Fig.~\ref{F:Draco_Segue_Fig} 
are shape-shifting 5G spherical-isotropic Jeans analysis predictions pertaining to various 
values of $m_0$ and $F$ (as indicated in the figure). The solid black curves pertain to 
predictions obtained with $m_0 = 1.5\times 10^{-22}$\,eV/c$^2$ and the total 
(non-radially resolved) $\sigma$ values reported by \citet{PozDwa24}. The corresponding 
gray regions pertain to predictions obtained when varying the observed $\sigma$ by its 
reported observational uncertainty limits \citep{PozDwa24}.

The aperture-weighted and binned Draco dSph galaxy results in Figs.~\ref{F:Draco_Segue_Fig}a 
and c reveal the good agreement between the observed points and the $m_0 = 1.5\times 10^{-22}$\,eV/c$^2$
predictions. The coloured-dashed curves pertain to predictions obtained with different values 
of $m_0$, thus indicating that only $m_0$ values of approximately 
$1$--$3\times 10^{-22}$\,eV/c$^2$ are consistent with the radially resolved velocity 
dispersion observations. The large discrepancy between the measured and predicted results 
obtained with $m_0 = 10$--$20\times 10^{-22}$\,eV/c$^2$ confirms the above
conclusion that such large $m_0$ predictions, and the associated tightly localized soliton distributions, produce physically unrealistic radially resolved velocity dispersion predictions. 

The corresponding Segue~I UFD galaxy results are shown in Fig.~\ref{F:Draco_Segue_Fig}b. 
In this case the radially resolved observations are clearly inconsistent with predictions 
obtained assuming $F=1$ (dashed curves), thus indicating that this galaxy contains a massive 
black hole, consistent with the conclusion reached in a recent independent analysis by 
\citet{Lujan25}. More specifically, the solid curves pertain to predictions obtained with 
$F<1$ and various values of $m_0$. Although the results are in this case quite insensitive 
to the dark matter particle mass, $0.1 \le m_0/10^{-22} \le 20$, they are consistent in 
requiring the presence of a black hole of mass 
$M_\bullet = 3.8^{+0.2}_{-1.5} \times 10^5$\,M$_\odot$, in good agreement with the 
value of $4.5^{+1.5}_{-1.5}\times 10^5$\,M$_\odot$ reported by \citet{Lujan25}. 
Implications of the results in Figs.~\ref{F:Expt_Layout_B} and 
\ref{F:Draco_Segue_Fig} with regard to the possible presence of black holes in other dwarf 
galaxies are further discussed in Section~\ref{S:Discussion}.

\section{Discussion and Implications}\label{S:Discussion}

The present 5G and Jeans analysis predictions describe the influence of a central black hole on the shape of 
ultralight dark matter solitons and the resulting stellar dynamics. Comparisons of the predictions with the observed 
velocity dispersions $\sigma$ and Wolf mass estimates 
$M_{1/2}$ of 48 dwarf galaxies are used to critically test the self-consistency of 
alternative models for both dSph and UFD galaxies. One such model is that 
proposed by \citet{PozDwa24}, who discovered that the stellar cores of these galaxies have 
soliton-like shapes with central densities that scale as the inverse fourth power of the 
stellar core width, but with a slope that is about 10 times larger for the UFD than the 
dSph galaxies. \citet{PozDwa24} further showed that, if the soliton and stellar core 
radii are assumed to be equal, then the observed slope difference implies that dSph and 
UFD galaxies contain different types of dark matter particles, whose masses are about 10 
times larger in UFD than in dSph galaxies \citep{PozDwa24}. The present results are 
consistent with this dual-$m_0$ model, as shown in Fig.~\ref{F:Expt_Layout} (and 
Tables~\ref{T:optimizem0} and \ref{T:Dual_m0}), and are 
additionally constrained to agree with the observed $M_{1/2}$, leading to optimal dark 
matter particle masses of $m_0 \approx 1.5\times 10^{-22}$\,eV/c$^2$ for the dSph 
galaxies and $m_0 \approx 20\times 10^{-22}$\,eV/c$^2$ for the UFD galaxies.

Alternatively, the present results demonstrate that the observed $\sigma$ and $M_{1/2}$ 
values of all the dSph and UFD galaxies are also consistent with a
single universal dark matter particle mass of $m_0 \approx 1.5^{+1.5}_{-0.8}\times 
10^{-22}$\,eV/c$^2$, as long as many of the UFD galaxies contain central black holes 
with masses of the order of $10^5$--$10^6$\,M$_\odot$, which are 
marginally consistent with recent CDM-based estimates of an upper bound of 
$10^5$--$10^6$\,M$_\odot$ on the mass of black holes in dSph  
galaxies \citep{Aditya26}.  However, the present predictions are 
not tightly constrained with regard to either the magnitude of the black hole mass or the 
number of dwarf galaxies that contain such black holes. For example, predictions consistent 
with observational uncertainties of $\sigma$ and $M_{1/2}$ indicate that 17 out of the 25 
UFD galaxies may contain black holes with $M_\bullet > 10^5$\,M$_\odot$ 
(see Fig.~\ref{F:Expt_Layout_B}b and Appendix 
Table~\ref{T:1x5m0_1SE}). However, if the uncertainty constraint is relaxed to between 2 
and 3 times the reported observational uncertainties, then the predicted number of UFD 
black holes decreases to 8 and 4, respectively (see Appendix 
Tables~\ref{T:1x5m0_2SE} and \ref{T:1x5m0_3SE}, respectively). The latter four UFD 
galaxies, Segue~I, Coma~Berenices, Reticulum~II, and Hydrus~I, are thus the most likely 
to contain massive black holes (assuming that the observationally derived values 
of $\sigma$, $R_{1/2}$, $M_{1/2}$ and $R_{\rm max}$, in Table~\ref{T:dSph_UFD_data} are 
self-consistent and correct).  Note that these galaxies include Segue~I, for which the 
present more detailed radially resolved velocity dispersion Jeans analysis has confirmed 
the presence of a black hole of mass $M_\bullet = 3.8^{+0.2}_{-1.5} \times 
10^5$\,M$_\odot$, in good agreement with the results of a recent independent analysis 
by \citet{Lujan25}.

The present predictions imply that some dSph galaxies may also contain supermassive black holes.
However, the predicted mass and number of black holes in dSph galaxies depend quite sensitively on 
the assumed value of $m_0$.  The predictions obtained assuming $m_0 = 1.5\times 10^{-22}$\,eV/c$^2$ 
(see Appendix Table~\ref{T:1x5m0_1SE}) indicate that the Crater II, Leo II, and Ursa Minor dSph galaxies may 
contain black holes with masses of  $3\times 10^6 \lesssim M_\bullet \lesssim 8\times10^6$\,M$_\odot$, 
which exceed the CDM-derived upper bound of $10^5$--$10^6$\,M$_\odot$ obtained by \citet{Aditya26} 
for classical dSph galaxies, and are thus in apparent tension with those bounds.

The predicted absence of black holes in other dSph galaxies does not preclude the 
possibility that they may nevertheless contain black holes,
since measurements of $\sigma$ and $M_{1/2}$ alone may not be 
sufficient to reveal the presence of a black hole. For example, the 
observed $\sigma$ and $M_{1/2}$ values of the Leo~I dSph galaxy are consistent with $F=1$ 
predictions, and thus do not require the presence of a massive black hole. However, a more 
detailed dynamical analysis by \citet{BusDyn21} has concluded that Leo~I contains a black 
hole of mass $M_\bullet = 3.3_{-2}^{+2}\times 10^6$\,M$_\odot$. Thus, although the 
observed average $\sigma \approx 9.2^{+1.2}_{-1.2}$\,(km/s) of Leo I \citep{MccObs12} does not require 
the presence of a black hole, the rise in $\sigma(R_{\rm ap})$ to approximately 12\,(km/s) at 
$R_{\rm ap} \sim 0.1$\,(kpc), seen in the radially resolved observations \citep{BusDyn21}, provides strong 
evidence of the influence of a large localized central mass.  As another example, the observed values of $\sigma$ and 
$M_{1/2}$ for the Draco dSph galaxy do not require the presence of a massive black hole, 
although the more detailed analysis of the radially resolved velocity measurements 
shown in Fig.~\ref{F:Draco_Segue_Fig}a indicates that Draco may contain a black 
hole of mass $M_\bullet \lesssim 10^6$\,M$_\odot$, as indicated by the dotted black 
curve in Fig.~\ref{F:Draco_Segue_Fig}a.

More broadly, the present predictions that many UFD and some dSph galaxies may host 
massive black holes are consistent with the growing observational census 
indicating that black holes may be considerably more common in low-mass galaxies than 
previously appreciated \citep{PucTri25}. Specifically, \citet{PucTri25} used early DESI 
data to identify $\sim$2,500 dwarf galaxies with active-galactic-nucleus (AGN) signatures 
and $\sim$300 intermediate-mass black hole 
candidates ($M_\bullet \lesssim 10^6\,M_\odot$), of which 
70 reside specifically in dwarf-mass hosts.

The present predictions also allow for
more exotic interpretations of dSph 
and UFD observations --- some of which may be physically unrealistic --- ranging from the 
complete absence of dark matter to the presence of sharply localised dark matter solitons 
in some dwarf galaxies. For example, the predictions in Appendix Table~\ref{T:F0} imply 
that the observed $\sigma$ and $M_{1/2}$ of many dSph and UFD galaxies may be consistent 
with their having a supermassive black hole and essentially no dark matter.  Alternatively,  
the $m_0 \sim 20\times 10^{-22}$\,eV/c$^2$ predictions shown in Figs.~\ref{F:Expt_Layout_B}e 
and Appendix  Table~\ref{T:20m0_1SE} are consistent with the absence of black holes in 
most of the dwarf galaxies, as long as the dSph galaxies have sharply localised dark matter 
solitons with radii of the order of $r_{c,S} \sim 0.01$\,(kpc), roughly 10 times smaller than the corresponding stellar $R_{1/2}$ and $r_c$ (as shown in Appendix Table~\ref{T:dSph_UFD_data}). 
Note that such sharply localized solitons with $r_{c,S} \ll r _c$ lead to a predicted increase
 in velocity dispersion with decreasing radius that can mimic the influence of a central black hole.
 However, as noted above, such large $m_0$ predictions are inconsistent with the radially
resolved velocity dispersion measurements of Draco (see Figs.~\ref{F:Draco_Segue_Fig}a and c).
Thus, if dwarf galaxies are in fact dark matter dominated with a universal dark matter $m_0$
and soliton width comparable to or larger than the stellar distribution, then the present predictions 
require that $m_0$ must be smaller than $m_0 \lesssim 3\times 10^{-22}$\,eV/c$^2$.

Although the present universal-$m_0$ predictions may provide a viable alternative interpretation of dSph and UFD observations, they also introduce the following apparent tensions, all of which can be reconciled within the present framework. The apparent tension between the observed $\sigma$ and $M_{1/2}$ of dSph and UFD galaxies and predictions obtained assuming a universal $m_0$ and $F=1$ (no black hole) is reconciled if many UFD galaxies host black holes ($F < 1$) with masses of order $10^5$--$10^6$\,M$_\odot$, marginally consistent with predicted bounds \citep{Aditya26}, or if the soliton and stellar radii differ significantly, $r_{c,S} \ne r_{c,\ast}$, and thus $\sigma_S \ne \sigma_\ast$ (relaxing the constraint implicit in Eq.~\ref{E:vsSigma}). The resulting universal-$m_0$ predictions imply that all the dSph and UFD galaxies share similar total masses of order $10^8$\,M$_\odot$ (stars, soliton, and black hole combined), which may appear to be in tension with the 10--100 times smaller Wolf $M_{1/2}$ masses of UFD relative to dSph galaxies, though this apparent tension is again resolved by relaxing the above constraints on $F$ and $r_{c,S}$ or $\sigma_S$. An additional potential tension is raised by the prediction that a central black hole can threaten the long-term survival of a fuzzy dark matter soliton core \citep{Cardoso22}. However, the present soliton plus black hole models reside deep in the ``slow-accretion'' regime, since $m_0 \lesssim 3\times 10^{-22}$\,eV/c$^2$ is far below the $8\times10^{-20}$\,eV/c$^2$ critical threshold predicted by \citet{Cardoso22}, and thus the predicted timescale for catastrophic accretion vastly exceeds the Hubble time.

Additionally, the present universal-$m_0$ upper bound of 
$m_0 \lesssim 3\times 10^{-22}$\,eV/c$^2$ is in apparent tension with mass bounds 
obtained from various other large- and small-scale galactic structure probes \citep{EbeUlt25}. More 
specifically, large-scale structure probes (Lyman-$\alpha$ forest, halo and subhalo mass 
functions) suggest that $m_0 \gtrsim \text{few}\times 10^{-21}$--$10^{-20}$\,eV/c$^2$, 
while small-scale probes (cores, heating, strong lensing) tend to favour 
$m_0 \gtrsim 10^{-21}$--$10^{-19}$\,eV/c$^2$. Resolving these tensions will require 
either (i) identifying alternative physical explanations of the above large- and 
small-scale observations that relax the constraints on dark matter particle mass, or (ii) accepting that all dark matter particles have $m_0 \gtrsim  10^{-21}$\,eV/c$^2$, 
and thus that dSph galaxies contain sharply localized dark matter soliton cores whose width 
is much smaller than $R_{1/2}$ (although that is apparently inconsistent with the Draco results
shown in Fig.~\ref{F:Draco_Segue_Fig}a) or (iii) entirely rejecting the hypothesis that dark matter 
is composed primarily of ultralight particles.

If dark matter is composed primarily of ultralight particles, then the present results 
provide a hierarchy of strategies for determining the structure of dwarf galaxies from 
their available observable properties. Given only $\sigma$, one may use 
Eqs.~\ref{E:vsSigma} and \ref{E:RcSigma}--\ref{E:MSigma}, to estimate the dark matter 
soliton radius $r_c$, central dark matter density $\rho_\odot$, 
and total galactic mass $M$, whose dependence on $m_0$ and $F$ is graphically represented 
in Fig.~\ref{F:Layout_vs_f}c--e.  A Jeans analysis, combined with the Wolf mass estimate 
(Eq~\ref{E:MhalfWolf}) may be used to narrow the range of $m_0$ and $F$ values that are 
observationally consistent, as illustrated by the results shown 
in Figs.~\ref{F:Expt_Layout} and \ref{F:Expt_Layout_B} (and Appendix Tables~\ref{T:optimizem0} to \ref{T:F0}). 
Beyond that, critically testing 
alternative galactic models and more definitively detecting the influence of black holes on 
soliton shape can be achieved using radially resolved velocity measurements, as illustrated 
in Fig.~\ref{F:Draco_Segue_Fig}.

\section*{Acknowledgement}

This work benefited from discussions with Lyudmila Slipchenko, Pierre-Henri Chavanis and K. Aditya. Perplexity AI was used to help navigate the relevant literature, Jeans programing, and proof reading -- the author is responsible for all of the analyses, results and conclusions.

\section*{Data Availability}

No new observational data were generated or analysed in this study. All observational quantities used in this work are taken from the published literature, as cited in the text. Polynomial coefficients used to generate the five-Gaussian soliton shapes and other $F$-dependent functions plotted in the figures are provided in the Appendix along with tables containing observations and predictions pertaining to dSph and UFD galaxies described by \citet{PozDwa24}. Python code for regenerating the present 5G soliton shapes as a function of $F$ is available from the author.

\bibliographystyle{mnras}
\bibliography{dark-light}

@ARTICLE{MatSho24,
 AUTHOR={Matos, Tonatiuh  and Ureña-López, Luis A.  and Lee, Jae-Weon },
TITLE={Short review of the main achievements of the scalar field, fuzzy, ultralight, wave, BEC dark matter model},
JOURNAL={Frontiers in Astronomy and Space Sciences},
VOLUME={Volume 11},
PAGES={1347518},
YEAR={2024},
URL={https://www.frontiersin.org/journals/astronomy-and-space-sciences/articles/10.3389/fspas.2024.1347518},
DOI={10.3389/fspas.2024.1347518},
ISSN={2296-987X}
}

@article{HuFuz20,
   author = {Hu, Wayne and Barkana, Rennan and Gruzinov, Andrei},
   title = {Fuzzy cold dark matter: The wave properties of ultralight particles},
   journal = {Phys. Rev. Lett.},
   volume = {85},
   number = {6},
   pages = {1158-1161},
   year = {2000}
}

@article{MemHar89,
   author = {Membrado, M. and Pacheco, A. F. and Sanudo, J.},
   title = {Hartree solutions for the self-{Y}ukawian boson sphere},
   journal = {Phys Rev A Gen Phys},
   volume = {39},
   number = {8},
   pages = {4207-4211},
   year = {1989}
}

@article{MemBos18,
   author = {Membrado, M. and Pacheco, A. F.},
   title = {{Bose-Einstein} condensate haloes embedded in dark energy},
   journal = {Astronomy \& Astrophysics},
   volume = {611},
   pages = {A81},
   year = {2018}
}

@article{SchCos14,
   author = {Schive, Hsi-Yu and Chiueh, Tzihong and Broadhurst, Tom},
   title = {Cosmic structure as the quantum interference of a coherent dark wave},
   journal = {Nat. Phys.},
   volume = {10},
   number = {7},
   pages = {496-499},
   year = {2014}
}

@article{Rinpar23,
   author = {Rindler-Daller, Tanja},
   title = {On particle scattering in {Gross-Pitaevskii} theory and implications for dark matter halos},
   journal = {Frontiers in Astronomy and Space Sciences},
   volume = {10},
   year = {2023},
   pages = {1121920},
  DOI={10.3389/fspas.2023.1121920}
}

@article{Rufsys69,
   author = {Ruffini, Remo and Bonazzola, Silvano},
   title = {Systems of self-gravitating particles in general relativity and the concept of an equation of state},
   journal = {Physical Review},
   volume = {187},
   number = {5},
   pages = {1767-1783},
   year = {1969}
}

@article{PasAct18,
   author = {Pascale, Raffaele and Posti, Lorenzo and Nipoti, Carlo and Binney, James},
   title = {Action-based dynamical models of dwarf spheroidal galaxies: Application to {Fornax}},
   journal = {Monthly Notices of the Royal Astronomical Society},
   volume = {480},
   number = {1},
   pages = {927-946},
   year = {2018}
}

@article{BarGal18,
   author = {Bar, Nitsan and Blas, Diego and Blum, Kfir and Sibiryakov, Sergey},
   title = {Galactic rotation curves versus ultralight dark matter: Implications of the soliton-host halo relation},
   journal = {Phys. Rev. D},
   volume = {98},
   number = {8},
   pages = {083027},
   year = {2018}
}

@article{GolVia22,
   author = {Goldstein, Isabelle S. and Koushiappas, Savvas M. and Walker, Matthew G.},
   title = {Viability of ultralight {Bosonic} dark matter in dwarf galaxies},
   journal = {Phys. Rev. D},
   volume = {106},
   number = {6},
   pages = {063010},
   year = {2022}
}

@article{PozDet24,
   author = {Pozo, Alvaro and Broadhurst, Tom and de Martino, Ivan and Chiueh, Tzihong and Smoot, George F. and Bonoli, Silvia and Angulo, Raul},
   title = {Detection of a universal core-halo transition in dwarf galaxies as predicted by {Bose-Einstein} dark matter},
   journal = {Phys. Rev. D},
   volume = {110},
   number = {4},
   pages = {043534},
   year = {2024}
}

@article{PozDwa24,
   author = {Pozo, Alvaro and Broadhurst, Tom and Smoot, George F. and Chiueh, Tzihong and Luu, Hoang Nhan and Vogelsberger, Mark and Mocz, Philip},
   title = {Dwarf galaxies united by dark bosons},
   journal = {Phys. Rev. D},
   volume = {109},
   number = {8},
   pages = {083532},
   year = {2024}
}

@article{RelDar19,
   author = {Relatores, Nicole C. and Newman, Andrew B. and Simon, Joshua D. and Ellis, Richard S. and Truong, Phuongmai and Blitz, Leo and Bolatto, Alberto and Martin, Christopher and Matuszewski, Matt and Morrissey, Patrick and Neill, James D.},
   title = {The dark matter distributions in low-mass disk galaxies. {II}. The inner density profiles},
   journal = {The Astrophysical Journal},
   volume = {887},
   number = {1},
   pages = {94},
   year = {2019}
}

@article{NavStr96,
   author = {Navarro, Julio F. and Frenk, Carlos S. and White, Simon D. M.},
   title = {The structure of cold dark matter halos},
   journal = {The Astrophysical Journal},
   volume = {462},
   pages = {563},
   year = {1996}
}

@book{BenUnd14,
   author = {Ben-Amotz, D.},
   title = {Understanding physical chemistry},
   publisher = {John Wiley and Sons},
   address = {New York},
   year = {2014}
}

@article{LazAcc20,
   author = {Lazar, Alexandres and Bullock, James S.},
   title = {Accurate mass estimates from the proper motions of dispersion-supported galaxies},
   journal = {Monthly Notices of the Royal Astronomical Society},
   volume = {493},
   number = {4},
   pages = {5825-5837},
   year = {2020}
}

@article{ChaQua11,
doi = {10.1088/1742-5468/2011/08/P08003},
url = {https://doi.org/10.1088/1742-5468/2011/08/P08003},
year = {2011},
month = {aug},
publisher = {},
volume = {2011},
number = {08},
pages = {P08003},
author = {Chavanis, Pierre-Henri},
title = {The quantum HMF model: II. Bosons},
journal = {Journal of Statistical Mechanics: Theory and Experiment}
}

@article{WolAcc10,
   author = {Wolf, Joe and Martinez, Gregory D. and Bullock, James S. and Kaplinghat, Manoj and Geha, Marla and Munoz, Ricardo R. and Simon, Joshua D. and Avedo, Frank F.},
   title = {Accurate masses for dispersion-supported galaxies},
   journal = {Monthly Notices of the Royal Astronomical Society},
   volume = {406},
   pages = {1220-1237},
   year = {2010}
}

@article{ChaDar21,
   author = {Chang, Laura J. and Necib, Lina},
   title = {Dark matter density profiles in dwarf galaxies: Linking jeans modelling systematics and observation},
   journal = {Monthly Notices of the Royal Astronomical Society},
   volume = {507},
   number = {4},
   pages = {4715-4733},
   year = {2021}
}

@article{NieSma20,
author = {Jens C. Niemeyer},
title = {Small-scale structure of fuzzy and axion-like dark matter},
journal = {Progress in Particle and Nuclear Physics},
volume = {113},
pages = {103787},
year = {2020},
issn = {0146-6410},
doi = {https://doi.org/10.1016/j.ppnp.2020.103787},
url = {https://www.sciencedirect.com/science/article/pii/S014664102030034X},
}

@article{SerLin16,
   author = {Serra, Paolo and Oosterloo, Tom and Cappellari, Michele and den Heijer, Milan and Jozsa, Gyula I. G.},
   title = {Linear relation between HI circular velocity and stellar velocity dispersion in early-type galaxies, and slope of the density profiles},
   journal = {Monthly Notices of the Royal Astronomical Society},
   volume = {460},
   number = {2},
   pages = {1382-1389},
   year = {2016}
}

@article{ChaPre19,
   author = {Chavanis, Pierre-Henri},
   title = {Predictive model of bec dark matter halos with a solitonic core and an isothermal atmosphere},
   journal = {Phys. Rev. D},
   volume = {100},
   number = {8},
   pages = {083022},
   year = {2019}
}

@article{DavFuz20,
   author = {Davies, Elliot Y and Mocz, Philip},
   title = {Fuzzy dark matter soliton cores around supermassive black holes},
   journal = {Monthly Notices of the Royal Astronomical Society},
   volume = {492},
   number = {4},
   pages = {5721-5729},
   year = {2020}
}

@article{MooEvi94,
   author = {Moore, Ben},
   title = {Evidence against dissipation-less dark matter from observations of galaxy haloes},
   journal = {Nature},
   volume = {370},
   number = {6491},
   pages = {629-631},
   year = {1994}
}

@article{KauFor93,
   author = {Kauffmann, G. and White, S. D. M. and Guiderdoni, B.},
   title = {The formation and evolution of galaxies within merging dark matter haloes},
   journal = {Monthly Notices of the Royal Astronomical Society},
   volume = {264},
   number = {1},
   pages = {201-218},
   year = {1993}
}

@article{KorVor25,
  title = {Vortex lines in ultralight bosonic dark matter around rotating supermassive black holes},
  author = {Korshynska, K. and Prykhodko, O. O. and Gorbar, E. V. and Jia, Junji and Yakimenko, A. I.},
  journal = {Phys. Rev. D},
  volume = {111},
  issue = {2},
  pages = {023006},
  numpages = {13},
  year = {2025},
  month = {Jan},
  publisher = {American Physical Society},
  doi = {10.1103/PhysRevD.111.023006},
  url = {https://link.aps.org/doi/10.1103/PhysRevD.111.023006}
}

@article{Aditya26,
doi = {10.3847/1538-4357/ae2d4f},
url = {https://doi.org/10.3847/1538-4357/ae2d4f},
year = {2026},
month = {jan},
publisher = {The American Astronomical Society},
volume = {997},
number = {2},
pages = {194},
author = {Aditya, K. and Mangalam, A.},
title = {Can Dwarf Spheroidal Galaxies Host a Central Black Hole?},
journal = {The Astrophysical Journal}
}

@article{ReiHun22,
   author = {Reines, Amy E.},
   title = {Hunting for massive black holes in dwarf galaxies},
   journal = {Nature Astronomy},
   volume = {6},
   number = {1},
   pages = {26-34},
   year = {2022}
}

@article{BusDyn21,
doi = {10.3847/1538-4357/ac0c79},
url = {https://doi.org/10.3847/1538-4357/ac0c79},
year = {2021},
month = {nov},
publisher = {The American Astronomical Society},
volume = {921},
number = {2},
pages = {107},
author = {Bustamante-Rosell, M. J. and Noyola, Eva and Gebhardt, Karl and Fabricius, Maximilian H. and Mazzalay, Ximena and Thomas, Jens and Zeimann, Greg},
title = {Dynamical Analysis of the Dark Matter and Central Black Hole Mass in the Dwarf Spheroidal Leo I},
journal = {The Astrophysical Journal}
}

@article{MccObs12,
doi = {10.1088/0004-6256/144/1/4},
url = {https://doi.org/10.1088/0004-6256/144/1/4},
year = {2012},
month = {jun},
publisher = {The American Astronomical Society},
volume = {144},
number = {1},
pages = {4},
author = {McConnachie, Alan W.},
title = {THE OBSERVED PROPERTIES OF DWARF GALAXIES IN AND AROUND THE LOCAL GROUP},
journal = {The Astronomical Journal}
}

@article{LiaDec25,
   author = {Liao, P. Y. and Su, G. M. and Schive, H. Y. and Kunkel, A. and Huang, H. and Chiueh, T.},
   title = {Deciphering the soliton-halo relation in fuzzy dark matter},
   journal = {Phys Rev Lett},
   volume = {135},
   number = {6},
   pages = {061002},
   year = {2025}
}

@article{MayUpd25,
  title={Updated bounds on ultra-light dark matter from the tiniest galaxies},
  author={May, Simon and Dalal, Neal and Kravtsov, Andrey},
  journal={arXiv preprint arXiv:2509.02781},
  year={2025}
}

@article{SipFuz25,
   author = {Sipple, Jackson and Lidz, Adam and Grin, Daniel and Sun, Guochao},
   title = {Fuzzy dark matter constraints from the hubble frontier fields},
   journal = {Monthly Notices of the Royal Astronomical Society},
   volume = {538},
   number = {3},
   pages = {1830-1842},
   year = {2025}
}

@article{LiuWar25,
   author = {Liu, Rayne and Hu, Wayne and Xiao, Huangyu},
   title = {Warm and fuzzy dark matter: Free streaming of wave dark matter},
   journal = {Phys. Rev. D},
   volume = {111},
   number = {2},
   pages = {023535},
   year = {2025}
}

@article{RogStr21,
   author = {Rogers, K. K. and Peiris, H. V.},
   title = {Strong bound on canonical ultralight axion dark matter from the lyman-alpha forest},
   journal = {Phys Rev Lett},
   volume = {126},
   number = {7},
   pages = {071302},
   year = {2021}
}

@article{SimFai19,
   author = {Simon, Joshua D.},
   title = {The faintest dwarf galaxies},
   journal = {Annual Review of Astronomy and Astrophysics},
   volume = {57},
   number = {1},
   pages = {375-415},
   year = {2019}
}

@article{EbeUlt25,
   author = {Eberhardt, Andrew and Ferreira, Elisa GM},
   title = {Ultralight fuzzy dark matter review},
   journal = {arXiv preprint arXiv:2507.00705},
   year = {2025}
}

@article{ChaMas11,
   author = {Chavanis, Pierre-Henri},
   title = {Mass-radius relation of newtonian self-gravitating Bose-Einstein condensates with short-range interactions. I. Analytical results},
   journal = {Phys. Rev. D},
   volume = {84},
   number = {4},
   pages = {043531},
   year = {2011}
}

@article{Chavanis19,
  author       = {Chavanis, P.-H.},
  title        = {Jeans instability of self-gravitating Bose--Einstein condensates with a single Gaussian ansatz},
  journal      = {Eur. Phys. J. Plus},
  year         = {2019},
  volume       = {134},
  number       = {7},
  pages        = {352},
  doi          = {10.1140/epjp/i2019-12767-2},
  archivePrefix= {arXiv},
  eprint       = {1905.08137}
}

@article{Alcubierre18,
  author       = {Alcubierre, M. and Barranco, J. and Bernal, A. and Degollado, J. C.},
  title        = {Approximation methods in the study of boson stars},
  journal      = {Phys. Rev. D},
  year         = {2018},
  volume       = {98},
  pages        = {123013},
  doi          = {10.1103/PhysRevD.98.123013},
  archivePrefix= {arXiv},
  eprint       = {1808.01711}
}

@article{Hiyama03,
  author       = {Hiyama, E. and Kino, Y. and Kamimura, M.},
  title        = {Gaussian expansion method for few-body systems},
  journal      = {Prog. Part. Nucl. Phys.},
  year         = {2003},
  volume       = {51},
  pages        = {223--307},
  doi          = {10.1016/S0146-6410(03)90015-9}
}

@article{Perez96,
  author       = {Perez-Garcia, V. M. and Michinel, H. and Cirac, J. I. and Lewenstein, M. and Zoller, P.},
  title        = {Low energy excitations of a Bose--Einstein condensate: A time-dependent variational analysis},
  journal      = {Phys. Rev. Lett.},
  year         = {1996},
  volume       = {77},
  pages        = {5320--5323},
  doi          = {10.1103/PhysRevLett.77.5320}
}

@ARTICLE{Chavanis25,
AUTHOR={Chavanis, Pierre-Henri },     
TITLE={A review of basic results on the {Bose€-Einstein} condensate dark matter model},     
JOURNAL={Frontiers in Astronomy and Space Sciences},  
VOLUME={12},
PAGES={1538434},
YEAR={2025},
URL={https://www.frontiersin.org/journals/astronomy-and-space-sciences/articles/10.3389/fspas.2025.1538434},
DOI={10.3389/fspas.2025.1538434},
ISSN={2296-987X}
}

@article{ChaMas19,
   author = {Chavanis, Pierre-Henri},
   title = {Mass-radius relation of self-gravitating Bose-Einstein condensates with a central black hole},
   journal = {The European Physical Journal Plus},
   volume = {134},
   number = {7},
   pages = {352},
   year = {2019}
}

@article{Oguri26_arXiv,
    author = "Oguri, Masamune and Kubo, Naoi",
    title = "{Fuzzy dark matter soliton core hosting a supermassive black hole as a dense low-mass perturber in strong gravitational lensing}",
    doi = "10.1088/1475-7516/2026/07/088",
    journal = "JCAP",
    volume = "07",
    pages = "088",
    year = "2026"
}

@article{Vegetti26,
    author = "Vegetti, Simona and White, Simon D. M. and McKean, John P. and Powell, Devon M. and Spingola, Cristiana and Massari, Davide and Despali, Giulia and Fassnacht, Christopher D.",
    title = "{A possible challenge for cold and warm dark matter}",
    eprint = "2601.02466",
    archivePrefix = "arXiv",
    primaryClass = "astro-ph.CO",
    doi = "10.1038/s41550-025-02746-w",
    journal = "Nature Astron.",
    volume = "10",
    number = "3",
    pages = "440--447",
    year = "2026"
}

@article{Tan25,
   author = {Tan, Chen and Le Delliou, M. and Wang, Ke},
   title = {Diversity of fuzzy dark matter solitons},
   journal = {Phys. Rev. D},
   volume = {112},
   number = {6},
   pages = {063021},
   year = {2025}
}

@article{Lujan25,
doi = {10.3847/2041-8213/ae0b4f},
url = {https://doi.org/10.3847/2041-8213/ae0b4f},
year = {2025},
month = {oct},
publisher = {The American Astronomical Society},
volume = {992},
number = {2},
pages = {L25},
author = {Lujan, Nathaniel and Gebhardt, Karl and Anantua, Richard and Chase, Owen and Debski, Maya H. and Finley, Claire and Gomez, Loraine V. and Gupta, Om and Lawson, Alex J. and Marron, Izabella and Martinez, Zorayda and Painter, Connor A. and Sklansky, Yonatan and West, Hayley},
title = {Modeling the "Dark-matter Dominated" Dwarf Galaxy Segue 1 with a Supermassive Black Hole},
journal = {The Astrophysical Journal Letters},
}

@article{Simon11,
doi = {10.1088/0004-637X/733/1/46},
url = {https://doi.org/10.1088/0004-637X/733/1/46},
year = {2011},
month = {may},
publisher = {The American Astronomical Society},
volume = {733},
number = {1},
pages = {46},
author = {Simon, Joshua D. and Geha, Marla and Minor, Quinn E. and Martinez, Gregory D. and Kirby, Evan N. and Bullock, James S. and Kaplinghat, Manoj and Strigari, Louis E. and Willman, Beth and Choi, Philip I. and Tollerud, Erik J. and Wolf, Joe},
title = {A COMPLETE SPECTROSCOPIC SURVEY OF THE MILKY WAY SATELLITE SEGUE 1: THE DARKEST GALAXY*},
journal = {The Astrophysical Journal}
}

@article{Walker15,
   author = {Walker, Matthew G. and Olszewski, Edward W. and Mateo, Mario},
   title = {Bayesian analysis of resolved stellar spectra: Application to {MMT/Hectochelle} observations of the {Draco} dwarf spheroidal},
   journal = {Monthly Notices of the Royal Astronomical Society},
   volume = {448},
   number = {3},
   pages = {2717-2732},
   year = {2015}
}

@article{Massari20,
	author = {{Massari, D.} and {Helmi, A.} and {Mucciarelli, A.} and {Sales, L. V.} and {Spina, L.} and {Tolstoy, E.}},
	title = {Stellar 3D kinematics in the Draco dwarf spheroidal galaxy⋆},
	DOI= "10.1051/0004-6361/201935613",
	url= "https://doi.org/10.1051/0004-6361/201935613",
	journal = {Astronomy \& Astrophysics},
	year = 2020,
	volume = 633,
	pages = "A36",
}

@article{Martinez11,
  title={A complete spectroscopic survey of the Milky Way satellite Segue 1: dark matter content, stellar membership, and binary properties from a Bayesian analysis},
  author={Martinez, Gregory D and Minor, Quinn E and Bullock, James and Kaplinghat, Manoj and Simon, Joshua D and Geha, Marla},
  journal={The Astrophysical Journal},
  volume={738},
  number={1},
  pages={55},
  year={2011},
  publisher={The American Astronomical Society}
}

@article{Reines13,
doi = {10.1088/0004-637X/775/2/116},
url = {https://doi.org/10.1088/0004-637X/775/2/116},
year = {2013},
month = {sep},
publisher = {The American Astronomical Society},
volume = {775},
number = {2},
pages = {116},
author = {Reines, Amy E. and Greene, Jenny E. and Geha, Marla},
title = {DWARF GALAXIES WITH OPTICAL SIGNATURES OF ACTIVE MASSIVE BLACK HOLES},
journal = {The Astrophysical Journal},
}

@article{Cardoso22,
  title = {Parasitic black holes: The swallowing of a fuzzy dark matter soliton},
  author = {Cardoso, Vitor and Ikeda, Taishi and Vicente, Rodrigo and Zilh\~ao, Miguel},
  journal = {Phys. Rev. D},
  volume = {106},
  issue = {12},
  pages = {L121302},
  numpages = {7},
  year = {2022},
  month = {Dec},
  publisher = {American Physical Society},
  doi = {10.1103/PhysRevD.106.L121302},
  url = {https://link.aps.org/doi/10.1103/PhysRevD.106.L121302}
}

@article{PucTri25,
   author  = {Pucha, Ragadeepika and Juneau, S. and Dey, Arjun and Siudek, M. and Mezcua, M. and Moustakas, J. and BenZvi, S. and Hainline, K. and Hviding, R. and Mao, Yao-Yuan and Alexander, D. M. and Martini, P. and Weaver, B. A.},
   title   = {Tripling the Census of Dwarf {AGN} Candidates Using {DESI} Early Data},
   journal = {The Astrophysical Journal},
   volume  = {982},
   number  = {1},
   pages   = {10},
   year    = {2025},
   doi     = {10.3847/1538-4357/adb1dd},
   eprint  = {2411.00091},
   archivePrefix = {arXiv}
}

\clearpage

\appendix
\onecolumn

\section {Shape-Shifting Theoretical Details and Numerical Results}\label{app:A}

The following is a more detailed description of the connection between Eqs.~\ref{E:Sch} and \ref{E:SP}, and the Poisson equation relation between $\rho$ and the self-gravitating contribution to the soliton potential energy.  This is followed by an explanation of the proper application of the virial theorem as a constraint when solving Eq.~\ref{E:SP}, and a description of the associated $\Psi_{5G}$ optimisation procedure. The resulting optimised $\Psi_{5G}$ coefficients are provided at $F=$ 1, 0.9, 0.8\,\dots\,0, as well as polynomial fits to the coefficients as a function of $F$ that may be used to regenerate $\Psi_{5G}$ and all other soliton properties at any value of $F$.  Additionally, polynomial fits to various $F$-dependent soliton properties are provided, some of which are plotted in Fig.~\ref{F:Layout_vs_f}a and used in evaluating Eqs.~\ref{E:m0SigmaRc}-\ref{E:MSigma}.

The zero angular momentum ground state wavefunction in a system with a spherically symmetric gravitational potential energy may be obtained by solving the following Schr\"{o}dinger equation.
\begin{equation}\label{E:SPr}
 	\hat{K}\Psi(r)+ \hat{V}\Psi(r)	 = E\Psi(r)=
    \frac{-\hbar^2}{2 m_0}\left(\frac{1}{r}\frac{\partial^2}{\partial r^2}r\right)\Psi(r)+ V(r) \Psi(r)
\end{equation}
For such a spherically symmetric galactic system, the Poisson equation may be used to obtain the following relation between the dark matter probability density $\rho(r)$ and the self-gravitating contribution to the dark matter potential energy, $V_S(r)$, associated with adding a dark matter particle of mass $m_0$ to the soliton of mass $M$.
\begin{equation}\label{E:PE}
	-4\pi G M m_0 \rho(r) = \nabla^2V_S(r) =
    \frac{1}{r} \frac{\partial^2}{\partial r^2} \left[r V_S(r)\right]
	=  \frac{1}{r^2} \frac{\partial}{\partial r} \left[r^2 \frac{\partial}{\partial r}V_S(r)\right]
\end{equation}
Upon integration, one obtains the following expression for $V_S(r)$ in terms of $\rho(r)$.
\begin{equation}\label{E:V_DM1}
V_S(r)=-G M m_0 \left[\frac{1}{r}\int_0^r \rho(r) 4\pi r^2 dr
	+ \int_r^\infty \rho(r) 4\pi r dr\right]
\end{equation}
Note that  $\rho(r)$ has units of probability per unit volume and thus $\rho(x) = \rho(r)a_0^3$ is dimensionless, while $V_S$ has units of energy and, after substituting $G M m_0/a_0 = 2\epsilon_0$ (from Eq.~\ref{E:a0e0}), becomes,
\begin{equation}\label{E:V_DM2}
	V_S(x)= -2\epsilon_0\left[\frac{1}{x}\int_0^x \rho(x) 4\pi x^2 dx + \int_x^\infty \rho(x) 4\pi x dx\right]
\end{equation}

The potential energy of a dark matter particle due its interaction with a central point mass $M$ is \mbox{$V_\bullet(r) = -G M m_0/r = -(G M m_0/a_0) (1/x)$} and thus,
\begin{equation}\label{E:V_CP}
	V_\bullet(x) = -\frac{2\epsilon_0}{x}
\end{equation}
Equations~\ref{E:V_DM2} and \ref{E:V_CP} may be combined to obtain the following potential energy of a dark matter particle in a model galaxy of total mass $M$ consisting of a central point mass $M_\bullet = (1-F) M$ and a dark matter soliton mass $M_S = F M$, so that $V(r) = F V_S + (1-F) V_\bullet$, and thus,
\begin{equation}\label{E:V_total}
	\frac{V(x)}{\epsilon_0} 
    = -2 F \left[\frac{1}{x}\int_0^x \rho(x) 4\pi x^2 dx + \int_x^\infty \rho(x) 4\pi x dx\right]
     -(1-F)  \frac{2}{x} 
\end{equation}
The following equation, equivalent to Eq.~\ref{E:SP}, is obtained by combining Eqs.~\ref{E:V_total}, \ref{E:SPr} and \ref{E:a0e0}.
\begin{equation}\label{E:SPx}
	-\epsilon_0 \left(\frac{1}{x}\frac{\partial^2}{\partial x^2}x\right)\Psi(x)+ V(x) \Psi(x)
	 = E \Psi(r)
\end{equation}

The exact solution of Eq.~\ref{E:SP} when $F=1$ is quardratic at small $r \rightarrow 0$, with no linear or cubic term \citep{MemHar89}, which is consistent with Eqs.~\ref{E:SPiSchive}, \ref{E:SPiGaussian} and \ref{E:5G}, as well as its generalization to a sum of any number of Gaussians which, in normalised form, is given by Eq.~\ref{E:nG}, below, and is equivalent to Eq.~\ref{E:SPiGaussian} when $n=1$ and to Eq.~\ref{E:5G} and $n=5$.
\begin{equation}\label{E:nG}
\Psi_{\rm{nG}}(x)=\frac{\sum_{j=0}^{n-1} c_{2j} e^{-(x/c_{2j+1})^2}}
{\sqrt{\sum_{i=0}^{n-1}\sum_{j=0}^{n-1}c_{2i} c_{2j} \left[\pi/(1/c_{2i+1}^2+1/c_{2j+1}^2)\right]^{3/2}}}\\
\end{equation}
Note that the double sum in the denominator of Eq.~\ref{E:nG} is the normalisation constant, which runs over all the indices $i$ and $j$, including those for which they are equal to each other.

The $\Psi_{\rm{nG}}(x)$ coefficients (re-scaled to hold $c_0=1$) are numerically optimised to minimise the ground state energy $\epsilon$ while maintaining consistency with the virial theorem. Properly applying the virial theorem to the above system involves subtleties that are well explained in Section~4 of ref. \citep{MemBos18}.  This is done using a multivariate optimisation algorithm, implemented in IgorPro, with a linear search option used for selecting the next step and the Secant (BFGS) method used for calculating the Hessian (matrix of second derivatives) with multivariate functions. Further details regarding the optimisation procedure are described below.

Briefly, $V_S(x)$ is the potential energy associated with adding one dark matter particle to the soliton, while the average self-gravitating energy of the soliton per dark matter particle is $\frac{1}{2}V_S(x)$ -- it is this potential energy to which the virial theorem pertains \citep{MemBos18}.  On the other hand, for the interaction of a dark matter particle with a central point mass, the virial theorem pertains to the entire interaction potential energy $V_\bullet(x)$.  Thus, a self-consistent solution of Eq.~\ref{E:SP} must satisfy the following virial theorem constraint.
\begin{equation}\label{E:VirialTheorem}
\frac{F\frac{1}{2}\langle V_S\rangle + (1-F) \langle V_\bullet\rangle}{\langle K\rangle} = -2
\end{equation}
In numerically solving Eq.~\ref{E:SP}, the coefficients of $\Psi_{nG}$ are optimised so as to both minimise the single particle energy $\epsilon=( \langle K\rangle+\langle V\rangle)/\epsilon_0$ and satisfy the virial theorem Eq.~\ref{E:VirialTheorem}.  In practice this is done by iteratively optimising the coefficients of $\Psi_{nG}$ to first minimise $\epsilon$ and then minimise the absolute value of $|2+[F\frac{1}{2}\langle V_S\rangle + (1-F) \langle V_\bullet\rangle]/\langle K\rangle|$. The latter two minimisation steps are sequentially repeated until they both converge to within $\pm 2\times 10^{-6}$.

The above optimisation process was performed for $0 \le F \le 1$ at increments of $\Delta F = 0.01$.  The following are a subset of the resulting optimised coefficients, pertaining to the curves in Fig.~\ref{F:Rhovsf}. The values in each array are the coefficents $c_i$, beginning with $i=0$ (for which $c_0 = 1$).    
\begin{eqnarray}\label{E:ci_5G}
	F=1~~&:& c_i=\{1,7.7949,3.03628,5.76119,2.87643,4.3888,1.86957,3.84334,0.841674,2.82223\}\\\nonumber
	F=0.9&:& c_i=\{1,6.68632,2.99852,4.7932,2.85032,3.5811,1.84389,2.87254,0.896186,1.55042\}\\\nonumber
	F=0.8&:& c_i=\{1,5.85336,2.95521,4.05241,2.81332,2.97984,1.81435,2.04906,0.948583,0.909529\}\\\nonumber
	F=0.7&:& c_i=\{1,5.20492,2.92424,3.4949,2.78436,2.49237,1.80217,1.51469,0.976133,0.613506\}\\\nonumber
	F=0.6&:& c_i=\{1,4.67946,2.90533,3.08098,2.7662,2.08247,1.80043,1.17886,0.996041,0.467135\}\\\nonumber
	F=0.5&:& c_i=\{1,4.2601,2.89517,2.74161,2.75832,1.75948,1.80426,0.944413,1.00655,0.367853\}\\\nonumber
	F=0.4&:& c_i=\{1,3.92206,2.88907,2.45619,2.75535,1.50108,1.80803,0.771708,1.01414,0.300781\}\\\nonumber
	F=0.3&:& c_i=\{1,3.63965,2.88348,2.21347,2.75189,1.29096,1.80911,0.642564,1.01642,0.248465\}\\\nonumber
	F=0.2&:& c_i=\{1,3.40499,2.88075,2.006,2.74919,1.12351,1.80977,0.549004,1.01483,0.21336\}\\\nonumber
	F=0.1&:& c_i= \{1,3.2062,2.88051,1.82543,2.74931,0.985045,1.80962,0.47477,1.01543,0.184914\}\\\nonumber
	F=0~~&:& c_i= \{1,3.00466,2.85791,1.67889,2.73448,0.902884,1.79766,0.404774,1.02145,0.193796\}
\end{eqnarray}
At any $F$ values between 0 and 1 one may use the following polynomials to approximate the optimised coefficients using $c_i = \sum_{j=0}^7 c_{ij} F^j$.  For example, $c_0=1$,  $c_1 = \sum_{j=0}^7 c_{1j }F^j$, $c_2 = \sum_{j=0}^7 c_{2j }F^j$ and similarly for the remaining coefficients up to $c_9$.  The resulting predictions agree with those obtained using the fully optimised $c_i$ coefficients, to within better than $\pm1\%$ (typically to within approximately $\pm 0.1\%$). 
%
%\small
\begin{eqnarray}\label{E:cxi_5G}
	c_{0}~&=& \{1, 0, 0, 0, 0, 0, 0, 0\}\\\nonumber
	c_{1j}&=& \{372122, 2.3133858, -5.9434899, 34.8439173, -88.5037257, 128.9757794, -95.8888281, 28.9923170\}\\\nonumber
	c_{2j}&=& \{2.8544898, 0.6243870, -5.6187478, 24.7438891, -57.4462063, 72.3256619, -46.1170779, 11.6723397\}\\\nonumber
	c_{3j}&=& \{1.6764439, 1.2428033, 3.5236704, -14.3248993, 47.0679488, -78.7709619, 66.4485485, -21.1028547\}\\\nonumber	
	c_{4j}&=& \{2.7307875, 0.4820703, -4.7640212, 22.9482783, -57.8578534, 78.0899871, -52.7739018, 14.0231475\}\\\nonumber
	c_{5j}&=& \{0.9016938, 0.5158374, 3.7172093, -3.9150409, -4.7724430, 27.4232936, -31.5804322, 12.0992837\}\\\nonumber
	c_{6j}&=& \{1.7952613, 0.3144760, -2.4999979, 9.6119585, -19.1040367, 18.5290315, -7.2364860, 0.4610386\}\\\nonumber
	c_{7j}&=& \{0.4044538, 0.4642604, 5.1244290, -40.3985263, 152.8980358, -275.5637542, 240.3019992, -79.3890755\}\\\nonumber
	c_{8j}&=& \{1.0200932, -0.0156436, -0.9910412, 9.7805889, -35.7928150, 61.6720394, -51.0310362, 16.1995692\}\\\nonumber
	c_{9j}&=& \{0.1919893, -0.6861140, 10.1596207, -53.9368316, 154.5268033, -231.0283053, 169.1617779, -45.5722875\}\\\nonumber
\end{eqnarray}
At precisely $F=0$ the exact hydrogenic solution may be used to obtain slightly more accurate (but nearly identical) predictions. Similarly, for a self-gravitating soliton with $F=1$ the coefficients in the first line of Eq.~\ref{E:ci_5G} are slightly preferable to those obtained from the polynomial approximation to the coefficients obtained using Eq.~\ref{E:cxi_5G}.

The following polynomial fit functions, $\sum_{i=0}^{n} a_i F^i$, may be used to regenerate the $F$-dependent soliton properties plotted in Fig.~\ref{F:Layout_vs_f}a, some of which are needed when evaluating Eqs~\ref{E:m0SigmaRc}-\ref{E:MSigma}.  These also include polynomial fits to the radii $x_{50\%} = r_{50\%}/a_0$ and $x_{99\%} = r_{99\%}/a_0$ that contains $50\%$ and $99\%$ of the soliton mass,  respectively, as well as polynomial fits to $\langle K\rangle/\epsilon_0$ and $\langle V\rangle/\epsilon_0$. Note that the purple curve in in Fig.~\ref{F:Layout_vs_f}a is obtained by multiplying $\rho_0$ by $F$.

\begin{eqnarray}\label{E:PolyF}
	&:&a_i = \{1.33271,0.918572,-0.0248566,2.61163,-3.66712,3.96499,-2.11364,0.903772\} \\\nonumber
		x_{99\%}\,(F)%= \sum_{i=0}^7 a_i F^i
	&:& a_i = \{4.13757,2.91761,-8.89622,44.3523,-98.2394,127.496,-86.7387,24.9225\}  \\\nonumber
	x_c\,(F)%= \sum_{i=0}^7 a_i F^i
	&:& a_i = \{0.394264,0.329775,3.37004,-21.2423,74.0841,-125.707,104.11,-32.6579\}\\\nonumber
	x_{{-}2}\,(F) % =\sum_{i=0}^7 a_i F^i
	&:&a_i = \{1,1.21053,-5.66103,38.869,-112.962,167.096,-119.841,33.6795\}\\\nonumber
	x_{{-}3}\,(F)  %=\sum_{i=0}^5 a_i F^i
	&:& a_i = \{1.5,0.509061,7.07851,-41.9223,135.537,-217.071,169.252,-50.5574\}\\\nonumber
	\nu_{{-}3}\,(F)  %=\sum_{i=0}^5 a_i F^i 
	&:& a_i = \{0.820383,-0.457724,0.161377,-0.516626,0.617144,-0.258997\}\\\nonumber
	\nu_S\,(F) % =\sum_{i=0}^4 a_i F^i
	&:& a_i = \{1,-0.69486,0.0515527,-0.114797,0.135807,-0.0486224\}\\\nonumber
	\epsilon\,(F)  %=\sum_{i=0}^5 a_i F^i
	&:& a_i = \{-1,0.757244,-0.140615,0.127519,-0.104019,0.034529\}\\\nonumber
	\langle K\rangle/\epsilon_{0}\,(F) % =\sum_{i=0}^5 a_i F^i
	&:&a_i = \{1,-1.38972,0.578584,-0.242451,0.273682,-0.111977\}\\\nonumber
	\langle V\rangle/\epsilon_{0}\,(F)  %=\sum_{i=0}^5 a_i F^i
	&:&a_i = \{-2,2.14696,-0.719199,0.369971,-0.377701,0.146506\}\\\nonumber
	\rho_0\,(F) %= \sum_{i=0}^7 a_i F^i
	&:&a_i = \{0.288543,-0.732719,0.716746,-0.682885,1.43297,-2.07395,1.43355,-0.377825\}
\end{eqnarray}

\section{Jeans analysis: numerical implementation}
\label{app:jeans}

This appendix describes the numerical implementation of the spherical
Jeans analysis introduced in Section~\ref{S:Jeans}, applied to the
full sample of dSph and UFD galaxies described by \citet{PozDwa24}.

\subsection*{Mass model}

The total enclosed mass entering Eq.~\ref{E:jeans} consists of a
dark matter soliton of mass $M_S = FM$ and a central point mass
$M_\bullet = (1-F)M$, so that
\begin{equation}
   M(<r) \;=\; M\,f_T(r;\,a_0,\,F),
\end{equation}
where $f_T(r)=1+ F [f_S(r)-1]$ is the dimensionless cumulative mass fraction defined in
Eq.~\ref{E:Mx} and $F \in (0,1]$ is the soliton mass fraction.
The stellar mass contribution to $M(<r)$ is omitted as it is negligible for
in the dwarf galaxies of interest.

\subsection*{Observational constraints and aperture}

For each galaxy two scalar constraints are imposed: the observed
aperture-averaged line-of-sight dispersion $\sigma=\sigma_\text{los,obs}$,
and the half-mass $M_{1/2,\text{obs}} \approx 4\sigma^2 R_{1/2}/G$
from the Wolf estimator, Eq.~\ref{E:MhalfWolf}.  The model
prediction compared to $\sigma_\text{los,obs}$ is the
tracer-weighted aperture average $\langle\sigma_\text{los}\rangle(R_\text{max})$
from Eq.~\ref{E:Jeans_sigma_ave}, where $R_\text{max}$ is the
approximate observed stellar extent tabulated by \citet{PozDwa24}.
The model prediction compared to $M_{1/2,\text{obs}}$ is the
total enclosed mass $M(<r_{1/2})$ evaluated directly from $f_T$. 
The above observationally derived input parameters obtained 
from \citet{PozDwa24} are provided in Appendix Table~\ref{T:dSph_UFD_data}.

\subsection*{Numerical grids}

The 3D Jeans solver uses 400 logarithmically spaced radial points
over $[10^{-4},\,\max(5,\,3R_\text{max})]$\,kpc.  The line-of-sight
projection uses 160 logarithmically spaced projected radii over
$[10^{-4},\,\max(0.5,\,1.1\,R_\text{max})]$\,kpc.  The stellar
tracer $\nu_\star(r)$ is the Pozo et al.\ profile evaluated on these
grids.

\subsection*{Optimisation at fixed $m_0$}

At each fixed particle mass $m_0$, the best-fit parameters $(F, M)$
for each galaxy are determined through a joint optimisation procedure
designed to identify the largest physically allowed soliton mass fraction
$F \in (0, 1]$ that yields a formally converged fit. A solution is
considered fully converged to within the observational uncertainties
 if it satisfies the primary criterion
\begin{equation}
   \max\!\bigl[|\Delta_\sigma(M)|,\,|\Delta_M(M)|\bigr] \leq 1,
\end{equation}
where the signed normalised residuals are
\begin{equation}
   \Delta_\sigma = \frac{\sigma_\text{model} - \sigma_\text{los,obs}}
                        {|\delta\sigma_\text{obs}|},
   \qquad
   \Delta_M = \frac{M(<r_{1/2}) - M_{1/2,\text{obs}}}
                   {|\delta M_{1/2,\text{obs}}|},
\end{equation}
and $\delta\sigma_\text{obs}$, $\delta M_{1/2,\text{obs}}$ are the
asymmetric observational uncertainties reported by \citet{PozDwa24}.

The search algorithm proceeds in three stages.

\textit{Stage~1: evaluation at $F=1$.}
The pure-soliton case is evaluated first.  Two characteristic
total masses are computed: $M_1$ is found by bisecting the
closed-form $M(<r_{1/2})$ to exactly match $M_{1/2,\text{obs}}$,
and $M_2$ is found via Brent's method so that the Jeans-predicted
aperture dispersion exactly matches $\sigma_\text{los,obs}$.
A single full Jeans evaluation is then performed at each of these
masses to obtain both residuals.  If either $(F=1,\,M_1)$ or
$(F=1,\,M_2)$ satisfies the primary criterion, that solution is
immediately accepted.  For example, $F=1$ results for $M_1$ and $M_2$
correspond to the $M$ values in
Tables~\ref{T:model3_F1} and \ref{T:F1}, respectively,
for $m_0 = 1.5\times10^{-22}$\,eV/c$^2$.

\textit{Stage~2: bisection over $F$.}
If $F=1$ fails, a lower bound $F_\text{min}$ is derived from the minimum
$F$ value consistent with $M_1$ and $M_2$, and the algorithm bisects
over $F \in [F_\text{min},\,1]$ to locate the largest $F$ satisfying
the primary criterion.  At each trial $F$, the optimal total mass is
determined by bisecting in $\log_{10}M$ until $M(<r_{1/2})=M_{1/2,\text{obs}}$
exactly, using the closed-form enclosed-mass evaluator (no Jeans call
inside the inner loop); one full Jeans evaluation is then performed at
the converged $M$ to obtain $\sigma_\text{model}$ and hence both
residuals.  The convergence score $\max(|\Delta_\sigma|,|\Delta_M|)$ is
thus reduced to a function of $F$ alone, and bisection narrows the
$F$-interval until the boundary between passing and failing is isolated
to the required tolerance.

\textit{Stage~3: fallback and final evaluation.}
If no $F \in [F_\text{min},1]$ satisfies the primary criterion, the
algorithm falls back to a secondary criterion, selecting the $F$ that
minimises the total absolute residual $|\Delta_\sigma|+|\Delta_M|$.  
In all cases, a final
full Jeans evaluation is performed at the selected $(F,\,M)$ to produce
the reported $\sigma_\text{model}$ and $M_{1/2,\text{model}}$.

\section{Velocity Dispersion Profiles for Draco and Segue~1}
\label{app:sigma_profiles}

Radially binned velocity dispersion profiles $\sigma_{\rm LOS}(R_{\rm bin})$
and cumulative aperture-integrated profiles $\sigma_{\rm LOS}(<R_{\rm max})$
are derived for Segue~I and Draco from published stellar-kinematic catalogues,
as described below. Both Segue~I and Draco use the moment-based estimator 
described below for all binned and cumulative profiles.
Consecutive cumulative apertures share all
interior stars and are therefore strongly correlated. The quoted uncertainties
represent marginal errors on each aperture rather than independent constraints.

%------------------------------------------------------------------
\subsection*{Moment-Based Estimator}
%------------------------------------------------------------------

For a subsample of $N$ stars with velocities $\{v_i\}$ and measurement
uncertainties $\{\epsilon_i\}$, the intrinsic line-of-sight dispersion is
estimated via
\begin{equation}
    \hat{\sigma}^2 = \frac{1}{N-1}\sum_{i=1}^{N}(v_i - \bar{v})^2
    - \frac{1}{N}\sum_{i=1}^{N}\epsilon_i^2,
\end{equation}
with $\hat{\sigma} = \sqrt{\max(\hat{\sigma}^2,\,0)}$ and uncertainty
$\sigma_{\rm err} = \hat{\sigma}/\sqrt{2(N-1)}$.  This estimator
marginalises individual measurement errors in quadrature and is the same 
in form as that used by \citet{Walker15} for Draco and \citet{Simon11} 
for Segue~I.  Note that the global binary-corrected dispersion of 
Segue~I adopted below, $\sigma_{\rm LOS,corr} = 3.7^{+1.4}_{-1.1}$~km~s$^{-1}$, 
was derived by \citet{Simon11} using the independent Bayesian framework of 
\citet{Martinez11} and is adopted directly here rather than recomputed.

%------------------------------------------------------------------
\subsection*{Draco}
%------------------------------------------------------------------

\paragraph*{Sample.}
The stellar catalogue of \citet{Walker15}, comprising 1565 unique
stars (2813 observations) from MMT/Hectochelle spectroscopy, is
used to pre-select likely Draco member stars whose velocities are within
$\pm 30$\,km\,s$^{-1}$ of the systemic velocity $v_{\rm sys} =
-291.0$\,km\,s$^{-1}$, and which satisfy [Fe/H]\,$<-1.5$ and projected
radius $R < 0.9$\,kpc.

\paragraph*{Dispersion estimator.}
The moment-based estimator described above is applied within each
radial bin and cumulative aperture.

\paragraph*{Cumulative profile.}
The cumulative profile $\sigma_{\rm LOS}(<R_{\rm max})$ is constructed
by including all member stars within projected radius $R \leq R_{\rm max}$,
evaluated at ten equally spaced values of $R_{\rm max}$ between
$0.09$\,kpc and $0.90$\,kpc, retaining only apertures containing at
least five stars.  The outermost aperture, encompassing all selected
members within $R < 0.9$\,kpc, yields a global dispersion consistent
with the line-of-sight value $\sigma_{\rm LOS} = 9.0 \pm
1.1$~km~s$^{-1}$ reported by \citet{Massari20} from a joint
HST$+$\textit{Gaia} proper-motion and line-of-sight analysis.  This
value is adopted as the preferred normalisation anchor for the Draco
cumulative profile shown in Fig.~\ref{F:Draco_Segue_Fig}.

\paragraph*{Binned profile.}
The binned profile $\sigma_{\rm LOS}(R_{\rm bin})$ for Draco is constructed 
by dividing the member-selected stars into 10 equal-width radial bins spanning 
$0$--$0.9$\,kpc, with bin edges at $R = 0.09\,i$\,kpc for $i = 0, 1, \ldots, 10$, 
giving a bin width of $\Delta R = 0.09$\,kpc. Stars are assigned to the unique bin 
satisfying $R_{\rm min} \leq R < R_{\rm max}$, so each star appears in exactly one 
bin and adjacent bins are statistically independent. The bin radius $R_{\rm bin}$ is 
taken as the mean projected radius of all member stars within the bin. The 
moment-based estimator described above is applied within each bin, and bins 
containing fewer than five stars are excluded. The same membership criteria 
(velocity window, metallicity cut, and outer radius limit) described in the Sample 
paragraph above are applied prior to binning.

%------------------------------------------------------------------
\subsection*{Segue~1}
%------------------------------------------------------------------

\paragraph*{Sample.}
The Segue~1 velocities are taken from the machine-readable Table~3 of
\citet{Simon11}, comprising 394 stars observed with DEIMOS on Keck~II.
Where multiple epochs exist, observations are combined via the
inverse-variance-weighted mean
\begin{equation}
    \bar{v} = \frac{\sum_k v_k/\epsilon_k^2}{\sum_k \epsilon_k^{-2}},
    \qquad
    \bar{\epsilon} = \!\left(\sum_k \epsilon_k^{-2}\right)^{-1/2}.
\end{equation}
The subjective membership flag (\texttt{Mem}~$=1$) of \citet{Simon11}
is adopted, and the two RR~Lyrae variables and one confirmed red-giant
binary identified by those authors are excluded, leaving $N = 68$
member stars at projected radii $0.0'$--$13.5'$ (0--90~pc at 23~kpc,
where $1' = 6.66$~pc).

\paragraph*{Binned and cumulative profiles.}
The binned profile $\sigma_{\rm LOS}(R_{\rm bin})$ is obtained by sorting 
the 68 member stars by projected radius and dividing them into contiguous 
equal-membership bins, using two independent binning schemes: five bins of 
15 stars each (with the outermost bin retaining the 8 remaining stars) and 
three bins of approximately 23 stars each (23, 23, and 22 stars, 
respectively), matching the two schemes shown in Fig.~11 of \citet{Simon11}. 
The bin radius $R_{\rm bin}$ is taken as the mean projected radius of stars 
within each bin. The moment-based estimator described above is applied within 
each bin; bins for which the resulting intrinsic variance $\hat{\sigma}^2 < 0$ 
are excluded from the figure. The cumulative profile 
$\sigma_{\rm LOS}(<R_{\rm max})$ is obtained by applying the same estimator 
to all member stars with $R_i < R_{\rm max}$, evaluated at each successive 
member radius for apertures containing at least five stars ($i = 5,\ldots,68$). 
Binary-corrected dispersions for both the binned and cumulative profiles are 
obtained by applying 
Equations~(\ref{eq:binary_corr})--(\ref{eq:binary_corr_err}).

\paragraph*{Binary-star correction.}
The raw cumulative dispersions include an additive variance contribution
from unresolved binary orbital motion.  Because binary velocities are
isotropic and the binary fraction is not expected to vary with projected
radius in a system as dynamically relaxed as Segue~1, the binary
contribution $\sigma_{\rm binary}$ is spatially uniform, giving
\begin{equation}
    \sigma_{\rm LOS,obs}^2(<R_{\rm max}) =
        \sigma_{\rm LOS,corr}^2(<R_{\rm max}) + \sigma_{\rm binary}^2.
\label{eq:quadrature}
\end{equation}
The value of $\sigma_{\rm binary}$ is fixed by combining the raw global
dispersion $\sigma_{\rm LOS,obs} = 5.71 \pm 0.79$~km~s$^{-1}$ (from
the full 68-star sample) with the binary-corrected global value
$\sigma_{\rm LOS,corr} = 3.7^{+1.4}_{-1.1}$~km~s$^{-1}$ of
\citet{Simon11}, obtained via the Bayesian framework of
\citet{Martinez11}:
\begin{equation}
    \sigma_{\rm binary}
        = \sqrt{5.71^2 - 3.7^2}
        = \sqrt{18.80}
        \approx 4.33~{\rm km\,s}^{-1}.
\label{eq:sigbinary}
\end{equation}
The binary-corrected dispersion at each aperture is then
\begin{equation}
    \sigma_{\rm LOS,corr}(<R_{\rm max}) =
        \sqrt{\sigma_{\rm LOS,obs}^2(<R_{\rm max}) - \sigma_{\rm binary}^2},
\label{eq:binary_corr}
\end{equation}
with uncertainties propagated as
\begin{equation}
    \sigma_{\rm err,corr} =
        \sigma_{\rm err,obs}\,
        \frac{\sigma_{\rm LOS,obs}(<R_{\rm max})}
             {\sigma_{\rm LOS,corr}(<R_{\rm max})}.
\label{eq:binary_corr_err}
\end{equation}
Apertures for which $\sigma_{\rm LOS,obs}^2 \leq \sigma_{\rm binary}^2$
are excluded.  As a self-consistency check, applying
Equations~(\ref{eq:binary_corr})--(\ref{eq:binary_corr_err}) to the
full $N = 68$ aperture recovers $\sigma_{\rm LOS,corr} = 3.711 \pm
1.217$~km~s$^{-1}$, in agreement with the \citet{Simon11} published
value ($<0.02\sigma$ difference).

Because $\sigma_{\rm binary}^2$ is a constant offset in variance, the
\emph{shape} of the corrected cumulative profile is identical to that
of the raw profile.  The central kinematic elevation at
$R_{\rm max} \lesssim 0.02$~kpc, where
$\sigma_{\rm LOS,corr} \sim 7$--$9$~km~s$^{-1}$, is therefore a
genuine dynamical signal and cannot be attributed to binary orbital motion.

\clearpage

\section{dSph and UFD Observed and Predicted Properties}\label{app:B}

The following tables contain galactic observational and 5G Jeans analysis predictions 
used in generating the dSph and UFD results shown in 
Figs.~\ref{F:Layout_vs_f}--\ref{F:Draco_Segue_Fig}. The last two columns in 
Tables~\ref{T:optimizem0}--\ref{T:F0} list $\Delta_\sigma$ and $\Delta_M$, 
which quantify the normalised residuals between the predicted and observed values of 
$\sigma$ and $M_{1/2}$, relative to the corresponding observational uncertainty. More 
specifically, $\Delta_\sigma \equiv (\sigma_\text{model} - 
\sigma_\text{los,obs})/|\delta\sigma_\text{obs}|$ and $\Delta_M \equiv 
(M(<r_{1/2}) - M_{1/2,\text{obs}})/|\delta M_{1/2,\text{obs}}|$, where 
$\delta\sigma_\text{obs}$ and $\delta M_{1/2,\text{obs}}$ are the asymmetric 
observational uncertainties reported by \citet{PozDwa24} (and also provided in 
Table~\ref{T:dSph_UFD_data}, below). Thus, predictions with 
$|\Delta_\sigma| \le 1$ and $|\Delta_M| \le 1$ are within observational uncertainty. 
Note that the predicted black hole masses $M_\bullet = (1-F)M$ are consistent with the 
corresponding soliton mass fraction $F$ and total galactic mass $M = M_S + M_\bullet = F M +(1-F) M$. 
\vspace{0.1in}

\noindent Table~\ref{T:dSph_UFD_data} contains observed and 
observationally derived properties of dSph and UFD galaxies, compiled by \citet{PozDwa24} (and references therein). 
These results are 
used to produce the points in Fig.~\ref{F:Layout_vs_f}d--e.\vspace{0.1in}

\noindent Table~\ref{T:optimizem0} contains 5G predictions for dSph and UFD galaxies 
obtained by optimising $m_0$ to yield self-consistent $M_{1/2}$ predictions with $F=1$ 
and $\sigma_S = \sigma_\ast$. These results are plotted in 
Fig.~\ref{F:Expt_Layout}c.\vspace{0.1in}

\noindent Table~\ref{T:Dual_m0} contains 5G predictions obtained assuming that the dSph 
and UFD galaxies have different types of ultralight dark matter particles with masses of 
either $1.5\times 10^{-22}$\,eV/c$^2$ (for the dSph galaxies) or 
$20\times 10^{-22}$\,eV/c$^2$ (for the UFD galaxies), assuming $F=1$ (no black holes) 
and optimising $M$ to yield self-consistent $M_{1/2}$ predictions. These results are 
plotted in Fig.~\ref{F:Expt_Layout}d.\vspace{0.1in}

\noindent Tables~\ref{T:1x5m0_1SE}--\ref{T:20m0_1SE} contain 5G predictions obtained 
assuming several values of $m_0$ between $0.75\times 10^{-22}$\,eV/c$^2$ and 
$20\times 10^{-22}$\,eV/c$^2$, derived by optimising both $F$ and $M$, maximising $F$ 
subject to the constraint that the predicted $\sigma$ and $M_{1/2}$ are observationally 
consistent. The corresponding results are plotted in 
Figs.~\ref{F:Expt_Layout_B}a, b, c, and e.\vspace{0.1in}

\noindent Tables~\ref{T:1x5m0_2SE} and \ref{T:1x5m0_3SE} contain 5G predictions obtained assuming 
$m_0=1.5\times 10^{-22}$\,eV/c$^2$, derived by optimising both $F$ and $M$, maximising 
$F$, constraining the predicted $\sigma$ and $M_{1/2}$ to remain within 2 and 3 times the 
reported observational uncertainties, respectively. Comparison of these results with those 
in Table~\ref{T:1x5m0_1SE} illustrate the predicted decrease in the number of galaxies that may 
contain black holes as the observational uncertainty bounds are 
relaxed.\vspace{0.1in}

\noindent Table~\ref{T:model2} contains 5G predictions obtained 
assuming $m_0=1.5\times 10^{-22}$\,eV/c$^2$, derived by optimising $F$ to yield observationally 
consistent $M_{1/2}$ predictions (with minimum $|\Delta_M|$). Those galaxies for which 
the predicted $|\Delta_\sigma|\le 1$ may be observationally consistent with having a 
supermassive black hole. However, for all of the dSph galaxies and some of the UFD 
galaxies, the predicted black hole mass exceeds $10^6$\,M$_\odot$, which may be
unrealistically large. These results could be viewed as providing an upper bound on 
the observationally consistent black hole mass in each galaxy.\vspace{0.1in}

\noindent Tables~\ref{T:model3_F1} and \ref{T:F1} contain 5G predictions obtained 
assuming $m_0 = 1.5\times 10^{-22}$\,eV/c$^2$ and $F=1$: Table~\ref{T:model3_F1} is derived 
by optimising $M$ to minimise $|\Delta_M|$, and Table~\ref{T:F1} by minimising 
$|\Delta_\sigma|$. These are used to provide bounds on the possible values of $M$ when 
performing optimisations of both $F$ and $M$ (as in 
Tables~\ref{T:1x5m0_1SE}--\ref{T:1x5m0_3SE} above).\vspace{0.1in}

\noindent Table~\ref{T:F0} contains 5G predictions obtained assuming $F=0.0001$, 
thus demonstrating the degree to which the observed 
$\sigma$ and $M_{1/2}$ values are compatible with a model in which the galaxies contain 
a supermassive black hole and little or no dark matter. Note that 
these predictions are precisely self-consistent with the observed velocity dispersions 
$\sigma$, and $|\Delta_M| \le 1$ for most of the galaxies, thus implying that the associated 
$\sigma$ and $M_{1/2}$ observations are consistent with the 
galaxy harbouring a supermassive central black hole and no dark matter.

\clearpage

\begin{table*}
\centering
\begin{tabular}{l c c c c c c c c}
\hline
\multicolumn{9}{c}{Observed (and observationally derived) properties of dSph and UFD galaxies}\\
\hline
Galaxy & $\sigma$ (km/s) & $R_{1/2}$\,kpc & $M_{1/2}/10^6$ M$_\odot$ & $M_\ast/10^5$ M$_\odot$ & $r_c$\,kpc & $r_t$\,kpc & $r_s$\,kpc & $R_{\rm max}$\,kpc \\
\hline
Tucana & $13.3^{+2.7}_{-2.3}$ & $0.28^{+0.05}_{-0.05}$ & $46.7^{+16}_{-14}$ & $5.5$ & 0.25 & 0.78 & 1.05 & 2 \\
Cetus & $11.1^{+1.6}_{-1.3}$ & $0.60^{+0.01}_{-0.01}$ & $68.7^{+14}_{-11}$ & $28^{+8}_{-8}$ & 0.36 & 0.87 & 0.24 & 2 \\
Aquarius & $10.3^{+1.6}_{-1.3}$ & $0.34^{+0.01}_{-0.01}$ & $33.5^{+7.4}_{-6.1}$ & $17$ & 0.35 & 1.25 & 1.05 & 2 \\
Draco$^{(a)}$ & $9^{+1.1}_{-1.1}$ & $0.23^{+0.01}_{-0.01}$ & $17.3^{+3.1}_{-3.1}$ & $2.2$ & 0.17 & 0.56 & 0.1 & 0.9 \\
Leo I & $9.2^{+1.2}_{-1.2}$ & $0.26^{+0.01}_{-0.01}$ & $20.5^{+3.9}_{-3.9}$ & $34^{+11}_{-11}$ & 0.24 & 1.3 & 1.75 & 9 \\
Phoenix & $9.3^{+0.7}_{-0.7}$ & $0.29^{+0.01}_{-0.01}$ & $23.3^{+2.6}_{-2.6}$ & $6.2$ & 0.28 & 1.27 & 1.1 & 1.6 \\
Canes Ventici & $7.6^{+0.4}_{-0.4}$ & $0.47^{+0.02}_{-0.02}$ & $25.2^{+2.2}_{-2.2}$ & $2.3$ & 0.308 & 1.06 & 2.29 & 1.4 \\
Sextans & $7.9^{+1.3}_{-1.3}$ & $0.72^{+0.01}_{-0.01}$ & $41.5^{+9.7}_{-9.7}$ & $4.4^{+1.7}_{-1.7}$ & 0.48 & 1.31 & 1.61 & 4.5 \\
Crater II & $2.7^{+0.3}_{-0.3}$ & $1.1^{+0.08}_{-0.08}$ & $7.2^{+1.3}_{-1.3}$ & $0.83$ & 0.71 & 1.68 & 2.6 & 7 \\
Leo II & $7.4^{+0.4}_{-0.4}$ & $0.19^{+0.02}_{-0.02}$ & $9.7^{+1.3}_{-1.3}$ & $7.4^{+2.0}_{-2.0}$ & 0.17 & 0.66 & 3.76 & 1.6 \\
Carina & $6.6^{+1.2}_{-1.2}$ & $0.42^{+0.06}_{-0.04}$ & $17.2^{+5.0}_{-4.7}$ & $5.9$ & 0.21 & 0.81 & 1.17 & 2.25 \\
Ursa Minor & $11.5^{+0.9}_{-0.8}$ & $0.47^{+0.06}_{-0.06}$ & $57.5^{+9.7}_{-9.3}$ & $3.0$ & 0.28 & 0.96 & 0.52 & 1.6 \\
Sculptor & $10.1^{+0.3}_{-0.3}$ & $0.29^{+0.01}_{-0.01}$ & $27.4^{+1.5}_{-1.5}$ & $20^{+7.9}_{-7.9}$ & 0.21 & 0.72 & 0.12 & 3.5 \\
And I & $9.4^{+1.7}_{-1.5}$ & $0.66^{+0.07}_{-0.07}$ & $54.2^{+15}_{-14}$ & $24^{+0.57}_{-0.54}$ & 0.52 & 1.47 & 0.13 & 3 \\
And III & $11^{+1.9}_{-1.6}$ & $0.41^{+0.04}_{-0.04}$ & $46.1^{+12.}_{-11}$ & $4.8^{+0.11}_{-0.11}$ & 0.33 & 1.38 & 2.53 & 3 \\
And V & $11.2^{+1.1}_{-1.0}$ & $0.35^{+0.04}_{-0.04}$ & $40.8^{+7.3}_{-7.0}$ & $4.1^{+0.10}_{-0.09}$ & 0.25 & 0.81 & 2.59 & 3.5 \\
And IX & $10.9^{+2.0}_{-2.0}$ & $0.36^{+0.06}_{-0.05}$ & $39.8^{+12}_{-12}$ & $2.0^{+0.52}_{-0.41}$ & 0.22 & 0.7 & 3.22 & 3 \\
And XIV & $5.4^{+1.3}_{-1.3}$ & $0.39^{+0.19}_{-0.20}$ & $10.6^{+6.3}_{-6.5}$ & $2.0^{+0.52}_{-0.41}$ & 0.33 & 0.98 & 1.77 & 4 \\
And XV & $11^{+7}_{-5}$ & $0.23^{+0.03}_{-0.02}$ & $25.9^{+24}_{-17}$ & $1.3^{+0.74}_{-0.25}$ & 0.19 & 0.65 & 1.88 & 3 \\
And XVIII & $9.7^{+2.3}_{-2.3}$ & $0.33^{+0.02}_{-0.02}$ & $28.9^{+9.8}_{-9.8}$ & $4.0^{+2.3}_{-1.5}$ & 0.25 & 0.85 & 2.99 & 5 \\
And XXI & $6.1^{+1.0}_{-0.9}$ & $1.0^{+0.17}_{-0.17}$ & $34.8^{+10.1}_{-9.4}$ & $3.2^{+0.8}_{-0.7}$ & 0.51 & 1.32 & 3.16 & 4.5 \\
And XXIII & $7.1^{+1.0}_{-1.0}$ & $1.1^{+0.10}_{-0.10}$ & $55.8^{+12}_{-12}$ & $6.3^{+1.6}_{-1.3}$ & 0.8 & 2.2 & 3.92 & 5 \\
And XXV & $3^{+1.2}_{-1.1}$ & $0.55^{+0.10}_{-0.07}$ & $4.6^{+2.7}_{-2.5}$ & $3.2^{+0.82}_{-0.65}$ & 0.42 & 1.15 & 2.79 & 3.5 \\
\hline
Phoenix II & $11^{+9.4}_{-5.3}$ & $0.036^{+0.008}_{-0.008}$ & $4.1^{+5.0}_{-2.9}$ & $0.018^{+0.014}_{-0.008}$ & 0.024 & 0.082 & 1.37 & 0.25 \\
Segue I & $3.9^{+0.8}_{-0.8}$ & $0.032^{+0.003}_{-0.003}$ & $0.45^{+0.14}_{-0.14}$ & $0.0028^{+0.003}_{-0.0014}$ & 0.02 & 0.065 & 1.59 & 0.25 \\
Pegasus III & $5.4^{+3.0}_{-2.5}$ & $0.053^{+0.014}_{-0.014}$ & $1.4^{+1.2}_{-1.0}$ & $0.020$ & 0.024 & 0.083 & 1.16 & 0.25 \\
Wilman I & $4^{+0.8}_{-0.8}$ & $0.033^{+0.008}_{-0.008}$ & $0.49^{+0.18}_{-0.18}$ & $0.0087^{+0.009}_{-0.004}$ & 0.0235 & 0.064 & 0.64 & 0.25 \\
Horoligium I & $4.9^{+2.8}_{-0.9}$ & $0.041^{+0.010}_{-0.010}$ & $0.92^{+0.77}_{-0.33}$ & $0.022^{+0.015}_{-0.009}$ & 0.028 & 0.11 & 1.29 & 0.4 \\
Pisces II & $5.4^{+3.6}_{-2.4}$ & $0.062^{+0.010}_{-0.010}$ & $1.7^{+1.6}_{-1.1}$ & $0.042^{+0.018}_{-0.012}$ & 0.032 & 0.12 & 1.85 & 0.5 \\
Coma Berenices & $4.6^{+0.8}_{-0.8}$ & $0.069^{+0.005}_{-0.005}$ & $1.4^{+0.35}_{-0.35}$ & $0.048^{+0.012}_{-0.01}$ & 0.053 & 0.16 & 1.67 & 0.5 \\
Reticulum II & $3.22^{+1.64}_{-0.49}$ & $0.053^{+0.002}_{-0.002}$ & $0.51^{+0.37}_{-0.11}$ & $0.024^{+0.002}_{-0.002}$ & 0.0333 & 0.1 & 1.48 & 0.14 \\
Hydrus I & $2.69^{+0.51}_{-0.43}$ & $0.056^{+0.004}_{-0.004}$ & $0.38^{+0.10}_{-0.089}$ & $0.034$ & 0.041 & 0.13 & 1.27 & 0.35 \\
Grus I & $5.4^{+3.0}_{-2.5}$ & $0.070^{+0.025}_{-0.025}$ & $1.9^{+1.6}_{-1.4}$ & $0.021^{+0.015}_{-0.009}$ & 0.0475 & 0.13 & 1.79 & 0.35 \\
Leo IV & $3.4^{+1.3}_{-0.9}$ & $0.11^{+0.010}_{-0.010}$ & $1.2^{+0.67}_{-0.47}$ & $0.18^{+0.08}_{-0.08}$ & 0.084 & 0.28 & 1.34 & 0.35 \\
Canes Ventici II & $4.6^{+1.0}_{-1.0}$ & $0.070^{+0.010}_{-0.010}$ & $1.38^{+0.47}_{-0.47}$ & $0.10^{+0.03}_{-0.03}$ & 0.037 & 0.14 & 1.39 & 0.35 \\
Bootes I & $2.4^{+0.9}_{-0.5}$ & $0.22^{+0.010}_{-0.010}$ & $1.2^{+0.63}_{-0.35}$ & $0.22^{+0.056}_{-0.045}$ & 0.065 & 0.23 & 1.86 & 0.35 \\
Tucana II & $2.8^{+1.2}_{-0.7}$ & $0.12^{+0.030}_{-0.030}$ & $0.88^{+0.57}_{-0.38}$ & $0.028$ & 0.11 & 0.31 & 2.45 & 0.35 \\
Tucana IV & $4.3^{+1.7}_{-1.0}$ & $0.11^{+0.011}_{-0.009}$ & $1.9^{+1.1}_{-0.64}$ & $0.014^{+0.006}_{-0.003}$ & 0.03 & 0.1 & 1.9 & 0.35 \\
Leo V & $3.7^{+2.3}_{-1.4}$ & $0.055^{+0.020}_{-0.020}$ & $0.70^{+0.67}_{-0.45}$ & $0.049^{+0.019}_{-0.014}$ & 0.021 & 0.076 & 1.61 & 0.35 \\
And X & $3.9^{+1.2}_{-1.2}$ & $0.21^{+0.040}_{-0.070}$ & $3.0^{+1.4}_{-1.6}$ & $0.79^{+0.21}_{-0.16}$ & 0.1 & 0.36 & 6.18 & 0.5 \\
And XI & $4.6^{+0.0}_{-4.6}$ & $0.12^{+0.050}_{-0.040}$ & $2.4^{+3.6}_{-2.4}$ & $0.25^{+0.15}_{-0.09}$ & 0.059 & 0.21 & 2.67 & 0.35 \\
And XII & $2.6^{+5.1}_{-2.6}$ & $0.32^{+0.060}_{-0.070}$ & $2.0^{+5.6}_{-2.0}$ & $0.50^{+0.29}_{-0.19}$ & 0.1 & 0.34 & 2.62 & 0.5 \\
And XIII & $5.8^{+2.0}_{-2.0}$ & $0.13^{+0.080}_{-0.060}$ & $4.1^{+3.2}_{-2.7}$ & $0.32^{+0.19}_{-0.16}$ & 0.045 & 0.13 & 2.96 & 0.35 \\
And XVI & $3.8^{+2.9}_{-2.9}$ & $0.13^{+0.030}_{-0.020}$ & $1.7^{+1.9}_{-1.7}$ & $0.63^{+0.16}_{-0.13}$ & 0.12 & 0.5 & 2.84 & 0.5 \\
And XVII & $2.9^{+2.2}_{-1.9}$ & $0.29^{+0.060}_{-0.050}$ & $2.3^{+2.5}_{-2.1}$ & $1.0^{+0.26}_{-0.21}$ & 0.15 & 0.57 & 1.84 & 0.5 \\
And XX & $7.1^{+3.9}_{-2.5}$ & $0.090^{+0.040}_{-0.020}$ & $4.2^{+3.8}_{-2.3}$ & $0.25^{+0.15}_{-0.093}$ & 0.042 & 0.16 & 2.64 & 0.35 \\
And XXII & $2.8^{+2.9}_{-1.4}$ & $0.23^{+0.080}_{-0.080}$ & $1.7^{+2.5}_{-1.3}$ & $0.40^{+0.23}_{-0.20}$ & 0.078 & 0.2 & 4.57 & 0.35 \\
And XXVI & $8.6^{+2.8}_{-2.2}$ & $0.15^{+0.14}_{-0.080}$ & $10.3^{+10.7}_{-6.6}$ & $0.16^{+0.24}_{-0.1}$ & 0.021 & 0.07 & 3.65 & 0.25 \\
\hline
\end{tabular}
\caption{Observational data pertaining to dSph (top section) and UFD (bottom section) galaxies from Tables II and III of \citet{PozDwa24}, and references therein.  The values of $\sigma = \sigma_{\rm{los}}$  and $R_{1/2} = r_{\rm{half,obs}}$ pertain to stellar observations, the right side of the equality proves the corresponding notation in \citet{PozDwa24}; $M_{1/2}$ is obtained using Eq.~\ref{E:MhalfWolf} and $M_\ast$ is obtained assuming a mass-to-light ratio of 1 with the luminosities $L_{\rm{obs}}$ [L$_\odot$] compiled in \citet{PozDwa24}; $r_c$, $r_t$ and $r_s$ are the Plummer plus NFW stellar profile fit parameters obtained by  \citet{PozDwa24} and $R_{\rm max}$ is the extent of the stellar distribution obtained from the radial range of Figs.~10-25 in \citet{PozDwa24}. (a) The Draco $\sigma$ is here equated with $\sigma_{\rm LOS}$ from the \citet{Massari20}, instead of equating $\sigma_{\rm{los}} = \sigma_R$ as in \citet{PozDwa24}.}
\label{T:dSph_UFD_data}
\end{table*}

\begin{table}
\centering
\begin{tabular}{lcccccccccc}
\hline
\multicolumn{10}{c}{5G predictions with optimised $m_0$ and observatinoally--consistent $M_{1/2}$ and $\sigma_S=\sigma_\ast=\sqrt{3}\,\sigma_{los,obs}$} \\
\hline
Galaxy & $F$ & $M/10^8$ & $m_0/10^{-22}$ & $\sigma$ & $M_{1/2}/10^6$ & $r_{c,S}$ & $\sigma_S/\sigma_\ast$ & $\Delta_\sigma$ & $\Delta_M$ \\
  &  & M$_\odot$ & eV/c$^2$ & (km/s) & M$_\odot$ &\,kpc & & & \\
\hline
Tucana & 1 & 0.477 & 6.54 & 11.50 & 46.7 & 0.11 & 1.00 & -0.78 & 0.00 \\
Cetus & 1 & 1.971 & 1.32 & 10.41 & 68.7 & 0.67 & 1.00 & -0.53 & 0.00 \\
Aquarius & 1 & 0.962 & 2.51 & 10.06 & 33.5 & 0.38 & 1.00 & -0.18 & 0.00 \\
Draco & 1 & 0.177 & 11.9 & 8.84 & 17.3 & 0.09 & 1.00 & -0.15 & 0.00 \\
Leo I & 1 & 0.209 & 10.3 & 8.41 & 20.5 & 0.10 & 1.00 & -0.66 & 0.00 \\
Phoenix & 1 & 0.669 & 3.26 & 8.97 & 23.3 & 0.32 & 1.00 & -0.47 & 0.00 \\
Canes Ventici & 1 & 0.724 & 2.46 & 6.64 & 25.2 & 0.52 & 1.00 & -2.40 & 0.00 \\
Sextans & 1 & 1.190 & 1.56 & 7.09 & 41.5 & 0.79 & 1.00 & -0.62 & 0.00 \\
Crater II & 1 & 0.207 & 3.06 & 2.47 & 7.23 & 1.18 & 1.00 & -0.77 & 0.00 \\
Leo II & 1 & 0.279 & 6.22 & 6.76 & 9.72 & 0.21 & 1.00 & -1.60 & 0.00 \\
Carina & 1 & 0.493 & 3.14 & 4.93 & 17.2 & 0.47 & 1.00 & -1.39 & 0.00 \\
Ursa Minor & 1 & 1.649 & 1.64 & 9.55 & 57.5 & 0.52 & 1.00 & -2.44 & 0.00 \\
Sculptor & 1 & 0.786 & 3.01 & 8.60 & 27.4 & 0.32 & 1.00 & -5.00 & 0.00 \\
And I & 1 & 0.554 & 3.98 & 8.80 & 54.2 & 0.26 & 1.00 & -0.40 & 0.00 \\
And III & 1 & 1.323 & 1.95 & 9.91 & 46.1 & 0.46 & 1.00 & -0.68 & 0.00 \\
And V & 1 & 0.417 & 6.30 & 8.77 & 40.8 & 0.14 & 1.00 & -2.43 & 0.00 \\
And IX & 1 & 0.406 & 6.30 & 8.39 & 39.8 & 0.14 & 1.00 & -1.25 & 0.00 \\
And XIV & 1 & 0.303 & 4.18 & 4.61 & 10.6 & 0.43 & 1.00 & -0.61 & 0.00 \\
And XV & 1 & 0.264 & 9.77 & 8.63 & 25.9 & 0.09 & 1.00 & -0.47 & 0.00 \\
And XVIII & 1 & 0.295 & 7.72 & 7.11 & 28.9 & 0.13 & 1.00 & -1.13 & 0.00 \\
And XXI & 1 & 0.355 & 4.03 & 5.07 & 34.8 & 0.40 & 1.00 & -1.14 & 0.00 \\
And XXIII & 1 & 1.600 & 1.04 & 6.72 & 55.8 & 1.32 & 1.00 & -0.38 & 0.00 \\
And XXV & 1 & 0.047 & 15.0 & 2.38 & 4.60 & 0.22 & 1.00 & -0.56 & 0.00 \\
\hline
Phoenix II & 1 & 0.041 & 62.4 & 8.50 & 4.05 & 0.01 & 1.00 & -0.47 & 0.00 \\
Segue I & 1 & 0.005 & 198 & 2.58 & 0.453 & 0.01 & 1.00 & -1.65 & 0.00 \\
Pegasus III & 1 & 0.041 & 30.7 & 4.54 & 1.44 & 0.06 & 1.00 & -0.34 & 0.00 \\
Wilman I & 1 & 0.005 & 187 & 2.55 & 0.491 & 0.01 & 1.00 & -1.81 & 0.00 \\
Horoligium I & 1 & 0.009 & 123 & 4.13 & 0.915 & 0.02 & 1.00 & -0.86 & 0.00 \\
Pisces II & 1 & 0.048 & 26.3 & 4.12 & 1.68 & 0.07 & 1.00 & -0.53 & 0.00 \\
Coma Berenices & 1 & 0.014 & 77.9 & 3.25 & 1.36 & 0.03 & 1.00 & -1.69 & 0.00 \\
Reticulum II & 1 & 0.015 & 51.5 & 3.05 & 0.511 & 0.06 & 1.00 & -0.35 & 0.00 \\
Hydrus I & 1 & 0.004 & 164 & 2.12 & 0.377 & 0.02 & 1.00 & -1.33 & 0.00 \\
Grus I & 1 & 0.019 & 65.4 & 3.98 & 1.90 & 0.03 & 1.00 & -0.57 & 0.00 \\
Leo IV & 1 & 0.013 & 63.8 & 3.31 & 1.23 & 0.05 & 1.00 & -0.10 & 0.00 \\
Canes Ventici II & 1 & 0.014 & 76.7 & 4.26 & 1.38 & 0.03 & 1.00 & -0.34 & 0.00 \\
Bootes I & 1 & 0.034 & 16.7 & 1.75 & 1.18 & 0.24 & 1.00 & -1.30 & 0.00 \\
Tucana II & 1 & 0.025 & 26.2 & 2.83 & 0.875 & 0.13 & 1.00 & 0.03 & 0.00 \\
Tucana IV & 1 & 0.019 & 52.2 & 3.42 & 1.89 & 0.04 & 1.00 & -0.88 & 0.00 \\
Leo V & 1 & 0.007 & 121 & 2.68 & 0.700 & 0.02 & 1.00 & -0.73 & 0.00 \\
And X & 1 & 0.030 & 30.2 & 3.87 & 2.97 & 0.08 & 1.00 & -0.02 & 0.00 \\
And XI & 1 & 0.068 & 15.9 & 3.80 & 2.36 & 0.13 & 1.00 & -0.17 & 0.00 \\
And XII & 1 & 0.021 & 29.7 & 2.47 & 2.01 & 0.13 & 1.00 & -0.05 & 0.00 \\
And XIII & 1 & 0.042 & 32.8 & 4.85 & 4.07 & 0.05 & 1.00 & -0.48 & 0.00 \\
And XVI & 1 & 0.018 & 50.0 & 3.65 & 1.75 & 0.05 & 1.00 & -0.05 & 0.00 \\
And XVII & 1 & 0.065 & 10.5 & 2.17 & 2.27 & 0.32 & 1.00 & -0.38 & 0.00 \\
And XX & 1 & 0.043 & 38.7 & 6.65 & 4.22 & 0.04 & 1.00 & -0.18 & 0.00 \\
And XXII & 1 & 0.017 & 38.4 & 2.71 & 1.68 & 0.09 & 1.00 & -0.06 & 0.00 \\
And XXVI & 1 & 0.105 & 19.2 & 7.39 & 10.3 & 0.06 & 1.00 & -0.55 & 0.00 \\
\hline
\end{tabular}
\caption{The first 23 rows are dSph galaxies and the lower 25 rows are UFD galaxies. $F$ is the soliton mass fraction, $M= M_S+M_\bullet$ is the total galactic mass, $\sigma=\sigma_{\rm los}$ is the predicted (2D projected) stellar velocity dispersion, $M_{1/2} = M(<r_{1/2})$ is the predicted total mass contained within the stellar half-light radius, $r_{c,S}$ is the soliton characteristic radius and $\sigma_S/\sigma_\ast = \nu_S v_0/(\sqrt{3} \sigma_{los,obs})$ is obtained assuming that the stellar velocities are isotropic (as are the soliton velocities). Note that the predicted soliton radii $r_{c,S}$ are not constrained to be equal to the stellar $r_c$ in Table~\ref{T:dSph_UFD_data}, but the two radii are similar in magnitude (to within a factor of about 2-3). The relative errors $\Delta_\sigma$ and $\Delta_M$ quantify observational consistency (as further described in the introduction to this Appendix section). }
\label{T:optimizem0}
\end{table}

\begin{table}
\centering
\begin{tabular}{lccccccccccc}
\hline
\multicolumn{11}{c}{5G predictions with two $m_0$ values and $F=1$, optimising $M$ for observational consistency with $M_{1/2}$ } \\
\hline
Galaxy & $F$ & $M/10^8$ & $m_0/10^{-22}$ & $\sigma$ & $M_{1/2}/10^6$ & $M_\bullet/10^6$ & $r_{c,S}$ & $\sigma_S/\sigma_\ast$ & $\Delta_\sigma$ & $\Delta_M$ \\
  &  & M$_\odot$ & eV/c$^2$ & (km/s) & M$_\odot$ & M$_\odot$ &\,kpc &  &  &  \\
\hline
Tucana & 1 & 2.449 & 1.5 & 13.9 & 46.7 & 0.00 & 0.42 & 1.18 & 0.21 & 0.00 \\
Cetus & 1 & 1.668 & 1.5 & 10.1 & 68.7 & 0.00 & 0.61 & 0.96 & -0.77 & -0.00 \\
Aquarius & 1 & 1.956 & 1.5 & 11.2 & 33.5 & 0.00 & 0.52 & 1.21 & 0.56 & 0.00 \\
Draco & 1 & 2.149 & 1.5 & 9.61 & 17.3 & 0.00 & 0.47 & 1.53 & 0.55 & 0.00 \\
Leo I & 1 & 2.059 & 1.5 & 9.63 & 20.5 & 0.00 & 0.49 & 1.43 & 0.36 & 0.00 \\
Phoenix & 1 & 1.974 & 1.5 & 9.99 & 23.3 & 0.00 & 0.52 & 1.36 & 0.99 & 0.00 \\
Canes Ventici & 1 & 1.432 & 1.5 & 7.13 & 25.3 & 0.00 & 0.71 & 1.20 & -1.17 & 0.00 \\
Sextans & 1 & 1.251 & 1.5 & 7.18 & 41.5 & 0.00 & 0.81 & 1.01 & -0.55 & 0.00 \\
Crater II & 1 & 0.557 & 1.5 & 3.43 & 7.23 & 0.00 & 1.83 & 1.32 & 2.43 & 0.00 \\
Leo II & 1 & 2.108 & 1.5 & 9.77 & 9.73 & 0.00 & 0.48 & 1.82 & 5.92 & 0.00 \\
Carina & 1 & 1.379 & 1.5 & 5.58 & 17.2 & 0.00 & 0.74 & 1.34 & -0.85 & 0.00 \\
Ursa Minor & 1 & 1.852 & 1.5 & 9.62 & 57.5 & 0.00 & 0.55 & 1.03 & -2.35 & 0.00 \\
Sculptor & 1 & 2.072 & 1.5 & 9.80 & 27.4 & 0.00 & 0.49 & 1.31 & -1.00 & 0.00 \\
And I & 1 & 1.443 & 1.5 & 8.66 & 54.2 & 0.00 & 0.71 & 0.98 & -0.49 & 0.00 \\
And III & 1 & 1.889 & 1.5 & 10.2 & 46.1 & 0.00 & 0.54 & 1.10 & -0.50 & -0.00 \\
And V & 1 & 2.032 & 1.5 & 10.7 & 40.8 & 0.00 & 0.50 & 1.16 & -0.54 & 0.00 \\
And IX & 1 & 1.977 & 1.5 & 10.6 & 39.8 & 0.00 & 0.52 & 1.16 & -0.13 & 0.00 \\
And XIV & 1 & 1.280 & 1.5 & 7.02 & 10.6 & 0.00 & 0.80 & 1.51 & 1.25 & 0.00 \\
And XV & 1 & 2.402 & 1.5 & 12.1 & 25.9 & 0.00 & 0.42 & 1.40 & 0.16 & 0.00 \\
And XVIII & 1 & 1.913 & 1.5 & 9.36 & 28.9 & 0.00 & 0.53 & 1.26 & -0.15 & 0.00 \\
And XXI & 1 & 0.940 & 1.5 & 5.70 & 34.8 & 0.00 & 1.08 & 0.98 & -0.44 & -0.00 \\
And XXIII & 1 & 1.005 & 1.5 & 6.11 & 55.8 & 0.00 & 1.01 & 0.90 & -0.99 & -0.00 \\
And XXV & 1 & 0.795 & 1.5 & 4.84 & 4.60 & 0.00 & 1.28 & 1.69 & 1.53 & 0.00 \\
\hline
Phoenix II & 1 & 0.134 & 20 & 9.44 & 4.05 & 0.00 & 0.04 & 1.09 & -0.29 & 0.00 \\
Segue I & 1 & 0.077 & 20 & 6.19 & 0.453 & 0.00 & 0.07 & 2.11 & 2.86 & 0.00 \\
Pegasus III & 1 & 0.074 & 20 & 5.48 & 1.44 & 0.00 & 0.08 & 1.47 & 0.03 & 0.00 \\
Wilman I & 1 & 0.077 & 20 & 6.60 & 0.491 & 0.00 & 0.07 & 2.06 & 3.25 & 0.00 \\
Horoligium I & 1 & 0.078 & 20 & 4.98 & 0.916 & 0.00 & 0.07 & 1.70 & 0.03 & 0.00 \\
Pisces II & 1 & 0.070 & 20 & 4.57 & 1.68 & 0.00 & 0.08 & 1.39 & -0.35 & 0.00 \\
Coma Berenices & 1 & 0.061 & 20 & 4.54 & 1.36 & 0.00 & 0.09 & 1.42 & -0.07 & 0.00 \\
Reticulum II & 1 & 0.055 & 20 & 4.34 & 0.511 & 0.00 & 0.10 & 1.83 & 0.68 & 0.00 \\
Hydrus I & 1 & 0.049 & 20 & 3.84 & 0.377 & 0.00 & 0.12 & 1.95 & 2.25 & 0.00 \\
Grus I & 1 & 0.067 & 20 & 5.37 & 1.90 & 0.00 & 0.09 & 1.33 & -0.01 & 0.00 \\
Leo IV & 1 & 0.042 & 20 & 3.20 & 1.23 & 0.00 & 0.14 & 1.32 & -0.22 & 0.00 \\
Canes Ventici II & 1 & 0.060 & 20 & 4.00 & 1.38 & 0.00 & 0.09 & 1.40 & -0.60 & 0.00 \\
Bootes I & 1 & 0.027 & 20 & 1.70 & 1.18 & 0.00 & 0.22 & 1.21 & -1.40 & -0.00 \\
Tucana II & 1 & 0.036 & 20 & 3.07 & 0.875 & 0.00 & 0.16 & 1.38 & 0.23 & 0.00 \\
Tucana IV & 1 & 0.049 & 20 & 3.93 & 1.89 & 0.00 & 0.12 & 1.22 & -0.37 & -0.00 \\
Leo V & 1 & 0.059 & 20 & 4.42 & 0.700 & 0.00 & 0.10 & 1.71 & 0.31 & 0.00 \\
And X & 1 & 0.040 & 20 & 3.14 & 2.97 & 0.00 & 0.14 & 1.10 & -0.63 & -0.00 \\
And XI & 1 & 0.050 & 20 & 3.67 & 2.36 & 0.00 & 0.11 & 1.16 & -0.20 & -0.00 \\
And XII & 1 & 0.027 & 20 & 1.94 & 2.01 & 0.00 & 0.21 & 1.11 & -0.25 & -0.00 \\
And XIII & 1 & 0.060 & 20 & 4.83 & 4.07 & 0.00 & 0.10 & 1.11 & -0.48 & -0.00 \\
And XVI & 1 & 0.043 & 20 & 3.57 & 1.75 & 0.00 & 0.13 & 1.21 & -0.08 & -0.00 \\
And XVII & 1 & 0.030 & 20 & 2.40 & 2.27 & 0.00 & 0.19 & 1.11 & -0.26 & -0.00 \\
And XX & 1 & 0.075 & 20 & 5.33 & 4.22 & 0.00 & 0.08 & 1.13 & -0.71 & -0.00 \\
And XXII & 1 & 0.030 & 20 & 2.74 & 1.68 & 0.00 & 0.19 & 1.15 & -0.04 & -0.00 \\
And XXVI & 1 & 0.104 & 20 & 7.49 & 10.3 & 0.00 & 0.05 & 1.29 & -0.50 & 0.00 \\
\hline
\end{tabular}
\caption{The first 23 rows are dSph galaxies with $m_0 = 1.5\times 10^{-22}$\,eV/c$^2$, and the lower 25 rows are UFD galaxies with $m_0 = 1.5\times 10^{-22}$\,eV/c$^2$. $F$ is the soliton mass fraction, $M= M_S+M_\bullet$ is the total galactic mass, $\sigma=\sigma_{los}$ is the predicted (2D projected) stellar velocity dispersion, $M_{1/2} = M(<r_{1/2})$ is the predicted total mass contained within the stellar half-light radius, $r_{c,S}$ is the soliton characteristic radius and $\sigma_S/\sigma_\ast = \nu_S v_0/(\sqrt{3} \sigma_{los,obs})$ is obtained assuming that the stellar velocities are isotropic (as are the soliton velocities).  The relative errors $\Delta_\sigma$ and $\Delta_M$ quantify observational consistency (as further described in the introduction to this Appendix section). Note that those galaxies for which $|\Delta_\sigma|>1$ may contain massive black holes, as further evidenced by the prediction with $F<1$ described in the following tables. }
\label{T:Dual_m0}
\end{table}

\begin{table}
\centering
\begin{tabular}{lccccccccccc}
\hline
\multicolumn{11}{c}{5G predictions with $m_0= 1.5\times 10^{-22}$, optimising $F$ and $M$ so $\max[|\Delta_\sigma|$, $|\Delta_M|]\le1$ (if possible)} \\
\hline
Galaxy & $F$ & $M/10^8$ & $m_0/10^{-22}$ & $\sigma$ & $M_{1/2}/10^6$ & $M_\bullet/10^6$ & $r_{c,S}$ & $\sigma_S/\sigma_\ast$ & $\Delta_\sigma$ & $\Delta_M$ \\
  &  & M$_\odot$ & eV/c$^2$ & (km/s) & M$_\odot$ & M$_\odot$ &\,kpc &  &  &  \\
\hline
Tucana & 1 & 2.419 & 1.5 & 13.7 & 44.8 & 0.00 & 0.42 & 1.16 & 0.14 & -0.14 \\
Cetus & 1 & 1.724 & 1.5 & 10.5 & 75.4 & 0.00 & 0.59 & 0.99 & -0.48 & 0.48 \\
Aquarius & 1 & 1.917 & 1.5 & 10.9 & 31.3 & 0.00 & 0.53 & 1.19 & 0.37 & -0.37 \\
Draco & 1 & 2.113 & 1.5 & 9.37 & 16.3 & 0.00 & 0.48 & 1.50 & 0.34 & -0.34 \\
Leo I & 1 & 2.036 & 1.5 & 9.46 & 19.6 & 0.00 & 0.50 & 1.41 & 0.22 & -0.21 \\
Phoenix & 1 & 1.937 & 1.5 & 9.71 & 21.8 & 0.00 & 0.53 & 1.33 & 0.59 & -0.58 \\
Canes Ventici & 1 & 1.458 & 1.5 & 7.31 & 26.8 & 0.00 & 0.70 & 1.22 & -0.73 & 0.73 \\
Sextans & 1 & 1.288 & 1.5 & 7.39 & 45.3 & 0.00 & 0.79 & 1.04 & -0.39 & 0.39 \\
Crater II & 0.9295 & 0.405 & 1.5 & 3 & 5.96 & 2.86 & 2.07 & 1.08 & 1.00 & -0.99 \\
Leo II & 0.9095 & 0.869 & 1.5 & 7.83 & 8.37 & 7.86 & 0.91 & 0.94 & 1.07 & -1.07 \\
Carina & 1 & 1.441 & 1.5 & 5.92 & 20.0 & 0.00 & 0.71 & 1.40 & -0.57 & 0.57 \\
Ursa Minor & 0.9834 & 1.888 & 1.5 & 10.7 & 67.2 & 3.13 & 0.52 & 1.07 & -1.00 & 1.00 \\
Sculptor & 1 & 2.091 & 1.5 & 9.92 & 28.3 & 0.00 & 0.49 & 1.32 & -0.60 & 0.61 \\
And I & 1 & 1.484 & 1.5 & 8.95 & 58.8 & 0.00 & 0.69 & 1.01 & -0.30 & 0.30 \\
And III & 1 & 1.933 & 1.5 & 10.5 & 49.7 & 0.00 & 0.53 & 1.12 & -0.29 & 0.29 \\
And V & 1 & 2.069 & 1.5 & 10.9 & 43.4 & 0.00 & 0.49 & 1.18 & -0.35 & 0.35 \\
And IX & 1 & 1.993 & 1.5 & 10.7 & 40.9 & 0.00 & 0.51 & 1.17 & -0.09 & 0.09 \\
And XIV & 1 & 1.125 & 1.5 & 6.20 & 6.57 & 0.00 & 0.91 & 1.33 & 0.62 & -0.61 \\
And XV & 1 & 2.348 & 1.5 & 11.8 & 23.9 & 0.00 & 0.43 & 1.36 & 0.12 & -0.12 \\
And XVIII & 1 & 1.933 & 1.5 & 9.46 & 29.9 & 0.00 & 0.53 & 1.27 & -0.10 & 0.11 \\
And XXI & 1 & 0.966 & 1.5 & 5.84 & 37.7 & 0.00 & 1.05 & 1.01 & -0.29 & 0.29 \\
And XXIII & 1 & 1.060 & 1.5 & 6.45 & 63.6 & 0.00 & 0.96 & 0.95 & -0.65 & 0.65 \\
And XXV & 1 & 0.668 & 1.5 & 4.08 & 2.39 & 0.00 & 1.52 & 1.42 & 0.90 & -0.90 \\
\hline
Phoenix II & 0.9977 & 3.276 & 1.5 & 20.4 & 1.20 & 0.75 & 0.31 & 1.90 & 0.99 & -0.98 \\
Segue I & 0.9962 & 0.834 & 1.5 & 4.69 & 0.315 & 0.32 & 1.21 & 1.36 & 0.99 & -0.99 \\
Pegasus III & 0.9978 & 1.604 & 1.5 & 8.38 & 0.436 & 0.35 & 0.63 & 1.89 & 0.99 & -0.99 \\
Wilman I & 0.9969 & 0.992 & 1.5 & 4.79 & 0.309 & 0.31 & 1.02 & 1.58 & 0.99 & -1.00 \\
Horoligium I & 0.9962 & 1.477 & 1.5 & 7.69 & 0.591 & 0.56 & 0.68 & 1.92 & 1.00 & -0.99 \\
Pisces II & 0.9966 & 1.53 & 1.5 & 8.98 & 0.606 & 0.52 & 0.67 & 1.80 & 0.99 & -0.99 \\
Coma Berenices & 0.9879 & 0.824 & 1.5 & 5.39 & 1.01 & 1.00 & 1.20 & 1.14 & 0.99 & -1.00 \\
Reticulum II & 0.9962 & 1.022 & 1.5 & 4.85 & 0.400 & 0.39 & 0.99 & 2.02 & 0.99 & -1.00 \\
Hydrus I & 0.9955 & 0.638 & 1.5 & 3.20 & 0.288 & 0.29 & 1.58 & 1.51 & 1.00 & -1.00 \\
Grus I & 0.9969 & 1.328 & 1.5 & 8.38 & 0.501 & 0.41 & 0.76 & 1.57 & 0.99 & -0.99 \\
Leo IV & 0.9959 & 1.211 & 1.5 & 4.69 & 0.757 & 0.50 & 0.83 & 2.27 & 0.99 & -0.99 \\
Canes Ventici II & 0.9901 & 0.905 & 1.5 & 5.60 & 0.914 & 0.90 & 1.09 & 1.25 & 1.00 & -0.99 \\
Bootes I & 1 & 1.011 & 1.5 & 3.22 & 0.856 & 0.00 & 1.01 & 2.69 & 0.91 & -0.92 \\
Tucana II & 0.9950 & 0.848 & 1.5 & 3.99 & 0.499 & 0.42 & 1.18 & 1.93 & 0.99 & -0.99 \\
Tucana IV & 0.9845 & 0.783 & 1.5 & 5.99 & 1.25 & 1.21 & 1.25 & 1.16 & 0.99 & -0.99 \\
Leo V & 0.9978 & 1.064 & 1.5 & 5.98 & 0.250 & 0.23 & 0.95 & 1.83 & 0.99 & -0.99 \\
And X & 1 & 1.236 & 1.5 & 4.88 & 1.64 & 0.00 & 0.82 & 2.02 & 0.82 & -0.82 \\
And XI & 1 & 1.692 & 1.5 & 4.6 & 0.28 & 0.00 & 0.60 & 2.35 & 0 & -0.88 \\
And XII & 1 & 0.914 & 1.5 & 3.35 & 1.71 & 0.00 & 1.11 & 2.24 & 0.15 & -0.15 \\
And XIII & 0.9903 & 1.122 & 1.5 & 7.79 & 1.37 & 1.09 & 0.88 & 1.23 & 1.00 & -0.99 \\
And XVI & 1 & 1.652 & 1.5 & 4.61 & 1.26 & 0.00 & 0.62 & 2.78 & 0.28 & -0.28 \\
And XVII & 1 & 1.083 & 1.5 & 2.73 & 2.49 & 0.00 & 0.94 & 2.39 & -0.09 & 0.09 \\
And XX & 1 & 2.419 & 1.5 & 11.0 & 1.92 & 0.00 & 0.42 & 2.18 & 0.99 & -1.00 \\
And XXII & 1 & 0.882 & 1.5 & 5.22 & 0.569 & 0.00 & 1.15 & 2.01 & 0.83 & -0.84 \\
And XXVI & 0.9775 & 1.302 & 1.5 & 11.4 & 3.77 & 2.93 & 0.74 & 1.01 & 1.00 & -0.98 \\
\hline
\end{tabular}
\caption{The first 23 rows are dSph galaxies and the lower 25 rows are UFD galaxies. $F$ is the soliton mass fraction, $M= M_S+M_\bullet$ is the total galactic mass, $\sigma=\sigma_{los}$ is the predicted (2D projected) stellar velocity dispersion, $M_{1/2} = M(<r_{1/2})$ is the predicted total mass contained within the stellar half-light radius, $r_{c,S}$ is the soliton characteristic radius and $\sigma_S/\sigma_\ast = \nu_S v_0/(\sqrt{3} \sigma_{los,obs})$ is obtained assuming that the stellar velocities are isotropic (as are the soliton velocities).   The relative errors $\Delta_\sigma$ and $\Delta_M$ quantify observational consistency (as further described in the introduction to this Appendix section).}
\label{T:1x5m0_1SE}
\end{table}

\begin{table}
\centering
\begin{tabular}{lccccccccccc}
\hline
\multicolumn{11}{c}{5G predictions with $m_0 = 0.75\times 10^{-22}$, optimising $F$ and $M$ so $\max[|\Delta_\sigma|$, $|\Delta_M|]\le1$ (if possible)} \\
\hline
Galaxy & $F$ & $M/10^8$ & $m_0/10^{-22}$ & $\sigma$ & $M_{1/2}/10^6$ & $M_\bullet/10^6$ & $r_{c,S}$ & $\sigma_S/\sigma_\ast$ & $\Delta_\sigma$ & $\Delta_M$ \\
  &  & M$_\odot$ & eV/c$^2$ & (km/s) & M$_\odot$ & M$_\odot$ &\,kpc &  &  &  \\
\hline
Tucana & 1 & 6.035 & 0.75 & 15.9 & 33.1 & 0.00 & 0.67 & 1.45 & 0.97 & -0.97 \\
Cetus & 1 & 4.206 & 0.75 & 11.8 & 63.4 & 0.00 & 0.97 & 1.21 & 0.47 & -0.47 \\
Aquarius & 0.9917 & 4.779 & 0.75 & 11.9 & 27.5 & 3.97 & 0.83 & 1.50 & 0.99 & -0.99 \\
Draco & 0.958 & 3.021 & 0.75 & 10.1 & 14.3 & 12.69 & 1.20 & 1.13 & 0.99 & -0.99 \\
Leo I & 1 & 5.510 & 0.75 & 9.86 & 18.3 & 0.00 & 0.74 & 1.91 & 0.55 & -0.55 \\
Phoenix & 0.9692 & 3.983 & 0.75 & 10.0 & 20.7 & 12.27 & 0.94 & 1.42 & 1.00 & -0.99 \\
Canes Ventici & 1 & 3.797 & 0.75 & 7.97 & 23.3 & 0.00 & 1.07 & 1.60 & 0.93 & -0.92 \\
Sextans & 1 & 3.104 & 0.75 & 8.93 & 33.8 & 0.00 & 1.31 & 1.26 & 0.79 & -0.79 \\
Crater II & 0.9259 & 0.724 & 0.75 & 3.00 & 5.96 & 5.36 & 4.57 & 0.94 & 1.00 & -0.99 \\
Leo II & 0.9557 & 1.882 & 0.75 & 7.80 & 8.47 & 8.34 & 1.91 & 0.86 & 1.00 & -0.99 \\
Carina & 1 & 3.748 & 0.75 & 6.73 & 16.7 & 0.00 & 1.09 & 1.82 & 0.11 & -0.10 \\
Ursa Minor & 1 & 5.018 & 0.75 & 11.1 & 63.0 & 0.00 & 0.81 & 1.39 & -0.56 & 0.57 \\
Sculptor & 0.9066 & 2.314 & 0.75 & 11.0 & 23.0 & 21.61 & 1.35 & 0.86 & 2.93 & -2.95 \\
And I & 1 & 3.677 & 0.75 & 9.90 & 50.3 & 0.00 & 1.11 & 1.25 & 0.29 & -0.29 \\
And III & 1 & 5.027 & 0.75 & 11.2 & 45.2 & 0.00 & 0.81 & 1.46 & 0.09 & -0.09 \\
And V & 0.9154 & 3.129 & 0.75 & 12.3 & 33.9 & 26.47 & 1.03 & 1.02 & 0.99 & -0.99 \\
And IX & 0.9587 & 3.728 & 0.75 & 12.9 & 28.1 & 15.40 & 0.97 & 1.15 & 1.00 & -0.99 \\
And XIV & 0.9907 & 2.283 & 0.75 & 6.69 & 4.17 & 2.12 & 1.74 & 1.36 & 0.99 & -0.98 \\
And XV & 1 & 5.870 & 0.75 & 14.8 & 16.5 & 0.00 & 0.69 & 1.70 & 0.55 & -0.56 \\
And XVIII & 1 & 4.694 & 0.75 & 11.9 & 19.5 & 0.00 & 0.87 & 1.55 & 0.95 & -0.96 \\
And XXI & 1 & 2.245 & 0.75 & 7.08 & 25.5 & 0.00 & 1.81 & 1.18 & 0.98 & -0.98 \\
And XXIII & 1 & 2.455 & 0.75 & 7.37 & 52.5 & 0.00 & 1.66 & 1.10 & 0.27 & -0.27 \\
And XXV & 0.9901 & 1.388 & 0.75 & 4.19 & 2.17 & 1.37 & 2.85 & 1.48 & 0.99 & -0.99 \\
\hline
Phoenix II & 0.9986 & 7.355 & 0.75 & 20.4 & 1.19 & 1.03 & 0.55 & 2.13 & 0.99 & -0.98 \\
Segue I & 0.9985 & 2.124 & 0.75 & 4.70 & 0.314 & 0.32 & 1.91 & 1.74 & 1.00 & -1.01 \\
Pegasus III & 0.999 & 3.883 & 0.75 & 8.36 & 0.435 & 0.39 & 1.05 & 2.29 & 0.99 & -0.99 \\
Wilman I & 0.9988 & 2.562 & 0.75 & 4.79 & 0.308 & 0.31 & 1.58 & 2.04 & 0.99 & -1.00 \\
Horoligium I & 0.9984 & 3.554 & 0.75 & 7.68 & 0.591 & 0.57 & 1.14 & 2.31 & 0.99 & -0.99 \\
Pisces II & 0.9984 & 3.506 & 0.75 & 8.97 & 0.600 & 0.56 & 1.16 & 2.07 & 0.99 & -0.99 \\
Coma Berenices & 0.9952 & 2.078 & 0.75 & 5.39 & 1.01 & 1.00 & 1.93 & 1.45 & 0.99 & -0.99 \\
Reticulum II & 0.9985 & 2.590 & 0.75 & 4.88 & 0.398 & 0.39 & 1.57 & 2.56 & 1.01 & -1.01 \\
Hydrus I & 0.9983 & 1.677 & 0.75 & 3.21 & 0.285 & 0.29 & 2.42 & 1.99 & 1.02 & -1.03 \\
Grus I & 0.9986 & 3.197 & 0.75 & 8.37 & 0.492 & 0.45 & 1.27 & 1.89 & 0.99 & -0.99 \\
Leo IV & 0.9979 & 2.915 & 0.75 & 4.69 & 0.758 & 0.61 & 1.39 & 2.74 & 0.99 & -0.99 \\
Canes Ventici II & 0.9961 & 2.295 & 0.75 & 5.60 & 0.910 & 0.90 & 1.76 & 1.59 & 1.00 & -1.00 \\
Bootes I & 0.9834 & 0.452 & 0.75 & 3.50 & 0.75 & 0.75 & 8.61 & 0.61 & 1.22 & -1.22 \\
Tucana II & 0.9978 & 2.107 & 0.75 & 3.99 & 0.498 & 0.46 & 1.92 & 2.40 & 0.99 & -0.99 \\
Tucana IV & 0.9938 & 1.984 & 0.75 & 5.99 & 1.25 & 1.23 & 2.02 & 1.48 & 0.99 & -1.00 \\
Leo V & 0.9991 & 2.605 & 0.75 & 5.93 & 0.254 & 0.23 & 1.56 & 2.25 & 0.97 & -0.98 \\
And X & 0.9964 & 2.497 & 0.75 & 5.09 & 1.36 & 0.90 & 1.61 & 2.04 & 0.99 & -0.99 \\
And XI & 1 & 2.889 & 0.75 & 4.60 & 0.149 & 0.00 & 1.41 & 2.00 & 0 & -0.94 \\
And XII & 1 & 2.482 & 0.75 & 3.87 & 1.51 & 0.00 & 1.64 & 3.05 & 0.25 & -0.25 \\
And XIII & 0.9955 & 2.675 & 0.75 & 7.79 & 1.35 & 1.20 & 1.5 & 1.47 & 1.00 & -0.99 \\
And XVI & 1 & 4.456 & 0.75 & 4.94 & 1.06 & 0.00 & 0.91 & 3.75 & 0.39 & -0.39 \\
And XVII & 1 & 2.954 & 0.75 & 2.94 & 2.24 & 0.00 & 1.38 & 3.26 & 0.02 & -0.02 \\
And XX & 0.9969 & 4.777 & 0.75 & 11.0 & 1.94 & 1.48 & 0.85 & 2.16 & 0.99 & -0.99 \\
And XXII & 1 & 2.260 & 0.75 & 5.64 & 0.390 & 0.00 & 1.80 & 2.58 & 0.98 & -0.97 \\
And XXVI & 0.9882 & 2.905 & 0.75 & 11.4 & 3.75 & 3.43 & 1.36 & 1.10 & 1.00 & -0.99 \\
\hline
\end{tabular}
\caption{The first 23 rows are dSph galaxies and the lower 25 rows are UFD galaxies. $F$ is the soliton mass fraction, $M= M_S+M_\bullet$ is the total galactic mass, $\sigma=\sigma_{los}$ is the predicted (2D projected) stellar velocity dispersion, $M_{1/2} = M(<r_{1/2})$ is the predicted total mass contained within the stellar half-light radius, $r_{c,S}$ is the soliton characteristic radius and $\sigma_S/\sigma_\ast = \nu_S v_0/(\sqrt{3} \sigma_{los,obs})$ is obtained assuming that the stellar velocities are isotropic (as are the soliton velocities).   The relative errors $\Delta_\sigma$ and $\Delta_M$ quantify observational consistency (as further described in the introduction to this Appendix section).}
\label{T:0x75m0_1SE}
\end{table}

\begin{table}
\centering
\begin{tabular}{lccccccccccc}
\hline
\multicolumn{11}{c}{5G predictions with $m_0 = 3\times 10^{-22}$, optimising $F$ and $M$ so $\max[|\Delta_\sigma|$, $|\Delta_M|]\le1$ (if possible)} \\
\hline
Galaxy & $F$ & $M/10^8$ & $m_0/10^{-22}$ & $\sigma$ & $M_{1/2}/10^6$ & $M_\bullet/10^6$ & $r_{c,S}$ & $\sigma_S/\sigma_\ast$ & $\Delta_\sigma$ & $\Delta_M$ \\
  &  & M$_\odot$ & eV/c$^2$ & (km/s) & M$_\odot$ & M$_\odot$ &\,kpc &  &  &  \\
\hline
Tucana & 1 & 1.020 & 3 & 12.2 & 53.9 & 0.00 & 0.25 & 0.98 & -0.46 & 0.46 \\
Cetus & 1 & 0.838 & 3 & 10.4 & 76.6 & 0.00 & 0.30 & 0.96 & -0.56 & 0.56 \\
Aquarius & 1 & 0.779 & 3 & 9.95 & 35.5 & 0.00 & 0.33 & 0.97 & -0.27 & 0.27 \\
Draco & 1 & 0.812 & 3 & 8.66 & 18.3 & 0.00 & 0.31 & 1.15 & -0.31 & 0.31 \\
Leo I & 1 & 0.781 & 3 & 8.97 & 21.2 & 0.00 & 0.33 & 1.08 & -0.19 & 0.19 \\
Phoenix & 1 & 0.753 & 3 & 9.16 & 23.8 & 0.00 & 0.34 & 1.03 & -0.20 & 0.20 \\
Canes Ventici & 0.9502 & 0.534 & 3 & 7.20 & 27.4 & 2.66 & 0.41 & 0.93 & -1.00 & 0.99 \\
Sextans & 1 & 0.600 & 3 & 6.76 & 50.0 & 0.00 & 0.42 & 0.97 & -0.88 & 0.87 \\
Crater II & 1 & 0.219 & 3 & 2.55 & 7.87 & 0.00 & 1.16 & 1.04 & -0.50 & 0.50 \\
Leo II & 1 & 0.75 & 3 & 7.67 & 8.86 & 0.00 & 0.34 & 1.29 & 0.67 & -0.68 \\
Carina & 1 & 0.567 & 3 & 5.52 & 21.7 & 0.00 & 0.45 & 1.10 & -0.90 & 0.90 \\
Ursa Minor & 0.9993 & 0.854 & 3 & 10.7 & 67.2 & 0.06 & 0.30 & 0.95 & -1.00 & 0.99 \\
Sculptor & 0.864 & 0.638 & 3 & 9.80 & 28.9 & 8.68 & 0.27 & 0.99 & -1.00 & 0.99 \\
And I & 1 & 0.696 & 3 & 8.75 & 60.7 & 0.00 & 0.37 & 0.95 & -0.43 & 0.43 \\
And III & 1 & 0.809 & 3 & 10.2 & 51.9 & 0.00 & 0.31 & 0.94 & -0.48 & 0.48 \\
And V & 0.7836 & 0.684 & 3 & 9.96 & 49.9 & 14.80 & 0.20 & 0.97 & -1.24 & 1.23 \\
And IX & 1 & 0.867 & 3 & 9.02 & 51.2 & 0.00 & 0.29 & 1.02 & -0.94 & 0.93 \\
And XIV & 1 & 0.484 & 3 & 5.28 & 11.2 & 0.00 & 0.53 & 1.14 & -0.09 & 0.09 \\
And XV & 1 & 0.954 & 3 & 10.0 & 30.5 & 0.00 & 0.27 & 1.11 & -0.20 & 0.20 \\
And XVIII & 1 & 0.799 & 3 & 7.96 & 36.3 & 0.00 & 0.32 & 1.05 & -0.76 & 0.75 \\
And XXI & 1 & 0.481 & 3 & 5.35 & 43.2 & 0.00 & 0.53 & 1.01 & -0.83 & 0.83 \\
And XXIII & 1 & 0.604 & 3 & 6.76 & 59.9 & 0.00 & 0.42 & 1.09 & -0.34 & 0.34 \\
And XXV & 1 & 0.279 & 3 & 3.34 & 3.91 & 0.00 & 0.91 & 1.19 & 0.28 & -0.28 \\
\hline
Phoenix II & 1 & 1.570 & 3 & 19.5 & 1.39 & 0.00 & 0.16 & 1.82 & 0.90 & -0.92 \\
Segue I & 0.9908 & 0.341 & 3 & 4.69 & 0.315 & 0.31 & 0.73 & 1.13 & 0.99 & -1.00 \\
Pegasus III & 0.9964 & 0.704 & 3 & 8.39 & 0.439 & 0.25 & 0.36 & 1.66 & 1.00 & -0.98 \\
Wilman I & 0.9923 & 0.395 & 3 & 4.79 & 0.309 & 0.30 & 0.63 & 1.26 & 0.99 & -1.00 \\
Horoligium I & 0.9921 & 0.660 & 3 & 7.68 & 0.591 & 0.52 & 0.38 & 1.71 & 0.99 & -0.99 \\
Pisces II & 0.9964 & 0.726 & 3 & 8.98 & 0.598 & 0.26 & 0.35 & 1.71 & 0.99 & -0.99 \\
Coma Berenices & 0.9714 & 0.344 & 3 & 5.40 & 1.01 & 0.98 & 0.68 & 0.98 & 1.00 & -0.99 \\
Reticulum II & 0.9911 & 0.422 & 3 & 4.85 & 0.400 & 0.38 & 0.59 & 1.68 & 0.99 & -0.99 \\
Hydrus I & 0.9888 & 0.254 & 3 & 3.20 & 0.288 & 0.28 & 0.97 & 1.21 & 1.00 & -1.00 \\
Grus I & 0.995 & 0.589 & 3 & 8.38 & 0.509 & 0.29 & 0.43 & 1.39 & 0.99 & -0.98 \\
Leo IV & 1 & 0.572 & 3 & 4.65 & 0.769 & 0.00 & 0.45 & 2.15 & 0.96 & -0.97 \\
Canes Ventici II & 0.977 & 0.380 & 3 & 5.60 & 0.914 & 0.87 & 0.63 & 1.06 & 1.00 & -0.99 \\
Bootes I & 1 & 0.379 & 3 & 2.78 & 1.03 & 0.00 & 0.67 & 2.02 & 0.42 & -0.42 \\
Tucana II & 0.9906 & 0.363 & 3 & 3.99 & 0.498 & 0.34 & 0.68 & 1.66 & 0.99 & -0.99 \\
Tucana IV & 0.9648 & 0.329 & 3 & 5.99 & 1.25 & 1.16 & 0.70 & 0.99 & 0.99 & -0.99 \\
Leo V & 0.9952 & 0.454 & 3 & 5.99 & 0.253 & 0.22 & 0.55 & 1.57 & 1.00 & -0.99 \\
And X & 1 & 0.493 & 3 & 4.28 & 2.46 & 0.00 & 0.52 & 1.62 & 0.32 & -0.32 \\
And XI & 1 & 0.524 & 3 & 4.60 & 0.632 & 0.00 & 0.49 & 1.45 & 0 & -0.73 \\
And XII & 1 & 0.339 & 3 & 2.82 & 1.93 & 0.00 & 0.75 & 1.67 & 0.04 & -0.04 \\
And XIII & 0.9903 & 0.527 & 3 & 7.79 & 1.37 & 0.51 & 0.47 & 1.16 & 1.00 & -0.99 \\
And XVI & 1 & 0.609 & 3 & 4.32 & 1.43 & 0.00 & 0.42 & 2.05 & 0.18 & -0.18 \\
And XVII & 1 & 0.400 & 3 & 2.56 & 2.72 & 0.00 & 0.64 & 1.76 & -0.18 & 0.18 \\
And XX & 1 & 0.970 & 3 & 9.14 & 3 & 0.00 & 0.26 & 1.75 & 0.52 & -0.53 \\
And XXII & 1 & 0.348 & 3 & 4.61 & 0.841 & 0.00 & 0.73 & 1.59 & 0.62 & -0.63 \\
And XXVI & 1 & 0.763 & 3 & 10.8 & 5.01 & 0.00 & 0.33 & 1.13 & 0.80 & -0.80 \\
\hline
\end{tabular}
\caption{The first 23 rows are dSph galaxies and the lower 25 rows are UFD galaxies. $F$ is the soliton mass fraction, $M= M_S+M_\bullet$ is the total galactic mass, $\sigma=\sigma_{los}$ is the predicted (2D projected) stellar velocity dispersion, $M_{1/2} = M(<r_{1/2})$ is the predicted total mass contained within the stellar half-light radius, $r_{c,S}$ is the soliton characteristic radius and $\sigma_S/\sigma_\ast = \nu_S v_0/(\sqrt{3} \sigma_{los,obs})$ is obtained assuming that the stellar velocities are isotropic (as are the soliton velocities).   The relative errors $\Delta_\sigma$ and $\Delta_M$ quantify observational consistency (as further described in the introduction to this Appendix section)}
\label{T:3m0_1SE}
\end{table}

\begin{table}
\centering
\begin{tabular}{lccccccccccc}
\hline
\multicolumn{11}{c}{5G predictions with $m_0 = 20\times 10^{-22}$, optimising $F$ and $M$ so $\max[|\Delta_\sigma|$, $|\Delta_M|]\le1$ (if possible)} \\
\hline
Galaxy & $F$ & $M/10^8$ & $m_0/10^{-22}$ & $\sigma$ & $M_{1/2}/10^6$ & $M_\bullet/10^6$ & $r_{c,S}$ & $\sigma_S/\sigma_\ast$ & $\Delta_\sigma$ & $\Delta_M$ \\
  &  & M$_\odot$ & eV/c$^2$ & (km/s) & M$_\odot$ & M$_\odot$ &\,kpc &  &  &  \\
\hline
Tucana & 1 & 0.475 & 20 & 13.2 & 47.5 & 0.00 & 0.01 & 3.04 & -0.05 & 0.05 \\
Cetus & 1 & 0.600 & 20 & 12.3 & 60.0 & 0.00 & 0.01 & 4.60 & 0.77 & -0.77 \\
Aquarius & 1 & 0.332 & 20 & 10.4 & 33.2 & 0.00 & 0.02 & 2.74 & 0.06 & -0.05 \\
Draco & 1 & 0.153 & 20 & 9.70 & 15.4 & 0.00 & 0.04 & 1.45 & 0.64 & -0.64 \\
Leo I & 1 & 0.195 & 20 & 9.48 & 19.6 & 0.00 & 0.03 & 1.80 & 0.23 & -0.24 \\
Phoenix & 1 & 0.220 & 20 & 9.66 & 22.0 & 0.00 & 0.03 & 2.01 & 0.51 & -0.52 \\
Canes Ventici & 1 & 0.200 & 20 & 8.56 & 20.0 & 0.00 & 0.03 & 2.24 & 2.40 & -2.41 \\
Sextans & 1 & 0.414 & 20 & 7.91 & 41.4 & 0.00 & 0.01 & 4.46 & 0.01 & -0.01 \\
Crater II & 1 & 0.081 & 20 & 2.50 & 8.07 & 0.00 & 0.07 & 2.55 & -0.67 & 0.66 \\
Leo II & 1 & 0.103 & 20 & 7.23 & 10.3 & 0.00 & 0.06 & 1.18 & -0.42 & 0.42 \\
Carina & 1 & 0.127 & 20 & 7.74 & 12.7 & 0.00 & 0.05 & 1.64 & 0.95 & -0.95 \\
Ursa Minor & 1 & 0.421 & 20 & 13.0 & 42.1 & 0.00 & 0.01 & 3.12 & 1.66 & -1.66 \\
Sculptor & 1 & 0.247 & 20 & 10.6 & 24.7 & 0.00 & 0.02 & 2.08 & 1.80 & -1.79 \\
And I & 1 & 0.498 & 20 & 9.96 & 49.8 & 0.00 & 0.01 & 4.51 & 0.33 & -0.33 \\
And III & 1 & 0.411 & 20 & 11.9 & 41.1 & 0.00 & 0.01 & 3.18 & 0.48 & -0.48 \\
And V & 1 & 0.438 & 20 & 10.8 & 43.8 & 0.00 & 0.01 & 3.33 & -0.41 & 0.41 \\
And IX & 1 & 0.417 & 20 & 10.6 & 41.7 & 0.00 & 0.01 & 3.26 & -0.16 & 0.15 \\
And XIV & 1 & 0.125 & 20 & 5.01 & 12.5 & 0.00 & 0.05 & 1.97 & -0.30 & 0.30 \\
And XV & 1 & 0.288 & 20 & 10.4 & 28.8 & 0.00 & 0.02 & 2.23 & -0.13 & 0.13 \\
And XVIII & 1 & 0.326 & 20 & 8.82 & 32.7 & 0.00 & 0.02 & 2.86 & -0.38 & 0.38 \\
And XXI & 1 & 0.330 & 20 & 6.29 & 33.0 & 0.00 & 0.02 & 4.61 & 0.19 & -0.19 \\
And XXIII & 1 & 0.518 & 20 & 7.43 & 51.9 & 0.00 & 0.01 & 6.21 & 0.33 & -0.33 \\
And XXV & 1 & 0.052 & 20 & 2.78 & 5.16 & 0.00 & 0.11 & 1.48 & -0.20 & 0.20 \\
\hline
Phoenix II & 1 & 0.143 & 20 & 10.1 & 4.94 & 0.00 & 0.04 & 1.11 & -0.18 & 0.18 \\
Segue I & 0.9472 & 0.045 & 20 & 4.70 & 0.315 & 0.24 & 0.11 & 1.25 & 1.00 & -0.99 \\
Pegasus III & 1 & 0.074 & 20 & 5.46 & 1.42 & 0.00 & 0.08 & 1.17 & 0.02 & -0.02 \\
Wilman I & 0.9525 & 0.046 & 20 & 4.80 & 0.310 & 0.22 & 0.11 & 1.20 & 1.00 & -0.99 \\
Horoligium I & 1 & 0.078 & 20 & 4.97 & 0.908 & 0.00 & 0.07 & 1.35 & 0.02 & -0.02 \\
Pisces II & 1 & 0.074 & 20 & 4.85 & 2.05 & 0.00 & 0.08 & 1.17 & -0.23 & 0.23 \\
Coma Berenices & 1 & 0.061 & 20 & 4.56 & 1.38 & 0.00 & 0.09 & 1.13 & -0.05 & 0.05 \\
Reticulum II & 1 & 0.053 & 20 & 4.17 & 0.446 & 0.00 & 0.11 & 1.40 & 0.58 & -0.58 \\
Hydrus I & 0.9419 & 0.032 & 20 & 3.20 & 0.288 & 0.19 & 0.15 & 1.23 & 1.00 & -0.99 \\
Grus I & 1 & 0.067 & 20 & 5.38 & 1.91 & 0.00 & 0.09 & 1.06 & -0.01 & 0.01 \\
Leo IV & 1 & 0.043 & 20 & 3.29 & 1.31 & 0.00 & 0.13 & 1.08 & -0.12 & 0.12 \\
Canes Ventici II & 1 & 0.063 & 20 & 4.19 & 1.57 & 0.00 & 0.09 & 1.17 & -0.41 & 0.41 \\
Bootes I & 1 & 0.030 & 20 & 2.00 & 1.68 & 0.00 & 0.19 & 1.06 & -0.80 & 0.80 \\
Tucana II & 1 & 0.035 & 20 & 3 & 0.813 & 0.00 & 0.16 & 1.06 & 0.17 & -0.16 \\
Tucana IV & 1 & 0.051 & 20 & 4.07 & 2.14 & 0.00 & 0.11 & 1.01 & -0.23 & 0.23 \\
Leo V & 1 & 0.056 & 20 & 4.27 & 0.589 & 0.00 & 0.10 & 1.29 & 0.25 & -0.25 \\
And X & 1 & 0.044 & 20 & 3.47 & 3.48 & 0.00 & 0.13 & 0.96 & -0.36 & 0.36 \\
And XI & 1 & 0.054 & 20 & 3.78 & 2.53 & 0.00 & 0.11 & 1.00 & -0.18 & 0.18 \\
And XII & 1 & 0.031 & 20 & 2.32 & 2.62 & 0.00 & 0.19 & 1.01 & -0.11 & 0.11 \\
And XIII & 1 & 0.066 & 20 & 5.22 & 4.99 & 0.00 & 0.09 & 0.97 & -0.29 & 0.29 \\
And XVI & 1 & 0.044 & 20 & 3.66 & 1.84 & 0.00 & 0.13 & 0.99 & -0.05 & 0.05 \\
And XVII & 1 & 0.032 & 20 & 2.64 & 2.60 & 0.00 & 0.18 & 0.94 & -0.14 & 0.13 \\
And XX & 1 & 0.085 & 20 & 6.18 & 5.61 & 0.00 & 0.07 & 1.02 & -0.37 & 0.37 \\
And XXII & 1 & 0.030 & 20 & 2.77 & 1.73 & 0.00 & 0.19 & 0.91 & -0.02 & 0.02 \\
And XXVI & 1 & 0.119 & 20 & 8.28 & 11.9 & 0.00 & 0.05 & 1.18 & -0.15 & 0.14 \\
\hline
\end{tabular}
\caption{The first 23 rows are dSph galaxies and the lower 25 rows are UFD galaxies. $F$ is the soliton mass fraction, $M= M_S+M_\bullet$ is the total galactic mass, $\sigma=\sigma_{los}$ is the predicted (2D projected) stellar velocity dispersion, $M_{1/2} = M(<r_{1/2})$ is the predicted total mass contained within the stellar half-light radius, $r_{c,S}$ is the soliton characteristic radius and $\sigma_S/\sigma_\ast = \nu_S v_0/(\sqrt{3} \sigma_{los,obs})$ is obtained assuming that the stellar velocities are isotropic (as are the soliton velocities).    The relative errors $\Delta_\sigma$ and $\Delta_M$ quantify observational consistency (as further described in the introduction to this Appendix section).}
\label{T:20m0_1SE}
\end{table}

\begin{table}
\centering
\begin{tabular}{lccccccccccc}
\hline
\multicolumn{11}{c}{5G predictions with $m_0= 1.5\times 10^{-22}$, optimising $F$ and $M$ so $\max[|\Delta_\sigma|$, $|\Delta_M|]\le2$ (if possible)} \\
\hline
Galaxy & $F$ & $M/10^8$ & $m_0/10^{-22}$ & $\sigma$ & $M_{1/2}/10^6$ & $M_\bullet/10^6$ & $r_{c,S}$ & $\sigma_S/\sigma_\ast$ & $\Delta_\sigma$ & $\Delta_M$ \\
  &  & M$_\odot$ & eV/c$^2$ & (km/s) & M$_\odot$ & M$_\odot$ &\,kpc &  &  &  \\
\hline
Tucana & 1 & 2.419 & 1.5 & 13.7 & 44.8 & 0.00 & 0.42 & 1.16 & 0.14 & -0.14 \\
Cetus & 1 & 1.724 & 1.5 & 10.5 & 75.4 & 0.00 & 0.59 & 0.99 & -0.48 & 0.48 \\
Aquarius & 1 & 1.917 & 1.5 & 10.9 & 31.3 & 0.00 & 0.53 & 1.19 & 0.37 & -0.37 \\
Draco & 1 & 2.113 & 1.5 & 9.37 & 16.3 & 0.00 & 0.48 & 1.50 & 0.34 & -0.34 \\
Leo I & 1 & 2.036 & 1.5 & 9.46 & 19.6 & 0.00 & 0.50 & 1.41 & 0.22 & -0.21 \\
Phoenix & 1 & 1.937 & 1.5 & 9.71 & 21.8 & 0.00 & 0.53 & 1.33 & 0.59 & -0.58 \\
Canes Ventici & 1 & 1.458 & 1.5 & 7.31 & 26.8 & 0.00 & 0.70 & 1.22 & -0.73 & 0.73 \\
Sextans & 1 & 1.288 & 1.5 & 7.39 & 45.3 & 0.00 & 0.79 & 1.04 & -0.39 & 0.39 \\
Crater II & 1 & 0.510 & 1.5 & 3.16 & 5.28 & 0.00 & 2.00 & 1.20 & 1.53 & -1.53 \\
Leo II & 0.9662 & 1.407 & 1.5 & 8.20 & 7.22 & 4.76 & 0.66 & 1.47 & 2.00 & -1.99 \\
Carina & 1 & 1.441 & 1.5 & 5.92 & 20.0 & 0.00 & 0.71 & 1.40 & -0.57 & 0.57 \\
Ursa Minor & 1 & 1.973 & 1.5 & 10.5 & 69.6 & 0.00 & 0.52 & 1.10 & -1.25 & 1.25 \\
Sculptor & 1 & 2.091 & 1.5 & 9.92 & 28.3 & 0.00 & 0.49 & 1.32 & -0.60 & 0.61 \\
And I & 1 & 1.484 & 1.5 & 8.95 & 58.8 & 0.00 & 0.69 & 1.01 & -0.30 & 0.30 \\
And III & 1 & 1.933 & 1.5 & 10.5 & 49.7 & 0.00 & 0.53 & 1.12 & -0.29 & 0.29 \\
And V & 1 & 2.069 & 1.5 & 10.8 & 43.4 & 0.00 & 0.49 & 1.18 & -0.35 & 0.35 \\
And IX & 1 & 1.993 & 1.5 & 10.7 & 40.9 & 0.00 & 0.51 & 1.17 & -0.09 & 0.09 \\
And XIV & 1 & 1.125 & 1.5 & 6.20 & 6.57 & 0.00 & 0.91 & 1.33 & 0.62 & -0.61 \\
And XV & 1 & 2.348 & 1.5 & 11.8 & 23.9 & 0.00 & 0.43 & 1.36 & 0.12 & -0.12 \\
And XVIII & 1 & 1.933 & 1.5 & 9.46 & 29.9 & 0.00 & 0.53 & 1.27 & -0.10 & 0.11 \\
And XXI & 1 & 0.966 & 1.5 & 5.84 & 37.7 & 0.00 & 1.05 & 1.01 & -0.29 & 0.29 \\
And XXIII & 1 & 1.060 & 1.5 & 6.45 & 63.6 & 0.00 & 0.96 & 0.95 & -0.65 & 0.65 \\
And XXV & 1 & 0.668 & 1.5 & 4.08 & 2.39 & 0.00 & 1.52 & 1.42 & 0.90 & -0.90 \\
\hline
Phoenix II & 1 & 3.628 & 1.5 & 22.2 & 0.636 & 0.00 & 0.28 & 2.11 & 1.19 & -1.18 \\
Segue I & 0.9983 & 1.010 & 1.5 & 5.46 & 0.178 & 0.17 & 1.00 & 1.66 & 1.95 & -1.99 \\
Pegasus III & 1 & 1.846 & 1.5 & 9.32 & 0.137 & 0.00 & 0.55 & 2.18 & 1.31 & -1.28 \\
Wilman I & 0.999 & 1.164 & 1.5 & 5.54 & 0.127 & 0.12 & 0.87 & 1.86 & 1.93 & -1.99 \\
Horoligium I & 0.9992 & 2.123 & 1.5 & 10.5 & 0.271 & 0.17 & 0.48 & 2.77 & 1.99 & -1.98 \\
Pisces II & 1 & 1.803 & 1.5 & 10.4 & 0.199 & 0.00 & 0.56 & 2.13 & 1.39 & -1.36 \\
Coma Berenices & 0.9937 & 1.005 & 1.5 & 6.18 & 0.665 & 0.63 & 1.00 & 1.40 & 1.98 & -1.99 \\
Reticulum II & 0.9984 & 1.471 & 1.5 & 6.47 & 0.289 & 0.24 & 0.69 & 2.92 & 1.98 & -1.99 \\
Hydrus I & 0.9976 & 0.791 & 1.5 & 3.70 & 0.199 & 0.19 & 1.28 & 1.88 & 1.98 & -1.99 \\
Grus I & 1 & 1.473 & 1.5 & 9.23 & 0.128 & 0.00 & 0.69 & 1.74 & 1.28 & -1.25 \\
Leo IV & 1 & 1.488 & 1.5 & 5.21 & 0.567 & 0.00 & 0.68 & 2.80 & 1.39 & -1.40 \\
Canes Ventici II & 0.9979 & 1.57 & 1.5 & 6.59 & 0.457 & 0.33 & 0.67 & 2.18 & 1.99 & -1.97 \\
Bootes I & 1 & 1.011 & 1.5 & 3.22 & 0.856 & 0.00 & 1.01 & 2.69 & 0.91 & -0.92 \\
Tucana II & 1 & 1.108 & 1.5 & 4.94 & 0.205 & 0.00 & 0.92 & 2.53 & 1.78 & -1.77 \\
Tucana IV & 0.9974 & 1.281 & 1.5 & 7.68 & 0.625 & 0.33 & 0.79 & 1.90 & 1.99 & -1.98 \\
Leo V & 1 & 1.293 & 1.5 & 7.14 & 0.0369 & 0.00 & 0.79 & 2.23 & 1.5 & -1.46 \\
And X & 1 & 1.236 & 1.5 & 4.88 & 1.64 & 0.00 & 0.82 & 2.02 & 0.82 & -0.82 \\
And XI & 1 & 1.200 & 1.5 & 4.60 & 0.281 & 0.00 & 0.85 & 1.67 & 0 & -0.88 \\
And XII & 1 & 0.901 & 1.5 & 3.29 & 1.62 & 0.00 & 1.13 & 2.21 & 0.14 & -0.19 \\
And XIII & 1 & 1.320 & 1.5 & 8.44 & 0.520 & 0.00 & 0.77 & 1.45 & 1.32 & -1.30 \\
And XVI & 1 & 1.644 & 1.5 & 4.57 & 1.24 & 0.00 & 0.62 & 2.77 & 0.27 & -0.29 \\
And XVII & 1 & 1.083 & 1.5 & 2.73 & 2.49 & 0.00 & 0.94 & 2.39 & -0.09 & 0.09 \\
And XX & 1 & 2.419 & 1.5 & 11.0 & 1.92 & 0.00 & 0.42 & 2.18 & 0.99 & -1.00 \\
And XXII & 1 & 0.882 & 1.5 & 5.22 & 0.569 & 0.00 & 1.15 & 2.01 & 0.83 & -0.84 \\
And XXVI & 1 & 1.786 & 1.5 & 11.9 & 2.60 & 0.00 & 0.57 & 1.33 & 1.17 & -1.16 \\
\hline
\end{tabular}
\caption{The first 23 rows are dSph galaxies and the lower 25 rows are UFD galaxies. $F$ is the soliton mass fraction, $M= M_S+M_\bullet$ is the total galactic mass, $\sigma=\sigma_{los}$ is the predicted (2D projected) stellar velocity dispersion, $M_{1/2} = M(<r_{1/2})$ is the predicted total mass contained within the stellar half-light radius, $r_{c,S}$ is the soliton characteristic radius and $\sigma_S/\sigma_\ast = \nu_S v_0/(\sqrt{3} \sigma_{los,obs})$ is obtained assuming that the stellar velocities are isotropic (as are the soliton velocities).    The relative errors $\Delta_\sigma$ and $\Delta_M$ quantify observational consistency (as further described in the introduction to this Appendix section). These results are obtained upon relaxing the observational uncertainties of $\sigma$ and $M_{1/2}$ by a factor of 2. }
\label{T:1x5m0_2SE}
\end{table}

\begin{table}
\centering
\begin{tabular}{lccccccccccc}
\hline
\multicolumn{11}{c}{5G predictions with $m_0= 1.5\times 10^{-22}$, optimising $F$ and $M$ so $\max[|\Delta_\sigma|$, $|\Delta_M|]\le3$ (if possible)} \\
\hline
Galaxy & $F$ & $M/10^8$ & $m_0/10^{-22}$ & $\sigma$ & $M_{1/2}/10^6$ & $M_\bullet/10^6$ & $r_{c,S}$ & $\sigma_S/\sigma_\ast$ & $\Delta_\sigma$ & $\Delta_M$ \\
  &  & M$_\odot$ & eV/c$^2$ & (km/s) & M$_\odot$ & M$_\odot$ &\,kpc &  &  &  \\
\hline
Tucana & 1 & 2.419 & 1.5 & 13.7 & 44.8 & 0.00 & 0.42 & 1.16 & 0.14 & -0.14 \\
Cetus & 1 & 1.724 & 1.5 & 10.5 & 75.4 & 0.00 & 0.59 & 0.99 & -0.48 & 0.48 \\
Aquarius & 1 & 1.917 & 1.5 & 10.9 & 31.3 & 0.00 & 0.53 & 1.19 & 0.37 & -0.37 \\
Draco & 1 & 2.113 & 1.5 & 9.37 & 16.3 & 0.00 & 0.48 & 1.50 & 0.34 & -0.34 \\
Leo I & 1 & 2.036 & 1.5 & 9.46 & 19.6 & 0.00 & 0.50 & 1.41 & 0.22 & -0.21 \\
Phoenix & 1 & 1.937 & 1.5 & 9.71 & 21.8 & 0.00 & 0.53 & 1.33 & 0.59 & -0.58 \\
Canes Ventici & 1 & 1.458 & 1.5 & 7.31 & 26.8 & 0.00 & 0.70 & 1.22 & -0.73 & 0.73 \\
Sextans & 1 & 1.288 & 1.5 & 7.39 & 45.3 & 0.00 & 0.79 & 1.04 & -0.39 & 0.39 \\
Crater II & 1 & 0.510 & 1.5 & 3.16 & 5.28 & 0.00 & 2.00 & 1.20 & 1.53 & -1.53 \\
Leo II & 1 & 1.880 & 1.5 & 8.48 & 6.31 & 0.00 & 0.54 & 1.62 & 2.70 & -2.71 \\
Carina & 1 & 1.441 & 1.5 & 5.92 & 20.0 & 0.00 & 0.71 & 1.40 & -0.57 & 0.57 \\
Ursa Minor & 1 & 1.973 & 1.5 & 10.5 & 69.6 & 0.00 & 0.52 & 1.10 & -1.25 & 1.25 \\
Sculptor & 1 & 2.091 & 1.5 & 9.92 & 28.3 & 0.00 & 0.49 & 1.32 & -0.60 & 0.61 \\
And I & 1 & 1.484 & 1.5 & 8.95 & 58.8 & 0.00 & 0.69 & 1.01 & -0.30 & 0.30 \\
And III & 1 & 1.933 & 1.5 & 10.5 & 49.7 & 0.00 & 0.53 & 1.12 & -0.29 & 0.29 \\
And V & 1 & 2.069 & 1.5 & 10.8 & 43.4 & 0.00 & 0.49 & 1.18 & -0.35 & 0.35 \\
And IX & 1 & 1.993 & 1.5 & 10.7 & 40.9 & 0.00 & 0.51 & 1.17 & -0.09 & 0.09 \\
And XIV & 1 & 1.125 & 1.5 & 6.20 & 6.57 & 0.00 & 0.91 & 1.33 & 0.62 & -0.61 \\
And XV & 1 & 2.348 & 1.5 & 11.8 & 23.9 & 0.00 & 0.43 & 1.36 & 0.12 & -0.12 \\
And XVIII & 1 & 1.933 & 1.5 & 9.46 & 29.9 & 0.00 & 0.53 & 1.27 & -0.10 & 0.11 \\
And XXI & 1 & 0.966 & 1.5 & 5.84 & 37.7 & 0.00 & 1.05 & 1.01 & -0.29 & 0.29 \\
And XXIII & 1 & 1.060 & 1.5 & 6.45 & 63.6 & 0.00 & 0.96 & 0.95 & -0.65 & 0.65 \\
And XXV & 1 & 0.668 & 1.5 & 4.08 & 2.39 & 0.00 & 1.52 & 1.42 & 0.90 & -0.90 \\
\hline
Phoenix II & 1 & 3.628 & 1.5 & 22.2 & 0.636 & 0.00 & 0.28 & 2.11 & 1.19 & -1.18 \\
Segue I & 0.9997 & 1.162 & 1.5 & 6.24 & 0.0411 & 0.03 & 0.88 & 1.91 & 2.92 & -2.98 \\
Pegasus III & 1 & 1.846 & 1.5 & 9.32 & 0.137 & 0.00 & 0.55 & 2.18 & 1.31 & -1.28 \\
Wilman I & 1 & 1.267 & 1.5 & 6.05 & 0.00737 & 0.00 & 0.80 & 2.03 & 2.56 & -2.64 \\
Horoligium I & 1 & 2.328 & 1.5 & 11.4 & 0.160 & 0.00 & 0.44 & 3.03 & 2.33 & -2.31 \\
Pisces II & 1 & 1.803 & 1.5 & 10.4 & 0.199 & 0.00 & 0.56 & 2.13 & 1.39 & -1.36 \\
Coma Berenices & 0.9976 & 1.162 & 1.5 & 6.99 & 0.323 & 0.28 & 0.87 & 1.62 & 2.99 & -2.97 \\
Reticulum II & 0.9998 & 1.830 & 1.5 & 8.04 & 0.178 & 0.04 & 0.56 & 3.63 & 2.94 & -2.99 \\
Hydrus I & 0.9989 & 0.915 & 1.5 & 4.20 & 0.111 & 0.10 & 1.11 & 2.18 & 2.96 & -2.97 \\
Grus I & 1 & 1.473 & 1.5 & 9.23 & 0.128 & 0.00 & 0.69 & 1.74 & 1.28 & -1.25 \\
Leo IV & 1 & 1.488 & 1.5 & 5.21 & 0.567 & 0.00 & 0.68 & 2.80 & 1.39 & -1.40 \\
Canes Ventici II & 1 & 1.723 & 1.5 & 7.07 & 0.238 & 0.00 & 0.59 & 2.40 & 2.47 & -2.44 \\
Bootes I & 1 & 1.011 & 1.5 & 3.22 & 0.856 & 0.00 & 1.01 & 2.69 & 0.91 & -0.92 \\
Tucana II & 1 & 1.108 & 1.5 & 4.94 & 0.205 & 0.00 & 0.92 & 2.53 & 1.78 & -1.77 \\
Tucana IV & 1 & 1.413 & 1.5 & 8.24 & 0.416 & 0.00 & 0.72 & 2.10 & 2.32 & -2.30 \\
Leo V & 1 & 1.293 & 1.5 & 7.14 & 0.0369 & 0.00 & 0.79 & 2.23 & 1.5 & -1.46 \\
And X & 1 & 1.236 & 1.5 & 4.88 & 1.64 & 0.00 & 0.82 & 2.02 & 0.82 & -0.82 \\
And XI & 1 & 1.200 & 1.5 & 4.60 & 0.281 & 0.00 & 0.85 & 1.67 & 0 & -0.88 \\
And XII & 1 & 0.914 & 1.5 & 3.35 & 1.71 & 0.00 & 1.11 & 2.24 & 0.15 & -0.15 \\
And XIII & 1 & 1.320 & 1.5 & 8.44 & 0.520 & 0.00 & 0.77 & 1.45 & 1.32 & -1.30 \\
And XVI & 1 & 1.652 & 1.5 & 4.61 & 1.26 & 0.00 & 0.62 & 2.78 & 0.28 & -0.28 \\
And XVII & 1 & 1.083 & 1.5 & 2.73 & 2.49 & 0.00 & 0.94 & 2.39 & -0.09 & 0.09 \\
And XX & 1 & 2.419 & 1.5 & 11.0 & 1.92 & 0.00 & 0.42 & 2.18 & 0.99 & -1.00 \\
And XXII & 1 & 0.882 & 1.5 & 5.22 & 0.569 & 0.00 & 1.15 & 2.01 & 0.83 & -0.84 \\
And XXVI & 1 & 1.786 & 1.5 & 11.9 & 2.60 & 0.00 & 0.57 & 1.33 & 1.17 & -1.16 \\
\hline
\end{tabular}
\caption{The first 23 rows are dSph galaxies and the lower 25 rows are UFD galaxies. $F$ is the soliton mass fraction, $M= M_S+M_\bullet$ is the total galactic mass, $\sigma=\sigma_{los}$ is the predicted (2D projected) stellar velocity dispersion, $M_{1/2} = M(<r_{1/2})$ is the predicted total mass contained within the stellar half-light radius, $r_{c,S}$ is the soliton characteristic radius and $\sigma_S/\sigma_\ast = \nu_S v_0/(\sqrt{3} \sigma_{los,obs})$ is obtained assuming that the stellar velocities are isotropic (as are the soliton velocities).   The relative errors $\Delta_\sigma$ and $\Delta_M$ quantify observational consistency (as further described in the introduction to this Appendix section).  These results are obtained upon relaxing the observational uncertainties of $\sigma$ and $M_{1/2}$ by a factor of 3. }
\label{T:1x5m0_3SE}
\end{table}

\begin{table}
\centering
\begin{tabular}{lccccccccccc}
\hline
\multicolumn{11}{c}{5G predictions with $m_0 = 1.5\times 10^{-22}$ and $\sigma_S =\sigma_\ast$, optimising $F$ to minimize $|\Delta_M|$} \\
\hline
Galaxy & $F$ & $M/10^8$ & $m_0/10^{-22}$ & $\sigma$ & $M_{1/2}/10^6$ & $M_\bullet/10^6$ & $r_{c,S}$ & $\sigma_S/\sigma_\ast$ & $\Delta_\sigma$ & $\Delta_M$ \\
  &  & M$_\odot$ & eV/c$^2$ & (km/s) & M$_\odot$ & M$_\odot$ &\,kpc &  &  &  \\
\hline
Tucana & 0.7969 & 1.487 & 1.5 & 13.76 & 46.72 & 30.21 & 0.39 & 1.00 & 0.17 & 0.00 \\
Cetus & 0.7766 & 1.207 & 1.5 & 11.39 & 68.73 & 26.95 & 0.46 & 1.00 & 0.18 & 0.00 \\
Aquarius & 0.8060 & 1.167 & 1.5 & 10.90 & 33.55 & 22.65 & 0.51 & 1.00 & 0.37 & 0.00 \\
Draco & 0.8704 & 1.123 & 1.5 & 10.79 & 17.32 & 14.56 & 0.63 & 1.00 & 1.63 & 0.00 \\
Leo I & 0.8534 & 1.118 & 1.5 & 10.25 & 20.46 & 16.40 & 0.60 & 1.00 & 0.88 & 0.00 \\
Phoenix & 0.8387 & 1.106 & 1.5 & 10.14 & 23.32 & 17.84 & 0.58 & 1.00 & 1.20 & 0.00 \\
Canes Ventici & 0.8037 & 0.858 & 1.5 & 9.02 & 25.24 & 16.85 & 0.69 & 1.00 & 3.55 & 0.00 \\
Sextans & 0.6837 & 0.760 & 1.5 & 7.62 & 41.41 & 24.04 & 0.58 & 1.00 & -0.22 & -0.01 \\
Crater II & 0.8306 & 0.317 & 1.5 & 2.94 & 7.23 & 5.37 & 1.99 & 1.00 & 0.80 & 0.00 \\
Leo II & 0.9095 & 0.984 & 1.5 & 8.49 & 9.72 & 8.91 & 0.80 & 1.00 & 2.73 & 0.00 \\
Carina & 0.8340 & 0.779 & 1.5 & 8.72 & 17.18 & 12.93 & 0.82 & 1.00 & 1.77 & 0.00 \\
Ursa Minor & 0.7510 & 1.207 & 1.5 & 12.91 & 57.49 & 30.06 & 0.43 & 1.00 & 1.57 & 0.00 \\
Sculptor & 0.8288 & 1.183 & 1.5 & 11.28 & 27.41 & 20.25 & 0.53 & 1.00 & 3.93 & 0.00 \\
And I & 0.7819 & 1.029 & 1.5 & 9.41 & 54.22 & 22.45 & 0.54 & 1.00 & 0.01 & 0.00 \\
And III & 0.7769 & 1.196 & 1.5 & 11.75 & 46.13 & 26.69 & 0.46 & 1.00 & 0.39 & 0.00 \\
And V & 0.7926 & 1.245 & 1.5 & 10.90 & 40.83 & 25.82 & 0.46 & 1.00 & -0.30 & 0.00 \\
And IX & 0.7925 & 1.211 & 1.5 & 10.90 & 39.78 & 25.14 & 0.47 & 1.00 & 0.00 & 0.00 \\
And XIV & 0.8684 & 0.672 & 1.5 & 5.82 & 10.57 & 8.84 & 1.04 & 1.00 & 0.32 & -0.00 \\
And XV & 0.8463 & 1.323 & 1.5 & 11.34 & 25.91 & 20.34 & 0.50 & 1.00 & 0.05 & 0.00 \\
And XVIII & 0.8173 & 1.117 & 1.5 & 9.21 & 28.87 & 20.41 & 0.55 & 1.00 & -0.21 & 0.00 \\
And XXI & 0.7843 & 0.670 & 1.5 & 6.00 & 34.77 & 14.45 & 0.84 & 1.00 & -0.11 & -0.00 \\
And XXIII & 0.7920 & 0.789 & 1.5 & 6.58 & 55.77 & 16.40 & 0.72 & 1.00 & -0.52 & -0.00 \\
And XXV & 0.8947 & 0.389 & 1.5 & 3.55 & 4.61 & 4.10 & 1.94 & 1.00 & 0.46 & 0.00 \\
\hline
Phoenix II & 0.9750 & 1.641 & 1.5 & 13.63 & 4.13 & 4.10 & 0.58 & 1.00 & 0.28 & 0.02 \\
Segue I & 0.9925 & 0.601 & 1.5 & 3.97 & 0.45 & 0.45 & 1.66 & 1.00 & 0.09 & 0.00 \\
Pegasus III & 0.9824 & 0.816 & 1.5 & 7.10 & 1.44 & 1.44 & 1.19 & 1.00 & 0.57 & 0.01 \\
Wilman I & 0.9920 & 0.616 & 1.5 & 3.52 & 0.49 & 0.49 & 1.62 & 1.00 & -0.60 & -0.00 \\
Horoligium I & 0.9872 & 0.748 & 1.5 & 5.85 & 0.96 & 0.96 & 1.32 & 1.00 & 0.34 & 0.06 \\
Pisces II & 0.9791 & 0.812 & 1.5 & 6.97 & 1.70 & 1.69 & 1.18 & 1.00 & 0.44 & 0.01 \\
Coma Berenices & 0.9805 & 0.693 & 1.5 & 5.09 & 1.36 & 1.35 & 1.39 & 1.00 & 0.61 & 0.00 \\
Reticulum II & 0.9887 & 0.493 & 1.5 & 4.11 & 0.56 & 0.56 & 2.00 & 1.00 & 0.54 & 0.13 \\
Hydrus I & 0.9908 & 0.413 & 1.5 & 2.77 & 0.38 & 0.38 & 2.40 & 1.00 & 0.16 & 0.03 \\
Grus I & 0.9766 & 0.808 & 1.5 & 6.23 & 1.91 & 1.89 & 1.18 & 1.00 & 0.28 & 0.00 \\
Leo IV & 0.9758 & 0.508 & 1.5 & 4.18 & 1.24 & 1.23 & 1.88 & 1.00 & 0.60 & 0.02 \\
Canes Ventici II & 0.9802 & 0.693 & 1.5 & 6.20 & 1.38 & 1.37 & 1.39 & 1.00 & 1.60 & 0.00 \\
Bootes I & 0.9665 & 0.352 & 1.5 & 4.44 & 1.20 & 1.18 & 2.63 & 1.00 & 2.27 & 0.03 \\
Tucana II & 0.9790 & 0.421 & 1.5 & 3.16 & 0.89 & 0.88 & 2.28 & 1.00 & 0.30 & 0.02 \\
Tucana IV & 0.9701 & 0.636 & 1.5 & 6.48 & 1.92 & 1.90 & 1.48 & 1.00 & 1.28 & 0.03 \\
Leo V & 0.9874 & 0.565 & 1.5 & 4.69 & 0.71 & 0.71 & 1.74 & 1.00 & 0.43 & 0.02 \\
And X & 0.9482 & 0.554 & 1.5 & 5.84 & 2.96 & 2.87 & 1.59 & 1.00 & 1.62 & -0.00 \\
And XI & 0.9664 & 0.675 & 1.5 & 6.61 & 2.30 & 2.27 & 1.37 & 1.00 & Inf  & -0.02 \\
And XII & 0.9464 & 0.368 & 1.5 & 4.64 & 2.04 & 1.97 & 2.38 & 1.00 & 0.40 & 0.00 \\
And XIII & 0.9522 & 0.830 & 1.5 & 8.34 & 4.08 & 3.97 & 1.07 & 1.00 & 1.27 & 0.00 \\
And XVI & 0.9692 & 0.561 & 1.5 & 4.25 & 1.75 & 1.73 & 1.67 & 1.00 & 0.16 & 0.00 \\
And XVII & 0.9465 & 0.411 & 1.5 & 4.26 & 2.27 & 2.20 & 2.14 & 1.00 & 0.62 & 0.00 \\
And XX & 0.9594 & 1.029 & 1.5 & 10.51 & 4.26 & 4.18 & 0.88 & 1.00 & 0.87 & 0.01 \\
And XXII & 0.9589 & 0.405 & 1.5 & 4.25 & 1.70 & 1.67 & 2.24 & 1.00 & 0.50 & 0.01 \\
And XXVI & 0.9169 & 1.158 & 1.5 & 16.08 & 10.36 & 9.62 & 0.70 & 1.00 & 2.67 & 0.00 \\
\hline
\end{tabular}
\caption{The first 23 rows are dSph galaxies and the lower 25 rows are UFD galaxies. $F$ is the soliton mass fraction, $M= M_S+M_\bullet$ is the total galactic mass, $\sigma=\sigma_{los}$ is the predicted (2D projected) stellar velocity dispersion, $M_{1/2} = M(<r_{1/2})$ is the predicted total mass contained within the stellar half-light radius, $r_{c,S}$ is the soliton characteristic radius and $\sigma_S/\sigma_\ast = \nu_S v_0/(\sqrt{3} \sigma_{los,obs})$ is obtained assuming that the stellar velocities are isotropic (as are the soliton velocities).   The relative errors $\Delta_\sigma$ and $\Delta_M$ quantify observational consistency (as further described in the introduction to this Appendix section).  Note that those galaxies for which  $|\Delta_\sigma|\le 1$ are observationally consistent with having a massive black hole, although for all of the dSph galaxies and some of the UFD galaxies the predicted black hole mass exceeds $10^6$\,M$_\odot$, which may be unrealistically large. Thus, these results could be viewed as providing an upper bound to the observationally consistent black hole mass in each galaxy.}
\label{T:model2}
\end{table}

\begin{table}
\centering
\begin{tabular}{lccccccccccc}
\hline
\multicolumn{11}{c}{5G predictions with $m_0 = 1.5\times 10^{-22}$ and $F=1$, optimising $M$ to minimize $|\Delta_M|$} \\
\hline
Galaxy & $F$ & $M/10^8$ & $m_0/10^{-22}$ & $\sigma$ & $M_{1/2}/10^6$ & $M_\bullet/10^6$ & $r_{c,S}$ & $\sigma_S/\sigma_\ast$ & $\Delta_\sigma$ & $\Delta_M$ \\
  &  & M$_\odot$ & eV/c$^2$ & (km/s) & M$_\odot$ & M$_\odot$ &\,kpc &  &  &  \\
\hline
Tucana & 1 & 2.449 & 1.5 & 13.86 & 46.72 & 0.00 & 0.42 & 1.18 & 0.21 & 0.00 \\
Cetus & 1 & 1.668 & 1.5 & 10.10 & 68.73 & 0.00 & 0.61 & 0.96 & -0.77 & -0.00 \\
Aquarius & 1 & 1.956 & 1.5 & 11.20 & 33.55 & 0.00 & 0.52 & 1.21 & 0.56 & 0.00 \\
Draco & 1 & 2.149 & 1.5 & 9.61 & 17.32 & 0.00 & 0.47 & 1.52 & 0.55 & 0.00 \\
Leo I & 1 & 2.059 & 1.5 & 9.63 & 20.46 & 0.00 & 0.49 & 1.43 & 0.36 & 0.00 \\
Phoenix & 1 & 1.974 & 1.5 & 9.99 & 23.33 & 0.00 & 0.52 & 1.35 & 0.99 & 0.00 \\
Canes Ventici & 1 & 1.432 & 1.5 & 7.13 & 25.25 & 0.00 & 0.71 & 1.20 & -1.17 & 0.00 \\
Sextans & 1 & 1.251 & 1.5 & 7.18 & 41.50 & 0.00 & 0.81 & 1.01 & -0.55 & 0.00 \\
Crater II & 1 & 0.557 & 1.5 & 3.43 & 7.23 & 0.00 & 1.83 & 1.32 & 2.43 & 0.00 \\
Leo II & 1 & 2.108 & 1.5 & 9.77 & 9.73 & 0.00 & 0.48 & 1.82 & 5.92 & 0.00 \\
Carina & 1 & 1.379 & 1.5 & 5.58 & 17.17 & 0.00 & 0.74 & 1.33 & -0.85 & 0.00 \\
Ursa Minor & 1 & 1.852 & 1.5 & 9.62 & 57.49 & 0.00 & 0.55 & 1.03 & -2.35 & 0.00 \\
Sculptor & 1 & 2.072 & 1.5 & 9.80 & 27.41 & 0.00 & 0.49 & 1.31 & -1.00 & 0.00 \\
And I & 1 & 1.443 & 1.5 & 8.66 & 54.22 & 0.00 & 0.71 & 0.98 & -0.49 & 0.00 \\
And III & 1 & 1.889 & 1.5 & 10.20 & 46.12 & 0.00 & 0.54 & 1.10 & -0.50 & -0.00 \\
And V & 1 & 2.032 & 1.5 & 10.66 & 40.84 & 0.00 & 0.50 & 1.16 & -0.54 & 0.00 \\
And IX & 1 & 1.977 & 1.5 & 10.64 & 39.78 & 0.00 & 0.52 & 1.16 & -0.13 & 0.00 \\
And XIV & 1 & 1.280 & 1.5 & 7.02 & 10.58 & 0.00 & 0.80 & 1.51 & 1.25 & 0.00 \\
And XV & 1 & 2.402 & 1.5 & 12.11 & 25.88 & 0.00 & 0.42 & 1.39 & 0.16 & 0.00 \\
And XVIII & 1 & 1.913 & 1.5 & 9.36 & 28.88 & 0.00 & 0.53 & 1.26 & -0.15 & 0.00 \\
And XXI & 1 & 0.940 & 1.5 & 5.70 & 34.77 & 0.00 & 1.08 & 0.98 & -0.44 & -0.00 \\
And XXIII & 1 & 1.005 & 1.5 & 6.11 & 55.77 & 0.00 & 1.01 & 0.90 & -0.99 & -0.00 \\
And XXV & 1 & 0.795 & 1.5 & 4.84 & 4.60 & 0.00 & 1.28 & 1.69 & 1.53 & 0.00 \\
\hline
Phoenix II & 1 & 5.792 & 1.5 & 36.08 & 4.05 & 0.00 & 0.18 & 3.36 & 2.67 & 0.00 \\
Segue I & 1 & 3.638 & 1.5 & 25.37 & 0.45 & 0.00 & 0.28 & 5.92 & 26.84 & 0.00 \\
Pegasus III & 1 & 3.337 & 1.5 & 19.34 & 1.44 & 0.00 & 0.31 & 3.94 & 4.65 & 0.00 \\
Wilman I & 1 & 3.628 & 1.5 & 24.80 & 0.49 & 0.00 & 0.28 & 5.79 & 26.00 & 0.00 \\
Horoligium I & 1 & 3.608 & 1.5 & 18.42 & 0.92 & 0.00 & 0.28 & 4.69 & 4.83 & 0.00 \\
Pisces II & 1 & 3.088 & 1.5 & 18.42 & 1.68 & 0.00 & 0.33 & 3.65 & 3.62 & 0.00 \\
Coma Berenices & 1 & 2.701 & 1.5 & 18.15 & 1.36 & 0.00 & 0.38 & 3.75 & 16.94 & 0.00 \\
Reticulum II & 1 & 2.571 & 1.5 & 12.37 & 0.51 & 0.00 & 0.40 & 5.09 & 5.58 & 0.00 \\
Hydrus I & 1 & 2.285 & 1.5 & 13.40 & 0.38 & 0.00 & 0.45 & 5.42 & 21.00 & 0.00 \\
Grus I & 1 & 2.909 & 1.5 & 20.61 & 1.90 & 0.00 & 0.35 & 3.09 & 5.07 & 0.00 \\
Leo IV & 1 & 1.809 & 1.5 & 6.80 & 1.23 & 0.00 & 0.56 & 3.39 & 2.62 & 0.00 \\
Canes Ventici II & 1 & 2.682 & 1.5 & 12.13 & 1.38 & 0.00 & 0.38 & 3.72 & 7.53 & 0.00 \\
Bootes I & 1 & 1.097 & 1.5 & 3.61 & 1.18 & 0.00 & 0.93 & 2.92 & 1.34 & 0.00 \\
Tucana II & 1 & 1.599 & 1.5 & 7.96 & 0.87 & 0.00 & 0.64 & 3.65 & 4.30 & 0.00 \\
Tucana IV & 1 & 2.075 & 1.5 & 13.20 & 1.89 & 0.00 & 0.49 & 3.08 & 5.24 & 0.00 \\
Leo V & 1 & 2.707 & 1.5 & 17.57 & 0.70 & 0.00 & 0.38 & 4.39 & 6.03 & 0.00 \\
And X & 1 & 1.441 & 1.5 & 5.89 & 2.97 & 0.00 & 0.71 & 2.11 & 1.66 & 0.00 \\
And XI & 1 & 2.058 & 1.5 & 9.01 & 2.36 & 0.00 & 0.49 & 4.90 & Inf & 0.00 \\
And XII & 1 & 0.953 & 1.5 & 3.55 & 2.01 & 0.00 & 1.07 & 1.04 & 0.19 & 0.00 \\
And XIII & 1 & 2.230 & 1.5 & 15.66 & 4.07 & 0.00 & 0.46 & 3.66 & 4.93 & 0.00 \\
And XVI & 1 & 1.795 & 1.5 & 5.27 & 1.75 & 0.00 & 0.57 & 3.02 & 0.51 & 0.00 \\
And XVII & 1 & 1.058 & 1.5 & 2.63 & 2.27 & 0.00 & 0.96 & 2.33 & -0.14 & 0.00 \\
And XX & 1 & 2.958 & 1.5 & 13.70 & 4.22 & 0.00 & 0.34 & 2.66 & 1.69 & 0.00 \\
And XXII & 1 & 1.162 & 1.5 & 7.41 & 1.68 & 0.00 & 0.88 & 2.65 & 1.59 & 0.00 \\
And XXVI & 1 & 2.557 & 1.5 & 17.90 & 10.32 & 0.00 & 0.40 & 1.90 & 3.32 & 0.00 \\
\hline
\end{tabular}
\caption{The first 23 rows are dSph galaxies and the lower 25 rows are UFD galaxies. $F$ is the soliton mass fraction, $M= M_S+M_\bullet$ is the total galactic mass, $\sigma=\sigma_{los}$ is the predicted (2D projected) stellar velocity dispersion, $M_{1/2} = M(<r_{1/2})$ is the predicted total mass contained within the stellar half-light radius, $r_{c,S}$ is the soliton characteristic radius and $\sigma_S/\sigma_\ast = \nu_S v_0/(\sqrt{3} \sigma_{los,obs})$ is obtained assuming that the stellar velocities are isotropic (as are the soliton velocities).   The relative errors $\Delta_\sigma$ and $\Delta_M$ quantify observational consistency.}
\label{T:model3_F1}
\end{table}

\begin{table}
\centering
\begin{tabular}{lccccccccccc}
\hline
\multicolumn{11}{c}{5G predictions with $m_0 = 1.5\times 10^{-22}$ and $F=1$, optimising $M$ to minimize $|\Delta_\sigma|$} \\
\hline
Galaxy & $F$ & $M/10^8$ & $m_0/10^{-22}$ & $\sigma$ & $M_{1/2}/10^6$ & $M_\bullet/10^6$ & $r_{c,S}$ & $\sigma_S/\sigma_\ast$ & $\Delta_\sigma$ & $\Delta_M$ \\
  &  & M$_\odot$ & eV/c$^2$ & (km/s) & M$_\odot$ & M$_\odot$ &\,kpc &  &  &  \\
\hline
Tucana & 1 & 2.362 & 1.5 & 13.30 & 41.34 & 0.00 & 0.43 & 1.13 & 0.00 & -0.38 \\
Cetus & 1 & 1.816 & 1.5 & 11.10 & 87.04 & 0.00 & 0.56 & 1.04 & 0.00 & 1.30 \\
Aquarius & 1 & 1.839 & 1.5 & 10.30 & 27.11 & 0.00 & 0.55 & 1.14 & 0.00 & -1.06 \\
Draco & 1 & 2.055 & 1.5 & 9.00 & 14.69 & 0.00 & 0.50 & 1.46 & 0.00 & -0.85 \\
Leo I & 1 & 2.001 & 1.5 & 9.20 & 18.46 & 0.00 & 0.51 & 1.39 & 0.00 & -0.52 \\
Phoenix & 1 & 1.885 & 1.5 & 9.30 & 19.77 & 0.00 & 0.54 & 1.29 & 0.00 & -1.36 \\
Canes Ventici & 1 & 1.499 & 1.5 & 7.60 & 29.49 & 0.00 & 0.68 & 1.26 & 0.00 & 1.96 \\
Sextans & 1 & 1.377 & 1.5 & 7.90 & 55.14 & 0.00 & 0.74 & 1.11 & 0.00 & 1.41 \\
Crater II & 1 & 0.432 & 1.5 & 2.70 & 2.89 & 0.00 & 2.36 & 1.02 & 0.00 & -3.42 \\
Leo II & 1 & 1.679 & 1.5 & 7.40 & 4.09 & 0.00 & 0.61 & 1.45 & 0.00 & -4.47 \\
Carina & 1 & 1.562 & 1.5 & 6.60 & 26.49 & 0.00 & 0.65 & 1.51 & 0.00 & 1.85 \\
Ursa Minor & 1 & 2.108 & 1.5 & 11.50 & 84.44 & 0.00 & 0.48 & 1.17 & 0.00 & 2.77 \\
Sculptor & 1 & 2.119 & 1.5 & 10.10 & 29.68 & 0.00 & 0.48 & 1.34 & 0.00 & 1.52 \\
And I & 1 & 1.547 & 1.5 & 9.40 & 66.20 & 0.00 & 0.66 & 1.05 & 0.00 & 0.80 \\
And III & 1 & 1.994 & 1.5 & 11.00 & 54.78 & 0.00 & 0.51 & 1.16 & 0.00 & 0.71 \\
And V & 1 & 2.136 & 1.5 & 11.20 & 48.18 & 0.00 & 0.48 & 1.22 & 0.00 & 1.00 \\
And IX & 1 & 2.030 & 1.5 & 10.90 & 43.47 & 0.00 & 0.50 & 1.19 & 0.00 & 0.30 \\
And XIV & 1 & 0.975 & 1.5 & 5.40 & 3.83 & 0.00 & 1.04 & 1.15 & 0.00 & -1.04 \\
And XV & 1 & 2.188 & 1.5 & 11.00 & 18.50 & 0.00 & 0.47 & 1.27 & 0.00 & -0.44 \\
And XVIII & 1 & 1.982 & 1.5 & 9.70 & 32.64 & 0.00 & 0.51 & 1.30 & 0.00 & 0.38 \\
And XXI & 1 & 1.015 & 1.5 & 6.10 & 43.32 & 0.00 & 1.00 & 1.06 & 0.00 & 0.85 \\
And XXIII & 1 & 1.168 & 1.5 & 7.10 & 79.36 & 0.00 & 0.87 & 1.05 & 0.00 & 1.96 \\
And XXV & 1 & 0.501 & 1.5 & 3.00 & 0.79 & 0.00 & 2.03 & 1.07 & 0.00 & -1.55 \\
\hline
Phoenix II & 1 & 2.010 & 1.5 & 11.00 & 0.06 & 0.00 & 0.51 & 1.17 & 0.00 & -1.37 \\
Segue I & 1 & 0.840 & 1.5 & 3.90 & 0.00 & 0.00 & 1.21 & 1.37 & 0.00 & -3.27 \\
Pegasus III & 1 & 1.235 & 1.5 & 5.40 & 0.03 & 0.00 & 0.82 & 1.46 & 0.00 & -1.39 \\
Wilman I & 1 & 0.962 & 1.5 & 4.00 & 0.00 & 0.00 & 1.06 & 1.53 & 0.00 & -2.67 \\
Horoligium I & 1 & 1.192 & 1.5 & 4.90 & 0.01 & 0.00 & 0.85 & 1.55 & 0.00 & -2.77 \\
Pisces II & 1 & 1.064 & 1.5 & 5.40 & 0.02 & 0.00 & 0.96 & 1.26 & 0.00 & -1.52 \\
Coma Berenices & 1 & 0.875 & 1.5 & 4.60 & 0.02 & 0.00 & 1.16 & 1.21 & 0.00 & -3.86 \\
Reticulum II & 1 & 0.954 & 1.5 & 3.22 & 0.01 & 0.00 & 1.07 & 1.89 & 0.00 & -4.49 \\
Hydrus I & 1 & 0.701 & 1.5 & 2.69 & 0.00 & 0.00 & 1.45 & 1.66 & 0.00 & -4.18 \\
Grus I & 1 & 0.984 & 1.5 & 5.40 & 0.03 & 0.00 & 1.03 & 1.16 & 0.00 & -1.32 \\
Leo IV & 1 & 1.099 & 1.5 & 3.40 & 0.17 & 0.00 & 0.93 & 2.06 & 0.00 & -2.24 \\
Canes Ventici II & 1 & 1.243 & 1.5 & 4.60 & 0.06 & 0.00 & 0.82 & 1.72 & 0.00 & -2.81 \\
Bootes I & 1 & 0.822 & 1.5 & 2.40 & 0.38 & 0.00 & 1.24 & 2.19 & 0.00 & -2.28 \\
Tucana II & 1 & 0.737 & 1.5 & 2.80 & 0.04 & 0.00 & 1.38 & 1.68 & 0.00 & -2.20 \\
Tucana IV & 1 & 0.871 & 1.5 & 4.30 & 0.06 & 0.00 & 1.17 & 1.26 & 0.00 & -2.86 \\
Leo V & 1 & 0.805 & 1.5 & 3.70 & 0.01 & 0.00 & 1.26 & 1.31 & 0.00 & -1.53 \\
And X & 1 & 1.032 & 1.5 & 3.90 & 0.81 & 0.00 & 0.99 & 1.51 & 0.00 & -1.33 \\
And XI & 1 & 1.200 & 1.5 & 4.60 & 0.28 & 0.00 & 0.85 & 1.67 & 0 & -0.88 \\
And XII & 1 & 0.761 & 1.5 & 2.60 & 0.84 & 0.00 & 1.34 & 0.83 & 0.00 & -0.58 \\
And XIII & 1 & 0.986 & 1.5 & 5.80 & 0.16 & 0.00 & 1.03 & 1.62 & 0.00 & -1.43 \\
And XVI & 1 & 1.464 & 1.5 & 3.80 & 0.78 & 0.00 & 0.70 & 2.46 & 0.00 & -0.55 \\
And XVII & 1 & 1.125 & 1.5 & 2.90 & 2.88 & 0.00 & 0.91 & 2.48 & 0.00 & 0.25 \\
And XX & 1 & 1.670 & 1.5 & 7.10 & 0.44 & 0.00 & 0.61 & 1.50 & 0.00 & -1.64 \\
And XXII & 1 & 0.562 & 1.5 & 2.80 & 0.10 & 0.00 & 1.81 & 1.28 & 0.00 & -1.20 \\
And XXVI & 1 & 1.368 & 1.5 & 8.60 & 0.91 & 0.00 & 0.74 & 1.02 & 0.00 & -1.41 \\
\hline
\end{tabular}
\caption{The first 23 rows are dSph galaxies and the lower 25 rows are UFD galaxies. $F$ is the soliton mass fraction, $M= M_S+M_\bullet$ is the total galactic mass, $\sigma=\sigma_{los}$ is the predicted (2D projected) stellar velocity dispersion, $M_{1/2} = M(<r_{1/2})$ is the predicted total mass contained within the stellar half-light radius, $r_{c,S}$ is the soliton characteristic radius and $\sigma_S/\sigma_\ast = \nu_S v_0/(\sqrt{3} \sigma_{los,obs})$ is obtained assuming that the stellar velocities are isotropic (as are the soliton velocities).   The relative errors $\Delta_\sigma$ and $\Delta_M$ quantify observational consistency (as further described in the introduction to this Appendix section). }
\label{T:F1}
\end{table}

\begin{table}
\centering
\begin{tabular}{lccccccccccc}
\hline
\multicolumn{11}{c}{Black hole dominated ($F\approx 0$) 5G predictions with $m_0 = 1.5\times 10^{-22}$, optimising $M$ to minimize $|\Delta_\sigma|$} \\
\hline
Galaxy & $F$ & $M/10^8$ & $m_0/10^{-22}$ & $\sigma$ & $M_{1/2}/10^6$ & $M_\bullet/10^6$ & $r_{c,S}$ & $\sigma_S/\sigma_\ast$ & $\Delta_\sigma$ & $\Delta_M$ \\
  &  & M$_\odot$ & eV/c$^2$ & (km/s) & M$_\odot$ & M$_\odot$ &\,kpc &  &  &  \\
\hline
Tucana & 0.0001 & 0.12478 & 1.5 & 13.3 & 47.8 & 12.48 & 0.31 & 0.06 & 0.00 & 0.07 \\
Cetus & 0.0001 & 0.485 & 1.5 & 11.1 & 48.5 & 48.50 & 0.31 & 0.27 & 0.00 & -1.77 \\
Aquarius & 0.0001 & 0.324 & 1.5 & 10.3 & 32.4 & 32.40 & 0.46 & 0.20 & 0.00 & -0.18 \\
Draco & 0.0001 & 0.121 & 1.5 & 9.00 & 12.1 & 12.10 & 1.24 & 0.09 & 0.00 & -1.68 \\
Leo I & 0.0001 & 0.164 & 1.5 & 9.20 & 16.4 & 16.40 & 0.91 & 0.11 & 0.00 & -1.06 \\
Phoenix & 0.0001 & 0.200 & 1.5 & 9.30 & 20.0 & 20.00 & 0.75 & 0.14 & 0.00 & -1.27 \\
Canes Ventici & 0.0001 & 0.155 & 1.5 & 7.60 & 15.5 & 15.50 & 0.97 & 0.13 & 0.00 & -4.50 \\
Sextans & 0.0001 & 0.411 & 1.5 & 7.90 & 41.1 & 41.10 & 0.36 & 0.33 & 0.00 & -0.04 \\
Crater II & 0.0001 & 0.092 & 1.5 & 2.70 & 9.23 & 9.20 & 1.62 & 0.22 & 0.00 & 1.58 \\
Leo II & 0.0001 & 0.090 & 1.5 & 7.40 & 9.01 & 9.00 & 1.66 & 0.08 & 0.00 & -0.56 \\
Carina & 0.0001 & 0.085 & 1.5 & 6.60 & 8.50 & 8.50 & 1.76 & 0.08 & 0.00 & -1.84 \\
Ursa Minor & 0.0001 & 0.327 & 1.5 & 11.5 & 32.7 & 32.70 & 0.46 & 0.18 & 0.00 & -2.66 \\
Sculptor & 0.0001 & 0.213 & 1.5 & 10.1 & 21.3 & 21.30 & 0.70 & 0.13 & 0.00 & -4.07 \\
And I & 0.0001 & 0.443 & 1.5 & 9.40 & 44.3 & 44.30 & 0.34 & 0.30 & 0.00 & -0.74 \\
And III & 0.0001 & 0.347 & 1.5 & 11.0 & 34.7 & 34.70 & 0.43 & 0.20 & 0.00 & -1.08 \\
And V & 0.0001 & 0.464 & 1.5 & 11.2 & 46.4 & 46.40 & 0.32 & 0.26 & 0.00 & 0.76 \\
And IX & 0.0001 & 0.433 & 1.5 & 10.9 & 43.3 & 43.30 & 0.35 & 0.25 & 0.00 & 0.29 \\
And XIV & 0.0001 & 0.139 & 1.5 & 5.40 & 13.9 & 13.90 & 1.07 & 0.16 & 0.00 & 0.54 \\
And XV & 0.0001 & 0.312 & 1.5 & 11.0 & 31.2 & 31.20 & 0.48 & 0.18 & 0.00 & 0.23 \\
And XVIII & 0.0001 & 0.383 & 1.5 & 9.70 & 38.3 & 38.30 & 0.39 & 0.25 & 0.00 & 0.96 \\
And XXI & 0.0001 & 0.309 & 1.5 & 6.10 & 30.9 & 30.90 & 0.49 & 0.32 & 0.00 & -0.41 \\
And XXIII & 0.0001 & 0.473 & 1.5 & 7.10 & 47.3 & 47.30 & 0.32 & 0.42 & 0.00 & -0.70 \\
And XXV & 0.0001 & 0.055 & 1.5 & 3.00 & 5.45 & 5.50 & 2.75 & 0.12 & 0.00 & 0.31 \\
\hline
Phoenix II & 0.0001 & 0.045 & 1.5 & 11.0 & 4.50 & 4.50 & 3.33 & 0.03 & 0.00 & 0.09 \\
Segue I & 0.0001 & 0.007 & 1.5 & 3.90 & 0.685 & 0.70 & 21.86 & 0.01 & 0.00 & 1.69 \\
Pegasus III & 0.0001 & 0.010 & 1.5 & 5.40 & 1.02 & 1.00 & 14.71 & 0.01 & 0.00 & -0.41 \\
Wilman I & 0.0001 & 0.009 & 1.5 & 4.00 & 0.937 & 0.90 & 16.00 & 0.01 & 0.00 & 2.44 \\
Horoligium I & 0.0001 & 0.008 & 1.5 & 4.90 & 0.832 & 0.80 & 18.01 & 0.01 & 0.00 & -0.26 \\
Pisces II & 0.0001 & 0.015 & 1.5 & 5.40 & 1.46 & 1.50 & 10.28 & 0.02 & 0.00 & -0.21 \\
Coma Berenices & 0.0001 & 0.020 & 1.5 & 4.60 & 1.98 & 2.00 & 7.55 & 0.03 & 0.00 & 1.80 \\
Reticulum II & 0.0001 & 0.004 & 1.5 & 3.22 & 0.373 & 0.40 & 40.12 & 0.01 & 0.00 & -1.23 \\
Hydrus I & 0.0001 & 0.004 & 1.5 & 2.69 & 0.436 & 0.40 & 34.34 & 0.01 & 0.00 & 0.57 \\
Grus I & 0.0001 & 0.026 & 1.5 & 5.40 & 2.62 & 2.60 & 5.72 & 0.03 & 0.00 & 0.44 \\
Leo IV & 0.0001 & 0.009 & 1.5 & 3.40 & 0.875 & 0.90 & 17.11 & 0.02 & 0.00 & -0.74 \\
Canes Ventici II & 0.0001 & 0.008 & 1.5 & 4.60 & 0.847 & 0.80 & 17.70 & 0.01 & 0.00 & -1.14 \\
Bootes I & 0.0001 & 0.004 & 1.5 & 2.40 & 0.353 & 0.40 & 42.44 & 0.01 & 0.00 & -2.35 \\
Tucana II & 0.0001 & 0.008 & 1.5 & 2.80 & 0.815 & 0.80 & 18.38 & 0.02 & 0.00 & -0.16 \\
Tucana IV & 0.0001 & 0.010 & 1.5 & 4.30 & 1.03 & 1.00 & 14.48 & 0.01 & 0.00 & -1.34 \\
Leo V & 0.0001 & 0.006 & 1.5 & 3.70 & 0.570 & 0.60 & 26.26 & 0.01 & 0.00 & -0.29 \\
And X & 0.0001 & 0.014 & 1.5 & 3.90 & 1.42 & 1.40 & 10.56 & 0.02 & 0.00 & -0.95 \\
And XI & 0.0001 & 0.012 & 1.5 & 4.60 & 1.23 & 1.20 & 12.13 & 0.03 & 0 & -0.48 \\
And XII & 0.0001 & 0.006 & 1.5 & 2.60 & 0.645 & 0.60 & 23.22 & 0.01 & 0.00 & -0.68 \\
And XIII & 0.0001 & 0.029 & 1.5 & 5.80 & 2.92 & 2.90 & 5.14 & 0.05 & 0.00 & -0.42 \\
And XVI & 0.0001 & 0.014 & 1.5 & 3.80 & 1.44 & 1.40 & 10.40 & 0.07 & 0.00 & -0.17 \\
And XVII & 0.0001 & 0.010 & 1.5 & 2.90 & 1.04 & 1.00 & 14.40 & 0.07 & 0.00 & -0.57 \\
And XX & 0.0001 & 0.022 & 1.5 & 7.10 & 2.23 & 2.20 & 6.70 & 0.06 & 0.00 & -0.86 \\
And XXII & 0.0001 & 0.009 & 1.5 & 2.80 & 0.888 & 0.90 & 16.87 & 0.06 & 0.00 & -0.60 \\
And XXVI & 0.0001 & 0.035 & 1.5 & 8.60 & 3.53 & 3.50 & 4.25 & 0.08 & 0.00 & -1.02 \\
\hline
\end{tabular}
\caption{The first 23 rows are dSph galaxies and the lower 25 rows are UFD galaxies, $F$ is the soliton mass fraction, $M= M_S+M_\bullet$ is the total galactic mass, $\sigma=\sigma_{los}$ is the predicted (2D projected) stellar velocity dispersion, $M_{1/2} = M(<r_{1/2})$ is the predicted total mass contained within the stellar half-light radius, $r_{c,S}$ is the soliton characteristic radius and $\sigma_S/\sigma_\ast = \nu_S v_0/(\sqrt{3} \sigma_{los,obs})$ is obtained assuming that the stellar velocities are isotropic (as are the soliton velocities).    The relative errors $\Delta_\sigma$ and $\Delta_M$ quantify observational consistency (as further described in the introduction to this Appendix section). These results indicate that the galaxies with $|\Delta_M|\le 1$ have observed values of $\sigma$ and $M_{1/2}$ that are compatible with having a supermassive black hole and little or no dark matter. }
\label{T:F0}
\end{table}

%\bsp
\label{lastpage}

\end{document}